\documentclass[
reprint,
floatfix,
longbibliography,
superscriptaddress,
amsmath,
amssymb,
showkeys,
]{revtex4-2}

\usepackage{array}[=2016-10-06]
\usepackage{floatrow}
\usepackage{threeparttable}
\usepackage{booktabs}
\usepackage{setspace}
\usepackage{amsmath}
\usepackage{graphicx}
\usepackage{hyperref}
\usepackage[mathlines]{lineno}
\usepackage{textcomp}
\usepackage{gensymb}
\usepackage{url}
\usepackage[version=3]{mhchem}
\usepackage{multirow}
\usepackage[american]{babel}
\usepackage{textgreek}
\usepackage{amsfonts,amssymb}
\usepackage{miller}
\usepackage{xcolor}

\definecolor{revisionred}{rgb}{0.00,0.00,0.00}
\newcommand{\revision}[1]{{\color{revisionred}#1}}

\renewcommand{\arraystretch}{1.1}

\definecolor{esired}{rgb}{0.80,0.00,0.00}
\definecolor{missblue}{rgb}{0.00,0.25,0.75}
\definecolor{notegrey}{rgb}{0.40,0.40,0.40}
\newcommand{\esi}[1]{{\color{esired}#1}}

\usepackage{siunitx}
\newcommand{\sifig}[2]{\includegraphics[width=#1]{#2}}

\begin{document}


\title{A General-Purpose and Efficient Machine-Learned Potential for SiC from Ambient to Extreme Environments}

\author{Jintong Wu}
\thanks{These authors contributed equally.}
\affiliation{The Hong Kong Microelectronics Research and Development Institute, Hong Kong, 999077, China}
\affiliation{Department of Physics, University of Helsinki, P.O. Box 43, FI-00014, Finland}

\author{Zhuang Shao}
\thanks{These authors contributed equally.}
\affiliation{School of Energy and Power Engineering, Xi'an Jiaotong University, Xi'an 710049, China}
\affiliation{Department of Physics, University of Helsinki, P.O. Box 43, FI-00014, Finland}

\author{Junlei Zhao}
\email{junlei.zhao@mrdi.org.hk}
\affiliation{The Hong Kong Microelectronics Research and Development Institute, Hong Kong, 999077, China}

\author{Flyura Djurabekova}
\affiliation{Department of Physics, University of Helsinki, P.O. Box 43, FI-00014, Finland}
\affiliation{Helsinki Institute of Physics, University of Helsinki, P.O. Box 43, FI-00014, Finland}

\author{Kai Nordlund}
\affiliation{Department of Physics, University of Helsinki, P.O. Box 43, FI-00014, Finland}

\author{Fredric Granberg}
\affiliation{Department of Physics, University of Helsinki, P.O. Box 43, FI-00014, Finland}

\author{Qingmin Zhang}
\affiliation{School of Energy and Power Engineering, Xi'an Jiaotong University, Xi'an 710049, China}

\author{Jesper Byggm{\"a}star}
\affiliation{Department of Physics, University of Helsinki, P.O. Box 43, FI-00014, Finland}

\date{November 14, 2026}

\begin{abstract}
Silicon carbide (SiC) polymorphs are widely employed as nuclear materials, mechanical components, and wide-bandgap semiconductors.
The rapid advancement of SiC-based applications has been complemented by computational modeling studies, including both \textit{ab initio} and classical atomistic approaches.
In this work, we develop a computationally efficient, general-purpose machine-learned interatomic potential (ML-IAP) capable of multimillion-atom molecular dynamics simulations.
\revision{Building on the tabulated Gaussian approximation potential (tabGAP) framework, we introduce an additional embedding term that captures local atomic coordination, improving accuracy and generality, while retaining high computational efficiency.}
\revision{Using this ML-tabGAP model, we map a broad pressure-temperature phase diagram, simulate the threshold displacement energy distributions for the 2H and 3C polymorphs, and simulate primary irradiation damage in 3C.}
Across a benchmark covering conditions from ambient to extreme, including high-pressure/high-temperature states and high-energy cascade damage, tabGAP provides a favorable balance of accuracy, robustness, transferability, and computational cost among the tested empirical and ML-IAPs.
\end{abstract}

\maketitle


\section{Introduction} \label{sec:intro}

Silicon carbide (SiC) is a wide-bandgap semiconductor ($E_\mathrm{g}\approx2.3$--$3.3$~eV)~\cite{kimoto2014physical}.
As a strongly polytypic material, it has over 250 known polymorphs, mainly polytypes with different stacking sequences of identical close-packed Si-C bilayers~\cite{ramakers2022effects}.
Among these, the 2H, 3C, 4H and 6H polytypes under ambient conditions and the high-pressure rock-salt (RS) structure~\cite{yoshida1993pressure} have attracted intensive research and application interest~\cite{katoh2019silicon, wang2019review, xu2021recent}.
SiC combines ultrahigh hardness, high strength, high thermal conductivity, high-temperature stability, a high breakdown field, a high carrier saturation velocity and a low neutron absorption cross section~\cite{neudeck1995progress, li2019shock, zhou2025silicon}.
These characteristics make SiC useful for nuclear reactor components~\cite{katoh2012radiation, koyanagi2018recent} and cladding~\cite{kocevski2020understanding, koyanagi2020design}, protective armor~\cite{hogg2006composites, wu2023research}, heat-resistant aircraft engine components~\cite{spitsberg2004thermal, zhang2015ionization}, high-power electronics~\cite{eddy2009silicon, nielsen2025high} and quantum information platforms~\cite{castelletto2014silicon, he2024robust, nishikawa2025coherent, zhou2025silicon}.

During fabrication and service, SiC components often encounter high pressure, high temperature or intense irradiation, as well as combinations of these conditions~\cite{guo2014modeling, xu2018high}.
Under high dynamic loads, structural phase transformations and chemical decomposition alter the mechanical strength and failure behavior of SiC~\cite{wang2020radiation, cai2024grain}.
Irradiation generates lattice defects that affect its physical, chemical and mechanical properties, and hence its reliability in radiation environments~\cite{ochedowski2014graphitic, zhang2015ionization, wang2021effect}.
The pressure-induced transition from the hexagonal (2H, 4H, 6H) and zinc-blende (3C) polytypes to rock salt~\cite{yoo1991solid, sekine1997shock, xie2023uncertainty} is also of interest for carbon-rich exoplanets and superhard materials~\cite{daviau2017decomposition}.
At low pressure, the thermal decomposition of SiC provides a route to graphene on an insulating substrate~\cite{jokubavicius2016surface, choi_laser-induced_2016}, although the sequence of atomic processes under nanosecond laser heating remains difficult to observe~\cite{choi_laser-induced_2016}.

Experiments at ultrahigh temperatures and pressures are challenging, particularly when the kinetics of melting and decomposition must be resolved.
Molecular dynamics (MD) can access the relevant temperatures and strain rates on submicrometer length and nanosecond time scales~\cite{zhou2023device}.
The polymorphism, phase transformations and incongruent decomposition of SiC make it an interesting system for atomistic modeling as well as a demanding case for interatomic-potential development~\cite{liu2018distribution,cai2024grain,jiang2025effects,cai2025collision,li2025computational}.
Empirical interatomic potentials (IAPs) have contributed extensively to this work~\cite{tersoff1989modeling, tersoff1994chemical,devanathan1998displacement,gao2002empirical,vashishta2007interaction,jiang2012carbon,kang2014governing}, but their simple functional forms limit the accuracy of some properties.
Degenerate energies of different polytypes and questionable threshold displacement energies (TDEs), for example, affect the interpretation of MD simulations of irradiation in SiC~\cite{lucas2005comparison,samolyuk2015molecular,cowen2018point,li2019threshold,wang2021effect}.

Machine-learned (ML) IAPs trained on density functional theory (DFT) data can describe complex potential energy surfaces at much lower cost than direct DFT calculations~\cite{deringer2021gaussian, lim2023molecular, liu2024deep}.
Several have been developed for SiC~\cite{xie2023uncertainty, lim2023molecular, klawohn2023massively, liu2024deep, macisaac2024genetic, du2024construction, liu2025neural, hamedani2025sic}, often for particular crystal structures or applications.
Given the structural diversity of SiC and its responses under extreme conditions, a general-purpose potential is useful for studying several phases and dynamical processes within one model.

In this work, we develop such a ML-IAP using the tabulated Gaussian approximation potential (tabGAP) framework~\cite{byggmastar2022simple}.
The original tabGAP combines low-dimensional two-body, three-body and embedded-atom density descriptors with sparse Gaussian-process regression of DFT data.
Each contribution is tabulated, allowing energies and forces to be evaluated at close to empirical-IAP speed.
\revision{We extend this description with a second embedded-atom density term based on smooth first-nearest-neighbor (1NN) coordination.
In the conventional scalar density, neighbor number and bond lengths both change the sum of distance-weighted contributions, so different local environments can have the same density.
The added term provides a separate measure of coordination, helping distinguish environments with missing or extra neighbors near defects, in disordered networks and at surfaces, while retaining the simple tabulated form.
The resulting tabGAP offers a favorable balance of accuracy, generality, and computational efficiency, as demonstrated by comparative benchmarks and applications to the SiC pressure-temperature ($P$-$T$) phase diagram, TDE maps in 2H and 3C, and primary irradiation damage in 3C.}


\section{Results} \label{sec:res_dis}

\subsection{Training database} \label{subsec:database}

\begin{figure*}[ht!]
\includegraphics[width=\linewidth]{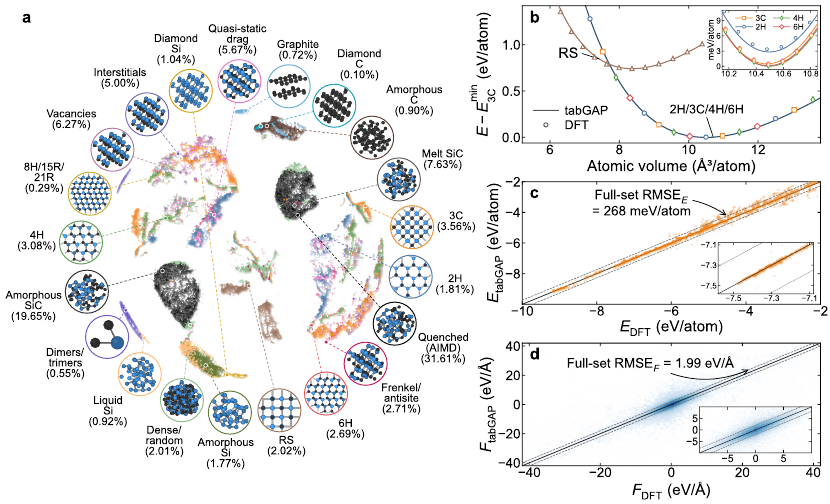}
\caption{\revision{\textbf{Training database and training accuracy of the tabGAP.}
(a) SOAP-based embedding of a stratified sample of training environments, colored by 21 structural classes; the class fractions refer to the full database.
The t-SNE projection~\cite{vandermaaten2008visualizing} (perplexity 400) is redistributed radially and angularly and density-equalized for display.
Isolated Si and C atoms, included as a global energy reference, are not shown.
(b) Equations of state of the 2H, 3C, 4H, 6H and RS polymorphs from DFT (symbols) and tabGAP (lines), relative to the minimum energy of 3C.
Inset: equations of state near the minimum, each method referenced to its own lowest-energy polytype.
(c) Cohesive energies and (d) force components from tabGAP against DFT, displayed over $-10$ to $-2$~eV/atom and $-42$ to $42$~eV/\r A, respectively; insets magnify the low-energy and low-force regions.
Solid lines denote equality and dashed lines $\pm$RMSE in both the main panels and insets.
The full-set RMSEs include all 5393 structures and all 1,018,863 Cartesian force components, including values outside the display windows. Per-class errors are given in \esi{Supplementary Table~S2}.}}
\label{fig:map}
\end{figure*}

The accuracy and range of applications of an ML-IAP depend on the environments represented in its training data.
\revision{Our \textit{ab initio} database contains 5393 structures and 339,621 local atomic environments, classified into 21 structural classes.
Fig.~\ref{fig:map}a visualizes a stratified sample of these atomic environments through a two-dimensional embedding based on smooth overlap of atomic positions (SOAP) descriptors~\cite{soap2013}.
Colors distinguish the structural classes, with the fraction of all database environments in each class shown next to a representative structure.
The database comprises five groups:
(\textit{i})~Crystalline SiC: the 2H, 3C, 4H, 6H and RS polymorphs and the long-period 8H, 15R and 21R polytypes, sampled by equations of state, independent lattice distortions along the three axes, strained cells for elastic constants, and finite-temperature \textit{ab initio} MD (AIMD).
(\textit{ii})~Point defects and recoil paths: 751 structures (14\% of the atomic environments) with vacancies, divacancies, interstitials and Frenkel pairs in 3C and 4H and Frenkel pairs in 2H, including relaxed, rattled and thermally disordered configurations and vacancy relaxation paths; 242 further structures follow quasi-static recoil paths in 2H and 3C (5.7\%).
(\textit{iii})~Disordered SiC: molten (7.6\%), amorphous (19.7\%) and quenched (31.6\%) configurations from AIMD melt-quench runs at densities corresponding to $\pm5\%$ linear strain, spanning hydrostatic pressures from about $+45$ to $-35$~GPa.
(\textit{iv})~Elemental phases: diamond, graphite and amorphous carbon, and diamond, liquid and amorphous silicon, included to describe the correct thermodynamic stability of SiC and the incongruent decomposition of SiC into its constituent phases.
(\textit{v})~High-energy configurations: dimers and trimers sampling short-range repulsion and angular interactions~\cite{byggmastar2019machine,byggmastar2022simple}, together with dense and random structures from random structure searches (2.0\%).
The DFT calculations and tabGAP fitting parameters are described in the Methods and \esi{Supplementary Notes~1 and 2}.}

\subsection{ML-tabGAP validation}
\label{subsec:valid}\label{subsec:eos}\label{subsec:phonons}\label{subsec:defects}\label{subsec:amorph}

Fig.~\ref{fig:map}c,d compares the tabGAP energies and forces with the DFT training data.
\revision{The full-set root-mean-square errors (RMSEs) are 268~meV/atom and 1.99~eV/\r A for the 5393 training structures.
These aggregate values combine near-equilibrium crystals with strongly repulsive configurations and dense extremely high-energy elemental cells contribute most of the squared error.
For 1595 distinct held-out structures, the corresponding RMSEs are 389~meV/atom and 2.78~eV/\r A \esi{(Supplementary Fig.~S3)}.
The energy errors of the held-out equations of state and strained crystals are 1.2--2.3 and 0.5--0.6~meV/atom, respectively, and those of vacancy, interstitial and Frenkel-pair configurations are 11.1--14.7~meV/atom.
Both training and held-out statistics include all configurations and force components, including values outside the displayed ranges. Errors divided by structure class and fitting weights are given in \esi{Supplementary Note~2, Table~S2}.}

\revision{Beyond these error statistics, we assess crystalline properties using energy-volume curves, which provide a static basis for comparing phase stability.
Fig.~\ref{fig:map}b compares the tabGAP curves with DFT from 6 to 13.5~\r A$^3$/atom, covering strong compression and about 30\% expansion, with a mean energy deviation of 11~meV/atom. 
The equilibrium lattice parameters of the four tetrahedral polytypes agree with our DFT within 0.1\%.
At zero pressure, tabGAP reproduces the DFT grouping of the tetrahedral polytypes: 2H lies above 3C, while the nearly degenerate 4H and 6H lie below it.
The internal order of stability for 3C, 4H and 6H is defined by sub-meV/atom energy differences from static calculations (no temperature) in DFT and tabGAP and hence not deemed significant \esi{(Supplementary Note~5 and Fig.~S6)}.
Under compression, the static 3C-RS enthalpy crossing is at 66.3~GPa, compared with 64.1~GPa in the DFT calculation of Ref.~\cite{pertierra2022pressure}.
For small strains around equilibrium, tabGAP also reproduces the 18 independent elastic constants of 2H, 3C, 4H and 6H with a mean deviation from DFT of 7.0\%, decreasing to 5.8\% when the bulk and shear moduli are included.
The largest component deviations, up to 17\%, occur in $C_{12}$ and $C_{13}$ of 2H \esi{(Supplementary Fig.~S7 and Table~S5)}.
}

\begin{figure*}[ht!]
\includegraphics[width=\linewidth]{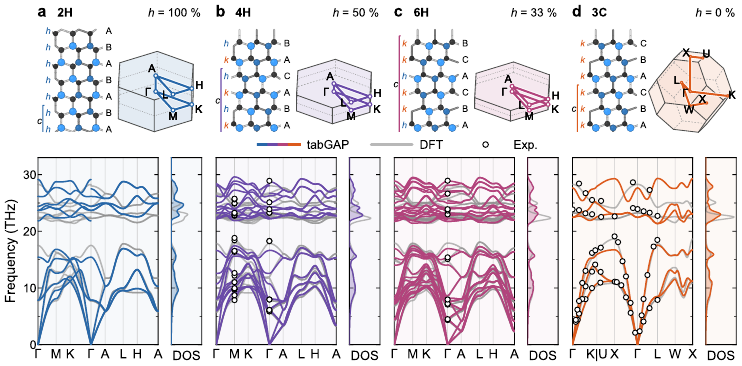}
\caption{\revision{\textbf{Phonons of the SiC polytypes.}
(a) 2H, (b) 4H, (c) 6H and (d) 3C, ordered from fully hexagonal to fully cubic stacking ($h$: fraction of hexagonal sites).
Top: side view of each polytype with its stacking sequence and its hexagonal ($h$) and cubic ($k$) sites, and the Brillouin zone with the band path.
Bottom: phonon dispersions and densities of states from tabGAP (colored lines) and DFT (grey lines), both including the long-range dipole correction. At $\Gamma$, each path segment uses its own directional limit. Circles are inelastic x-ray scattering for 3C~\cite{serrano2002determination}, Raman $\Gamma$ modes for 4H~\cite{feldman1968phonon,rao2023thermooptic} and 6H~\cite{Bacherikov2020}, and $M$-point phonon energies for 4H~\cite{klahold2020band}. The room-temperature Raman data of Rao et al. were measured along the $c$ axis. The $M$-point data were measured at 1.4--2~K, with reported energy uncertainties of approximately $\pm0.6$~meV.}}
\label{fig:phon}
\end{figure*}

\revision{Phonon dispersions further test the near-equilibrium response through collective atomic vibrations.
Fig.~\ref{fig:phon} compares tabGAP with DFT for 2H, 4H, 6H, and 3C, together with inelastic x-ray scattering for 3C~\cite{serrano2002determination}, Raman measurements for 4H~\cite{feldman1968phonon,rao2023thermooptic} and 6H~\cite{Bacherikov2020}, and $M$-point phonon energies for 4H from wavelength-modulated absorption and low-temperature photoluminescence~\cite{klahold2020band}.
The acoustic branches and lower optical manifold broadly follow DFT, including the folded acoustic branches of 4H and 6H.
The main differences remain in the upper optical branches, while the longitudinal acoustic mode of 3C at $X$ is softer than in DFT and experiment \esi{(Supplementary Note~6)}.}

\revision{Both tabGAP and DFT calculations include the long-range dipole interaction through a non-analytical correction based on Born effective charges and dielectric tensors from density functional perturbation theory.
For tabGAP, the dipole blocks inside a fixed 7~\r A matching radius are evaluated directly on the primitive lattice and subtracted before the full Ewald-summed dipole term is added at every wave vector.
The construction and its numerical checks are described in Methods and \esi{Supplementary Note~6}.}

\revision{To assess defect energetics relevant to irradiation damage simulations, we compare 39 vacancy, interstitial and antisite formation energies in 3C, 4H and 6H with our DFT calculations and five other IAPs (\esi{Supplementary Fig.~S11 and Table~S8}).
All values use the Si-rich limit with each method's self-consistent chemical potentials.
tabGAP gives a mean absolute deviation from our DFT results of 0.64~eV, with 34 of the 39 values within 20\%.
For carbon vacancy formation energies in 3C and 4H, the relative deviations from DFT are 20\%--23\%.
The corresponding Frenkel-pair energies at infinite separation, evaluated as vacancy--interstitial sums, have a mean absolute relative deviation of 10.7\% from our DFT results \esi{(Supplementary Note~9)}.
Comparisons of vacancy migration barriers in 3C with DFT are provided in \esi{Supplementary Fig.~S12}.}

\begin{figure*}[ht!]
\includegraphics[width=\linewidth]{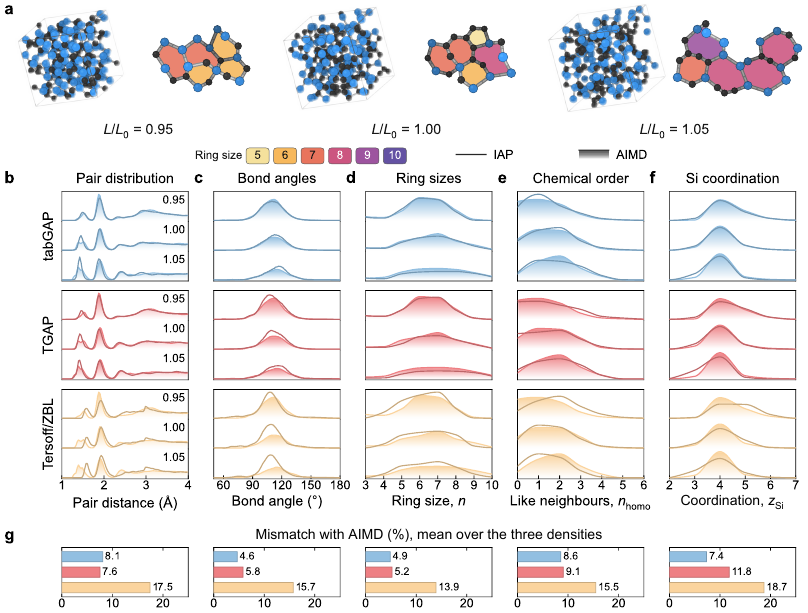}
\caption{\revision{\textbf{Local structure and network order in amorphous SiC predicted by tabGAP, TGAP, and Tersoff/ZBL against AIMD.}
(a) Melt-quenched 216-atom tabGAP cells at $L/L_0 = 0.95$, 1.00 and 1.05. 
Each cell is paired with five connected shortest-path rings from the same snapshot, with faces colored by ring size.
Si atoms are blue and C atoms black.
(b--f) Comparison of tabGAP, TGAP~\cite{hamedani2025sic} and Tersoff/ZBL~\cite{devanathan1998displacement} (lines) with AIMD (shading).
Within each potential block, curves correspond to $L/L_0 = 0.95$, 1.00, and 1.05 from top to bottom.
Columns show (b) RDFs, (c) bond-angle distributions, (d) shortest-path ring-size distributions per atom, $R(n)$, (e) homonuclear-neighbor distributions, $n_{\rm homo}$, and (f) Si coordination distributions, $z_{\rm Si}$.
All distributions use the final 0.6~ps of the quench and a common scale within each column.
The curves in (d--f) are shape-preserving interpolations of discrete averages.
(g) Mismatch with AIMD, $100\,(1-O)$, where $O$ is the overlap of area-normalized distributions, averaged over the three densities. RDF and bond-angle overlaps use common grids over 1.0--4.0~\r A and $50^\circ$--$180^\circ$, respectively, as in Fig.~\ref{fig:benchmark}.
Bond angles use a 2.00~\r A cutoff.
For (d--f), all sources use AIMD-derived bond cutoffs of 2.68~\r A (Si-Si), 2.37~\r A (Si-C) and 1.85~\r A (C-C).
Potential results are averaged over three melt-quench runs and AIMD uses one trajectory per density.}}
\label{fig:am_gr}
\end{figure*}

\revision{Disordered SiC provides a further validation of a general-purpose tabGAP as coordination and chemical order vary throughout the network.
We melt-quenched 216-atom cells at three densities corresponding to $-5\%$, 0 and $+5\%$ linear strain.
Fig.~\ref{fig:am_gr} compares tabGAP, TurboGAP (TGAP)~\cite{hamedani2025sic}, and Tersoff/ZBL~\cite{devanathan1998displacement} with AIMD through local structural and network statistics.
The tabGAP RDFs reproduce the AIMD peak positions and their evolution with density, including the shift and broadening of the second-nearest-neighbor shell on expansion.
RDFs and bond-angle distributions have overlaps of 0.88--0.97 with the corresponding AIMD distributions (Fig.~\ref{fig:am_gr}b,c).}

\revision{Shortest-path rings~\cite{franzblau1991computation}, homonuclear neighbors and Si coordination further characterize network connectivity and chemical order (Fig.~\ref{fig:am_gr}d--f).
Ring distributions peak at six- and seven-membered rings under compression and at seven-membered rings in the unstrained cell.
Expansion increases the homonuclear bond fraction in both tabGAP and AIMD, whereas Tersoff/ZBL predicts the opposite trend and forms almost no three-membered rings.
In the most expanded cell, tabGAP predicts fewer rings and lower Si coordination than AIMD, a tendency shared by TGAP.
Across the five measures, the density-averaged mismatch with AIMD ranges from 4.6\%--8.6\% for tabGAP, compared with 5.2\%--11.8\% for TGAP and 13.9\%--18.7\% for Tersoff/ZBL (Fig.~\ref{fig:am_gr}g).
Together, these validation results demonstrate the ability of ML-tabGAP to describe crystalline, defective and disordered SiC, providing a solid basis for further applications under conditions ranging from ambient to extreme.}

\subsection{Pressure-temperature phase diagram} \label{subsec:phase}

\begin{figure*}[ht!]
\includegraphics[width=\linewidth]{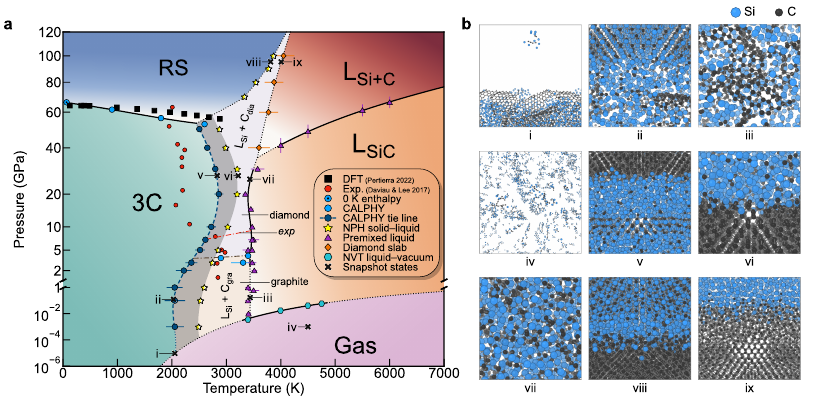}
\caption{\revision{\textbf{Pressure-temperature phase diagram of SiC from tabGAP.}
(a) Phase fields: tetrahedral SiC, RS, a Si-rich liquid coexisting with graphite ($\rm L_{Si}+C_{gra}$) or diamond ($\rm L_{Si}+C_{dia}$), a homogeneous 1:1 liquid ($\rm L_{SiC}$), a liquid demixed into Si-rich and C-rich regions ($\rm L_{Si+C}$), and gas.
Symbols identify \texttt{CALPHY} free-energy calculations, the 0~K 3C-RS enthalpy crossing, two-phase $NPH$ simulations, liquid mixing/demixing and diamond-slab tests, and liquid-vacuum $NVT$ simulations.
Relative stability among the tetrahedral polytypes is not resolved in this diagram.
The blue dashed line marks incongruent decomposition from a common-tangent construction, and the joint grey band extends to the metastable 3C superheating limits from $NPH$ simulations.
The black dash-dotted line is the tabGAP graphite-diamond boundary.
Error bars show method-specific uncertainties or bracketing intervals (Methods and \esi{Supplementary Note~8}).
Dotted segments indicate extrapolated or incompletely constrained boundaries.
The gas line is estimated from evaporation in liquid--vacuum simulations.
References are the DFT-PBE quasi-harmonic 3C-RS boundary~\cite{pertierra2022pressure}, experimental decomposition onsets~\cite{daviau2017decomposition}, and the experimental graphite-diamond line digitized from Fig.~1a of Ref.~\cite{donadio2025metastability} (red dash-dotted line, labeled \textit{exp}).
The pressure axis is logarithmic below 1~GPa and scales as $\sqrt{P}$ above.
(b) Representative tabGAP configurations at the states marked by crosses i--ix in (a); Si atoms are blue and C atoms grey.}}
\label{fig:PT}
\end{figure*}

\revision{Constructing the pressure-temperature ($P$-$T$) phase diagram tests the competition between SiC crystals, carbon phases and silicon-rich liquids, and helps interpret SiC phase transformations under experimentally challenging conditions.
We combine free-energy calculations and direct MD simulations at pressures up to 100~GPa to construct the $P$-$T$ phase diagram in Fig.~\ref{fig:PT}.
Frenkel-Ladd and reversible-scaling calculations with \texttt{CALPHY}~\cite{menon2021automated,freitas2016nonequilibrium} locate solid-solid crossings and incongruent decomposition, while coexistence and liquid simulations constrain melting limits, carbon dissolution and liquid demixing.
The resulting diagram reflects the PBE-DFT-trained tabGAP description.
The boundary construction and uncertainties are given in the Methods and \esi{Supplementary Note~8, Fig.~S10}.}

\revision{\textbf{Low-pressure phase evolution ($<1$~GPa).}
At low pressure, 3C decomposes incongruently into a Si-rich liquid and graphite (snapshot ii), consistent with the established C--Si phase diagram~\cite{olesinski1984c}.
Common-tangent free energies place the decomposition into liquid silicon and graphite at 2051--2055~K ($\pm150$--190~K) between $10^{-3}$~GPa and 1~GPa.
A crystal in contact with a homogeneous 1:1 melt instead survives to 2450--2600~K in two-phase $NPH$ simulations, defining a metastable congruent coexistence branch within the grey band.
These composition-constrained simulations therefore do not locate the equilibrium decomposition boundary.
Both temperature ranges lie below the experimental peritectic value of $2818 \pm 40$~K~\cite{olesinski1984c}.

Further heating dissolves the graphite precipitates at 3400--3500~K, producing a homogeneous liquid (snapshot iii).
Snapshots i and iv illustrate surface evaporation and a gas of atoms and small clusters, respectively.
}

\revision{\textbf{Pressure-dependent decomposition and liquid mixing at medium pressure (1--50~GPa).}
Pressure changes both the decomposition temperature and the form of precipitated carbon.
The decomposition temperature rises to a maximum of 2863~K at 20~GPa, then falls towards the junction with the 3C-RS boundary near 53~GPa and 2455~K.
Compared with the decomposition onsets measured by Daviau and Lee~\cite{daviau2017decomposition}, tabGAP reproduces the negative high-pressure trend but not the sharp drop near 8--10~GPa.
Its boundary lies below the measured onsets below 8~GPa and above them between 10 and 42~GPa.
Experimental onsets depend on the heating protocol and transformation kinetics, whereas the common-tangent construction compares equilibrium free energies.
Hysteresis is also reported in FLARE simulations~\cite{xie_incongruent_2026}, but this distinction alone does not explain the quantitative discrepancies.}

\revision{Pressure also changes the identity of the precipitated carbon.
The graphite-diamond boundary lies at 3.7--4.0~GPa between 2900 and 3400~K, separating graphite-containing mixtures (snapshot ii) from diamond-containing mixtures (snapshots v and vi).
This boundary is set by equality of the chemical potentials of the two carbon phases and lies below the experimental reference shown in Fig.~\ref{fig:PT}a~\cite{donadio2025metastability}.
Between 5 and 30~GPa, diamond dissolves at 3400--3500~K to produce a homogeneous liquid (snapshot vii), distinguishing SiC decomposition from complete carbon dissolution.
At higher pressures, loss of crystalline carbon need not produce a homogeneous liquid: Si-rich and C-rich liquid regions persist, with the demixing boundary rising from about 41~GPa at 4000~K to 66~GPa at 6000~K.
The high-temperature boundary of the diamond-containing field above 30~GPa remains incompletely constrained and is drawn dotted.
The sequence from SiC through a decomposed mixture to a homogeneous liquid is also found in the FLARE study~\cite{xie_incongruent_2026}.
The present diagram further distinguishes the graphite- and diamond-containing fields, carbon dissolution and liquid demixing.}

\revision{\textbf{Ultra-high pressure ($>50$~GPa).}
Compression favors the denser, sixfold-coordinated RS structure over fourfold-coordinated 3C~\cite{kidokoro2017}.
The calculated 3C-RS free-energy crossing decreases from 66.3~GPa at 0~K to 52.8~GPa at 2600~K.
The crossing boundary agrees with the DFT-PBE quasi-harmonic calculation~\cite{pertierra2022pressure} within about 2~GPa below 1000~K, but lies about 4~GPa lower at 1800--2600~K, with a steeper average Clapeyron slope ($-5.2$ versus $-2.6$~MPa/K).
Kidokoro \textit{et al.} measured the transition at 74~GPa and 2100~K and independently calculated a slope of $-4.0$~MPa/K at 2000~K~\cite{kidokoro2017}.
The calculations therefore agree on the negative thermal trend, while both tabGAP and the quasi-harmonic DFT-PBE calculation predict lower transition pressures than that experimental measurement.
At 70--100~GPa, coexistence simulations place upper bounds of approximately 3300--3900~K on RS decomposition into a Si-rich liquid and diamond.
Snapshots viii and ix show representative configurations near the high-pressure boundaries.}

\revision{Together, the $P$-$T$ phase diagram and snapshots distinguish changes in crystal coordination, incongruent decomposition, carbon dissolution and liquid demixing across a broad range of conditions.
We next turn from thermodynamic phase stability to the far-from-equilibrium atomic collisions and defect formation induced by irradiation.}

\subsection{Irradiation damage} \label{subsec:rad}

\begin{figure*}[ht!]
\includegraphics[width=\linewidth]{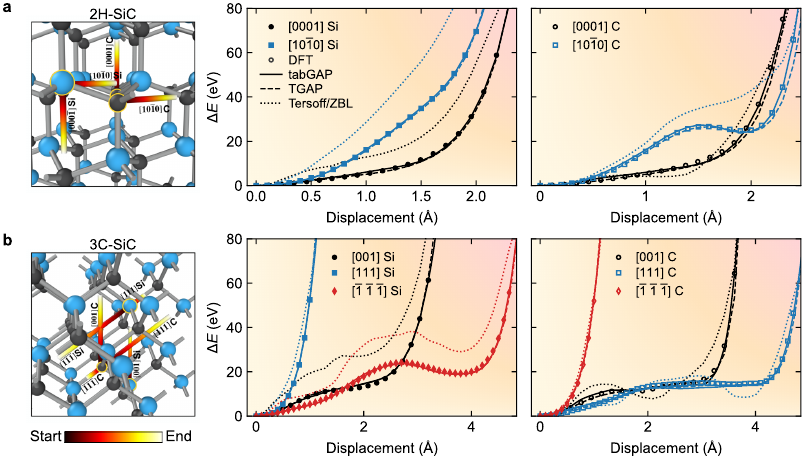}
\caption{\revision{\textbf{Quasi-static recoil paths in 2H- and 3C-SiC.}
Energy change along the path of a displaced Si or C atom in (a) 2H- and (b) 3C-SiC from DFT, tabGAP, TGAP~\cite{hamedani2025sic}, and Tersoff/ZBL~\cite{devanathan1998displacement}.
Colors distinguish the crystallographic directions.
The displacement is measured from the initial lattice site. }}
\label{fig:qsd}
\end{figure*}

\revision{Irradiation-damage simulations require accurate descriptions of both short-range atomic collisions and the defects they create.
We first probe the local energy landscape by displacing individual Si and C atoms along selected directions in 2H and 3C while keeping the surrounding lattice fixed (Fig.~\ref{fig:qsd}).
These paths test direction-dependent changes in bonding and short-range repulsion in local environments represented in the training database.
Across all paths, the mean absolute deviation from DFT is 0.53~eV for tabGAP, compared with 0.77~eV for TGAP and 7.1~eV for Tersoff/ZBL.
Particular improvement over the classical Tersoff/ZBL potential is evident in 3C, as the open Si \hkl[-1-1-1] path exhibits a maximum followed by a secondary minimum, whereas the C \hkl[111] path forms a broad plateau.
tabGAP reproduces these distinct DFT profiles, while Tersoff/ZBL overestimates the Si energies and distorts the C plateau. TGAP shows qualitatively and quantitatively similar results to tabGAP and DFT.}

\begin{figure*}[ht!]
\includegraphics[width=\linewidth]{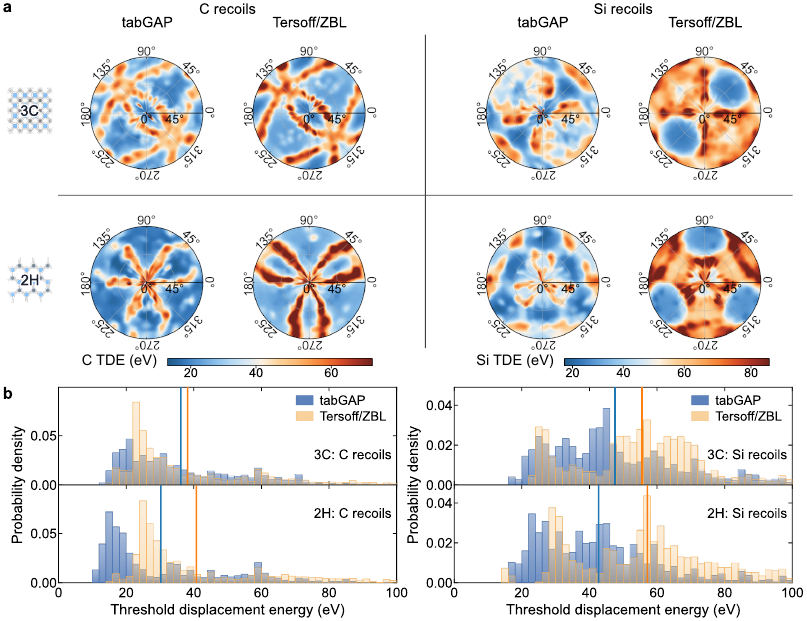}
\caption{\revision{\textbf{Directional dependence of TDEs in 2H- and 3C-SiC.}
(a) TDE maps for C (left) and Si (right) recoils in 3C (top row) and 2H (bottom row).
Each pair compares tabGAP with Tersoff/ZBL~\cite{devanathan1998displacement}.
Radial and angular coordinates denote the polar and azimuthal angles of the recoil direction, respectively.
Maps interpolate resolved directions by Gaussian kernel regression with a width of 3.5\degree.
Color scales span 10--70~eV for C and 15--85~eV for Si and are shared across potentials and polytypes for each recoil species.
(b) Distributions of the TDEs for C (left) and Si (right) recoils in 3C (top) and 2H (bottom), with vertical lines marking the corresponding means over resolved directions.
For C and Si, respectively, both potentials sample 1266 and 1222 directions in 2H and 2023 and 2079 in 3C.
Both IAPs use 30~K initial states, bisection and a Wigner-Seitz defect criterion (Methods). Histograms are normalized over the displayed range.}}
\label{fig:tde}
\end{figure*}

\revision{Recoil dynamics and subsequent recombination determine whether an atomic displacement leaves a surviving defect.
The resulting TDE depends strongly on the recoil species and direction, as established by \textit{ab initio} calculations~\cite{lucas2005ab, gao2011initio}.
We mapped Si and C TDEs over 1200--2100 directions per map in 2H and 3C (Fig.~\ref{fig:tde}), using trajectories initialized at 30~K and a Wigner-Seitz surviving-defect criterion.
The mean TDEs are higher for Si than for C: 42.7~eV \textit{vs.} 30.3~eV in 2H and 47.6~eV \textit{vs.} 36.2~eV in 3C.
As shown in Fig.~\ref{fig:tde}a, C recoils along close-packed atomic rows require higher TDEs to produce surviving defects \esi{(see also Supplementary Note~9 and Table~S9)}.
For Si recoils in 3C, tabGAP predicts alternating high- and low-threshold sectors. In both polytypes, Tersoff/ZBL gives higher mean thresholds, particularly for Si recoils (Fig.~\ref{fig:tde}b).
The maps compare directional patterns under the same simulation protocol (Methods).}

\revision{Low-index thresholds provide a complementary comparison with published DFT results \esi{(Supplementary Note~9 and Table~S9)}.
Across the six directions reported by Lucas and Pizzagalli~\cite{lucas2005ab}, tabGAP differs from DFT by less than 12\% along five directions and by 30\% along Si\hkl[-1-1-1], although we note that the exact simulation protocols differ and TDEs can be sensitive to the simulation methods.
For Si\hkl[100], tabGAP gives 42~eV, compared with the experimental estimates of 23, 25, and 45~eV~\cite{geiczy1971luminescence, honstvet1980determination, lefevre2009silicon} and the 46~eV DFT value of Lucas and Pizzagalli~\cite{lucas2005ab}. For direct comparison to our DFT, we selected the C\hkl[-1-1-1] direction in 3C and performed AIMD simulations. A surviving defect first appears at 34~eV in our AIMD. With 100 thermal seeds per energy, tabGAP gives 50\% survival thresholds of 37.6--37.9~eV across 64-, 216- and 1000-atom cells within the same 640~fs simulation time.
Both calculations exceed the experimental estimate of 22~eV~\cite{geiczy1971luminescence, honstvet1980determination, lefevre2009silicon}, but are consistent with the DFT value 38 eV from~\cite{lucas2005ab}.}

\revision{To assess how individual displacements develop into cascade damage, we next simulated 360 collision cascades in 3C-SiC with Si and C primary knock-on atoms from 0.1 to 2~keV in random directions.
Averaged over C and Si primary knock-on atoms, the mean number of surviving Frenkel pairs increases approximately linearly with recoil energy, from 1.4 at 0.1~keV to 28 at 2~keV \esi{(Supplementary Table~S11)}.
C vacancies outnumber Si vacancies by a factor of 2.1--2.9 over this range, consistent with the lower C TDEs.
The surviving populations also depend on recombination during cascade cooling, whose sensitivity to vacancy-interstitial recombination barriers has been demonstrated in earlier SiC simulations~\cite{samolyuk2015molecular}.
Therefore, these cascades probe the combined effects of defect production and recombination on the surviving damage.}

\section{Discussion}

\begin{figure*}[ht!]
\includegraphics[width=\linewidth]{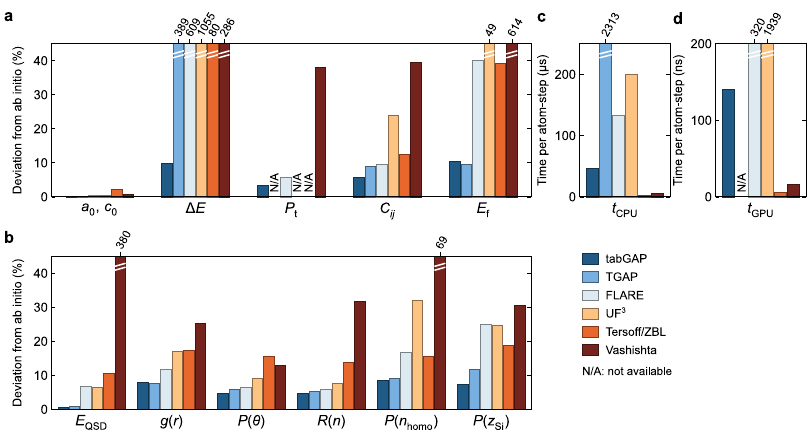}
\caption{\revision{\textbf{Accuracy and computational cost of tabGAP and five other SiC IAPs.}
(a,b)~Mean deviations from \textit{ab initio} references across eleven property categories.
In (a): $a_0$, $c_0$, lattice constants; $\Delta E$, energies of 2H and RS relative to 3C; $P_\mathrm{t}$, the $\rm 3C \leftrightarrow RS$ transition pressure; $C_{ij}$, elastic constants together with bulk and shear moduli; $E_\mathrm{f}$, point-defect formation energies.
In (b): $E_\mathrm{QSD}$, quasi-static recoil-path energies; $g(r)$, RDFs for melt-quenched amorphous SiC; $P(\theta)$, bond-angle distributions; $R(n)$, numbers of shortest-path rings of size $n=3$--10 per atom; $P(n_\mathrm{homo})$, distributions of the number of homonuclear neighbors per atom; $P(z_\mathrm{Si})$, Si coordination-number distributions.
Scalar deviations are $|x-x_\mathrm{ref}|/|x_\mathrm{ref}|$; recoil-path deviations are mean absolute errors normalized by the DFT energy rise; distribution mismatches are $1-O$, where $O$ is the overlap of the area-normalized model and AIMD distributions.
Bars exceeding the plotting limits are broken and labeled with their values.
N/A denotes unavailable results (Supplementary Note~11).
(c,d) Time per atom-step on one CPU core, in $\mu$s (c), and one GPU, in ns (d); benchmarking details are given in Supplementary Fig.~S14.
The DFT-PBE $\rm 3C \leftrightarrow RS$ transition pressure is from Ref.~\cite{pertierra2022pressure}.}}
\label{fig:benchmark}
\end{figure*}

\revision{Across the properties tested here, ML-tabGAP provides a favorable balance of accuracy, generality and computational efficiency.
Fig.~\ref{fig:benchmark} summarizes 100 quantities in 11 categories spanning crystalline properties, amorphous structure and irradiation response.
Among the available results, tabGAP has the smallest mean deviation from the \textit{ab initio} reference in 9 categories, including lattice constants, polytype energy differences, recoil paths and most amorphous-network measures.
TGAP gives slightly smaller deviations for defect formation energies and amorphous radial distribution functions.
The category definitions and individual comparisons are provided in the Methods and \esi{Supplementary Note~11 and Table~S10}.}

\revision{This accuracy is achieved at a substantially lower computational cost than the other tested ML-IAPs.
On one CPU core (Intel Xeon Platinum 8358P), tabGAP requires 47~$\mu$s per atom-step, making it approximately four times faster than UF$^{3}$~\cite{macisaac2024genetic}, three times faster than FLARE~\cite{xie_incongruent_2026} and fifty times faster than TGAP~\cite{hamedani2025sic}.
On one GPU (NVIDIA A800), it requires 0.14~$\mu$s per atom-step, approximately 2.3 times faster than FLARE and fourteen times faster than UF$^{3}$.
For a 64,000-atom cell, tabGAP retains 82\% parallel efficiency on 128 CPU cores \esi{(Supplementary Note~10 and Fig.~S14)}.
Although the empirical IAPs remain faster, tabGAP combines lower deviations across a broad range of properties with a cost suitable for large-scale simulations.
This balance of accuracy, generality and computational efficiency supports larger-scale simulations of phase transformations, decomposition and irradiation damage, including the coupled effects of thermal loading, mechanical loading and irradiation.}

\revision{Notably, the additional embedded-atom term introduced in this work describes 1NN-shell coordination separately from the conventional distance-weighted density, retaining the low-dimensional, tabulated representation~\cite{byggmastar2022simple}.
Explicitly resolving 1NN-shell coordination provides a physically motivated representation of local bonding environments in semiconductors and insulators dominated by covalent or ionic bonding.}

\revision{The mapped phase diagram distinguishes the decomposition of SiC from subsequent C dissolution and liquid mixing, clarifying the relationship between the resulting multiphase states~\cite{daviau2017decomposition, xie_incongruent_2026}.
Free-energy crossings locate equilibrium phase stability, while transformation onsets during heating also depend on pathways and timescales.
The small free-energy differences among tetrahedral polytypes require a careful uncertainty assessment (including DFT accuracy), their relative finite-temperature stability is therefore not resolved here~\cite{ramakers2022effects}.
Under irradiation, the surviving damage reflects both initial atomic displacements and recombination during cascade cooling~\cite{samolyuk2015molecular}.
Defect formation energies, displacement thresholds and recombination barriers therefore are all aspects of damage production that should be well reproduced by the interatomic potential.
Quantitative studies of long-term defect accumulation will require further validation of recombination and clustering pathways. 
In summary, the presented tabGAP provides an efficient and versatile tool for large-scale simulations of SiC across relevant phases and under extreme conditions (temperatures, pressures, irradiation).}


\section{Methods}

\textbf{DFT calculations.}
DFT calculations were performed with \texttt{VASP}~\cite{kresse1993ab, hafner2008ab}, the projector augmented-wave method~\cite{vasp1996} and the PBE generalized gradient approximation~\cite{perdew1996generalized}.
\revision{The plane-wave cutoff was 900~eV. Electronic convergence thresholds were $10^{-6}$~eV for training-data calculations and $10^{-8}$~eV for phonon and selected equilibrium-reference calculations. The force criterion for training-data ionic relaxations was 0.02~eV/\r A.
For the training database, Brillouin-zone sampling used $\Gamma$-centered meshes with $N_i=\max[1,\mathrm{round}(22~\text{\AA}\,|\mathbf{b}_i|)]$ points along each reciprocal lattice vector $\mathbf{b}_i$ defined without $2\pi$, corresponding to a spacing of about 0.29~\r A$^{-1}$ in the reciprocal-space convention including $2\pi$.
The 216-atom AIMD cells used $\Gamma$ only. Denser meshes are used for phonons and specified in \esi{Supplementary Note~6}.
Convergence tests are given in \esi{Supplementary Note~1}.}

\textbf{tabGAP training and tabulation.}
The GAP was trained with \texttt{QUIP}/\texttt{GAP}~\cite{bartok2010gaussian} and tabulated for use in \texttt{LAMMPS}~\cite{thompson2022lammps}.
\revision{As element-specific descriptors we use pair distances and an embedded-atom density with a 5.0~\r A cutoff, three-body clusters with a 4.1~\r A cutoff, and a first-shell coordination density, all supported by the standard tabGAP framework~\cite{byggmastar2022simple}.
The embedding contribution is $F_1(\rho)+F_2(\mathrm{CN})$, with separate functions of the distance-weighted density $\rho$ and coordination CN.
The coordination sums a neighbor-counting function that is unity below 2.40~\r A and switches smoothly to zero at 3.00~\r A.
The embedding energy functions $F_1$ and $F_2$ are optimised together with the other descriptor terms in a standard multi-descriptor GAP fit~\cite{bartok2010gaussian} (see Supplementary Note 2). The $F_2$ term is further split into a wide and a local Gaussian-process term.
Kernel parameters, sparse points and structure-class-dependent regularization are specified in \esi{Supplementary Tables~S1 and S2}.
The model is fitted on top of a screened Coulomb pairwise repulsion derived from all-electron data~\cite{byggmastar2019machine}.}

\revision{The pair terms and pair densities are tabulated versus distance, the two embedding functions versus their respective densities, and the three-body terms on $[r_{ij},r_{ik},\cos\theta_{jik}]$ grids~\cite{byggmastar2022simple}.
Energies and forces are evaluated by cubic splines, with the pair and embedding terms in two \texttt{eam/fs} files and the three-body terms in a \texttt{tabgap} file for \texttt{LAMMPS}. More details on the fitting, descriptors, and tabulation are given in Supplementary Notes 2--4.}

\textbf{Energy and force tests.}
\revision{We removed 653 training/validation overlaps and 13 repeated geometries from 2261 DFT structures, leaving 1595 held-out configurations. Matching used exact geometry hashes and a permutation- and periodicity-invariant RMS displacement tolerance of 0.05~\r A.
The set includes over-coordinated cells, elemental and hypothetical high-pressure phases, quenched amorphous cells and 100,000~K AIMD snapshots. Energy RMSE weights per-atom structure residuals equally; force RMSE weights Cartesian components equally. All retained configurations are included, without fitting-regularization weights \esi{(Supplementary Note~2)}.}

\textbf{Point defects.}
\revision{Formation energies are $E_\mathrm{f}=E_\mathrm{def}-E_\mathrm{bulk}-\sum_i\Delta n_i\mu_i$.
In the Si-rich limit, $\mu_\mathrm{Si}$ is the energy per atom of diamond Si and $\mu_\mathrm{C}=E_\mathrm{SiC}-\mu_\mathrm{Si}$, where $E_\mathrm{SiC}$ is the energy per SiC formula unit.
Each method uses its own bulk and elemental energies.
DFT used 64-atom (3C), 72-atom (4H) and 108-atom (6H) supercells.
tabGAP used the same reference supercell sizes.
DFT and tabGAP cells were fixed at their equilibrium lattice parameters, and ions relaxed to 0.02~eV/\r A.
The defect configurations and formation energies are listed in \esi{Supplementary Note~9}.}

\textbf{Phonons.}
\revision{Force constants were computed with \texttt{phonopy}~\cite{togo2023first} in 64--1728-atom tabGAP and 64--216-atom DFT supercells.
Finite displacements of 0.01~\r A were used, except for the 2H tabGAP result in Fig.~\ref{fig:phon}, for which a displacement-size check led to $9.765625\times10^{-6}$~\r A.
Born effective charges and dielectric tensors were computed by density functional perturbation theory for each polytype. The Born tensors were mapped to the primitive-cell sites, symmetrized over equivalent sites and constrained by charge neutrality.
The force constants were symmetrized under the primitive space group, permutation symmetry and the acoustic sum rule.
For DFT, the dipole term was folded on the actual force-constant supercell, subtracted before short-range interpolation and added back at each wave vector~\cite{gonze1997dynamical}.
For tabGAP, the near-field dipole blocks were evaluated directly on the primitive lattice inside a fixed 7~\r A matching radius, with their self terms set by the acoustic sum rule, and subtracted before adding the full real- and reciprocal-space Ewald sum \esi{(Supplementary Note~6)}.}

\textbf{Amorphous structures.}
\revision{Starting from 216-atom 3C cells, fixed-volume $NVT$ runs at $L/L_0=0.95$, 1.00 and 1.05 used the AIMD reference length $L_0$. Cells were melted at 5500~K, equilibrated at 3000~K and quenched to 300~K in 13.5~ps, with three runs per box factor and potential.
Tersoff/ZBL was melted at 6500~K because the crystal persists at 5500~K.
The same protocol was applied to FLARE, Vashishta and UF$^{3}$ for Fig.~\ref{fig:benchmark}, with AIMD melt-quench trajectories as the reference.
RDFs and angles between bonds shorter than 2.00~\r A were averaged over the final 0.6~ps of the quench.
King shortest-path rings of 3--10 members~\cite{king1967ring,franzblau1991computation}, homonuclear bond fractions and the distributions of homonuclear neighbors and Si coordination were evaluated on 300 frames from that window, using the bond cutoffs in the caption of Fig.~\ref{fig:am_gr}.}

\textbf{Phase diagram.}
\revision{We calculated Gibbs energies with \texttt{CALPHY}~\cite{menon2021automated}, using Frenkel--Ladd integration for solids~\cite{freitas2016nonequilibrium} and Uhlenbeck--Ford integration for liquids~\cite{paula2016the}.
The 3C--RS boundary follows the Gibbs-energy crossing. For decomposition, we compared 3C with the common-tangent mixture of carbon and a silicon-rich liquid at the overall 1:1 composition.
Reversible scaling used a pressure of $\lambda P$ and was checked against independent fixed-temperature integrations. Cell sizes and integration settings are given in \esi{Supplementary Note~8}.}

\revision{Two-phase $NPH$ simulations with pre-existing 3C--liquid interfaces followed the metastable congruent melting edge.
Carbon dissolution was estimated from the liquid composition and persistence of graphite or diamond in phase-separated 1:1 liquids. Diamond--liquid-Si slabs and premixed liquids were used to constrain the high-pressure diamond and demixing regions.
Liquid--vacuum $NVT$ simulations gave an evaporation estimate from atom counts, $P_\mathrm{atom}$, fitted to $\ln(P_\mathrm{atom}/\mathrm{bar})=13.05-31{,}644~\mathrm{K}/T$. The protocols, run lengths and temperature or pressure brackets are given in \esi{Supplementary Note~8}.}

\textbf{Irradiation damage.}
\revision{Quasi-static recoil paths used the DFT configurations of 108-atom 2H and 64-atom 3C cells.
The primary atom was displaced by 0.10--0.19~\r A per step while all other atoms remained at their perfect-crystal positions.
The tabGAP and Tersoff/ZBL TDE maps used 1000-atom 3C and 960-atom 2H cells at 30~K, with 51 thermal snapshots separated by 0.5~ps.
The atom closest to the cell center received the recoil, followed by 1~ps in $NVE$ with an adaptive step limiting displacements to 0.02~\r A, a conjugate-gradient quench and Wigner-Seitz analysis.
A defect survives when a lattice site is empty.
Bisection between 5 and 110~eV gives 1~eV resolution, although we note that for nonmonotonic survival it can return a higher energy than a computationally expensive brute-force scan over all energies, so the TDE maps mainly characterize directional patterns.
}

\revision{Statistical C\hkl[-1-1-1] scans for specific directional TDEs used 64-, 216- and 1000-atom cells at 300~K, 100 seeds per energy and 10~ps trajectories.
The recoil trajectories were integrated in the $NVE$ ensemble with an adaptive timestep capped at 1~fs and a maximum atomic displacement of 0.0025~\r A per step.
The Si\hkl[111], Si\hkl[-1-1-1] and C\hkl[111] scans used 216-atom cells with 100 seeds per energy, C\hkl[100] used a 1000-atom cell with 50 seeds per energy to avoid replacement sequences passing through the periodic boundary.
The TDE for these directional scans was defined as the half height of logistic fits to the surviving-defect fraction (probability), using the first rising branch when survival is nonmonotonic.
AIMD used 64- and 216-atom cells with $2\times2\times2$ and $\Gamma$-only sampling, respectively, one trajectory per energy in 4~eV steps from the perfect lattice, 640~fs trajectories and the same Wigner-Seitz criterion.}

\revision{Cascades were simulated at 300~K in 2744--39,304-atom cells for 0.1, 0.25, 0.5, 1 and 2~keV recoils, using each potential's 300~K lattice parameter.
60 random directions per primary knock-on atom element were simulated at 0.1~keV and 30 at each higher energy.
The cascade core evolved in $NVE$ with a Langevin thermostat on a 3~\r A border of the cell and an adaptive time step limiting displacements per time step to 0.021~\r A.
After 20~ps of cascade evolution, all atoms were coupled to a 300~K Langevin thermostat for 2~ps at fixed volume. The cells were then quenched by the conjugate-gradient method and analyzed by Wigner--Seitz occupancy.
Electronic stopping was not included.}

\textbf{Benchmark.}
\revision{For each category in Fig.~\ref{fig:benchmark}a,b, the reported deviation is the mean of $|x-x_\mathrm{ref}|/|x_\mathrm{ref}|$ for scalar quantities, $1-O$ for distributions, with $O$ their area-normalized overlap with AIMD, or the recoil-path mean absolute energy deviation divided by the DFT energy rise.
The 100 quantities comprise lattice constants (7), 2H and RS energies relative to 3C (2), the 3C-RS pressure (1), elastic constants and moduli (26), defect formation energies (39), recoil paths (10), and five amorphous distributions at three densities (15).
The scalar data and references are given in Supplementary Tables~S5 and S8 and Ref.~\cite{pertierra2022pressure}.
The amorphous distributions use the protocol above.

CPU and GPU times per atom-step are from \esi{Supplementary Fig.~S14}.
All values, comparison sources and category definitions are given in \esi{Supplementary Note~11 and Table~S10}.}

\section*{Data Availability}
The tabGAP files and the data underlying the figures will be made available in the figshare repository upon publication.
Additional data are available from the corresponding authors upon request.

\section*{Code Availability}
The codes used in this work are \texttt{LAMMPS}, \texttt{VASP}, \texttt{OVITO}, \texttt{Phonopy}, \texttt{CALPHY} and \texttt{QUIP}/\texttt{GAP}, which are available from their developers; \revision{the tabulation scripts, including the coordination embedding and the three-body correction, will be deposited with the data.}

%


\section*{Acknowledgments}
J.Z. acknowledges the Shenzhen Fundamental Research Program under Grant No. JCYJ20240813094508011.
F.D. acknowledges the SPATEC project (Grant No. 349690).
J.W. and F.G. acknowledge the DEVHIS project (Grant No. 340538) of the Research Council of Finland for financial support.
J.B. was supported by funding from the Research Council of Finland through the OCRAMLIP project (Grant No. 354234).
This work has partially been carried out within the framework of the EUROfusion Consortium, funded by the European Union via the Euratom Research and Training Programme (Grant Agreement No. 101052200 --- EUROfusion).
Views and opinions expressed are however those of the author(s) only and do not necessarily reflect those of the European Union or the European Commission. Neither the European Union nor the European Commission can be held responsible for them.
Computer time granted by the IT Center for Science -- CSC -- Finland is gratefully acknowledged.
J.Z. thanks Dr. Shouhang Li at MRDI for helpful discussions on the phonon calculations.

\section*{Author Contributions}
J.W. and Z.S. contributed equally to this work.
\revision{J.W., J.Z., and J.B. conceived the study and designed the methodology.}
J.W. and J.Z. developed the core DFT database.
\revision{J.W. fitted the tabGAP model.}
\revision{J.W. and J.B. refined the ML-IAP and evaluated its accuracy, reliability and performance.}
J.W. and Z.S. conducted and analyzed the MD simulations presented in this work.
J.W. and Z.S. prepared the first draft of the manuscript with input from J.Z. and J.B.
F.D., K.N., F.G., and Q.Z. revised the final version of the manuscript.
All authors contributed to the scientific discussion throughout the study.
All authors reviewed and approved the manuscript.
\revision{J.Z., K.N., F.G., F.D., Q.Z., and J.B. managed their parts of the project and contributed to funding acquisition.}

\section*{Competing Interests}
The authors declare no competing interests.

\section*{Disclosure of AI use}

Agentic AI workflows with careful human supervision and instructions were used for automating the potential fitting and validation, including testing the optimal set of descriptors and hyperparameters (sparse points, coordination-counting descriptor). All results are reproducible and verifiable (see Data and Code Availability) and were carefully reviewed and refined by the authors.

\clearpage
\onecolumngrid
\normalsize
\setstretch{1.10}
\makeatletter
\def\table@hook{\normalsize}
\def\ps@SIPageStyle{%
  \def\@oddhead{\hfil\thepage}%
  \let\@evenhead\@oddhead
  \let\@oddfoot\@empty
  \let\@evenfoot\@empty
}
\makeatother
\pagestyle{SIPageStyle}
\setcounter{page}{1}
\renewcommand{\thepage}{S\arabic{page}}
\renewcommand{\citenumfont}[1]{S#1}
\renewcommand{\bibnumfmt}[1]{[S#1]}
\setcounter{figure}{0}
\setcounter{table}{0}
\setcounter{equation}{0}
\renewcommand{\thefigure}{S\arabic{figure}}
\renewcommand{\thetable}{S\arabic{table}}
\renewcommand{\theequation}{S\arabic{equation}}
\renewcommand{\theHfigure}{SI.\arabic{figure}}
\renewcommand{\theHtable}{SI.\arabic{table}}
\renewcommand{\theHequation}{SI.\arabic{equation}}
\renewcommand{\bibsection}{\subsection*{References in Supplementary Information}}
\definecolor{red}{rgb}{0.0,0.0,0.0}

\begingroup
\centering
{\Large\bfseries Supplementary Materials:\\[0.35em]
A General-Purpose and Efficient Machine-Learned Potential for SiC from Ambient to Extreme Environments\par}
\vspace{1.1em}
{\normalsize
Jintong Wu\textsuperscript{1,2,*},
Zhuang Shao\textsuperscript{3,2,*},
Junlei Zhao\textsuperscript{1,\textdagger},
Flyura Djurabekova\textsuperscript{2,4},
Kai Nordlund\textsuperscript{2},
Fredric Granberg\textsuperscript{2},
Qingmin Zhang\textsuperscript{3},
Jesper Byggm{\"a}star\textsuperscript{2}\par}
\vspace{0.8em}
{\small
\textsuperscript{1}The Hong Kong Microelectronics Research and Development Institute, Hong Kong, 999077, China\par
\textsuperscript{2}Department of Physics, University of Helsinki, P.O. Box 43, FI-00014, Finland\par
\textsuperscript{3}School of Energy and Power Engineering, Xi'an Jiaotong University, Xi'an 710049, China\par
\textsuperscript{4}Helsinki Institute of Physics, University of Helsinki, P.O. Box 43, FI-00014, Finland\par}
\vspace{0.45em}
{\textsuperscript{*}These authors contributed equally.\par
\textsuperscript{\textdagger}\href{mailto:junlei.zhao@mrdi.org.hk}{junlei.zhao@mrdi.org.hk}\par}
\endgroup

\vspace{0.9em}
\begin{center}
{\bfseries CONTENTS}
\end{center}
\begingroup
\normalsize
\newcommand{\SIcontentsline}[2]{\noindent #1\nobreak\dotfill\nobreak\hbox{#2}\par}
\SIcontentsline{Supplementary Note 1: Convergence test of DFT calculations}{\pageref{SuppNote_1}}
\SIcontentsline{Supplementary Note 2: Training details of the tabGAP}{\pageref{SuppNote_2}}
\SIcontentsline{Supplementary Note 3: High-energy repulsive potential}{\pageref{SuppNote_3}}
\SIcontentsline{Supplementary Note 4: \revision{Energy terms and descriptor coverage}}{\pageref{SuppNote_4}}
\SIcontentsline{Supplementary Note 5: Fundamental physical properties}{\pageref{SuppNote_5}}
\SIcontentsline{Supplementary Note 6: \revision{Phonons and the non-analytical correction for a short-ranged potential}}{\pageref{SuppNote_6}}
\SIcontentsline{Supplementary Note 7: Robustness of tabGAP in the ultrahigh-temperature regime}{\pageref{SuppNote_7}}
\SIcontentsline{Supplementary Note 8: Phase diagram calculation}{\pageref{SuppNote_8}}
\SIcontentsline{Supplementary Note 9: \revision{Point defects and threshold displacement energies}}{\pageref{SuppNote_9}}
\SIcontentsline{Supplementary Note 10: Computational speed}{\pageref{SuppNote_10}}
\SIcontentsline{Supplementary Note 11: \revision{Comparison of accuracy and computational cost}}{\pageref{SuppNote_11}}

\vspace{0.9em}
\SIcontentsline{References in Supplementary Information}{\pageref{SIReferences}}
\endgroup
\clearpage

\subsection*{Supplementary Note 1: Convergence test of DFT calculations} \label{SuppNote_1}

\begin{figure}[ht!]
\sifig{10cm}{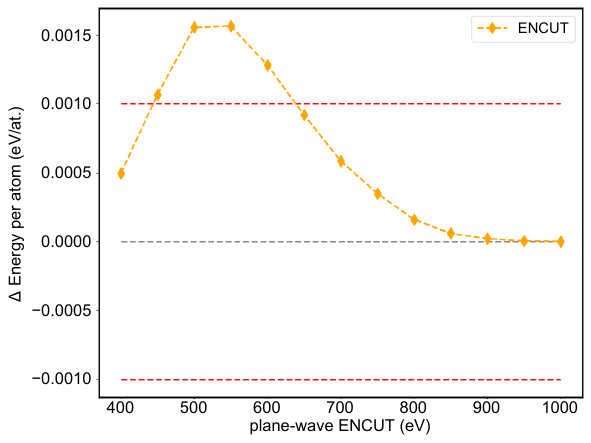}
\caption{Total energy convergence tests on plane-wave kinetic energy cutoff, ``ENCUT''.
 The zero-energy lines are aligned for the $\mathrm{ENCUT}=1000$~eV and $12\times12\times12$ $k$-mesh grid to emphasize the energy difference per atom.
 The red dashed lines are drawn at $\pm10^{-3}$~eV/atom as a threshold of energy convergence.}
\label{sfig:ENCUT}
\end{figure}

The plane-wave cutoff convergence was tested for cubic 3C-SiC (Fig.~\ref{sfig:ENCUT}).
Below 650~eV the cutoff is the main source of the energy variation.
\revision{We use a cutoff of 900~eV for the DFT calculations.}

\subsection*{Supplementary Note 2: Training details of the tabGAP} \label{SuppNote_2}

Fitting a tabGAP consists of two steps: (\textit{i}) training a GAP with low-dimensional descriptors, and (\textit{ii}) creating a tabulated version for fast evaluation with negligible loss in accuracy~\cite{SI:byggmastar2022simple}.
GAPs can be trained with multiple descriptors as a single model using Gaussian process regression.
The low-dimensional two-body (2b), three-body (3b) and embedded-atom (EAM) descriptors allow the fitted energy to be tabulated for large-scale MD~\cite{SI:byggmastar2022simple}.
They describe bond distances, the geometry of triplets and the local density, respectively.
The 2b descriptor, $\mathbf{q}_{\mathrm{2b}}$, is the interatomic distance between the central atom $i$ and its neighbor $j$:
\begin{equation} \label{Eq:S1}
\mathbf{q}_{i,\mathrm{2b}} = r_{ij}.
\end{equation}
The 3b descriptor, $\mathbf{q}_{\mathrm{3b}}$, is a three-component permutation-invariant vector of the three interatomic distances, containing all information about bond lengths and bond angles:
\begin{equation} \label{Eq:S2}
\mathbf{q}_{i,\mathrm{3b}} = [r_{ik} + r_{ij},\, (r_{ik} - r_{ij})^2,\, r_{jk}].
\end{equation}
The EAM descriptors use a scalar density to represent the local many-body environment. In the generalized many-element formulation (see Refs.~\cite{SI:byggmastar2022simple,SI:foiles1986embedded}),
\begin{equation} \label{Eq:S3}
\mathbf{q}_{i,\mathrm{EAM}} = \sum_{j \ne i} \kappa_{ij} \phi(r_{ij}),
\end{equation}
where $\kappa_{ij}$ is a prefactor that depends on the chemical species of $i$ and $j$ and $\phi(r_{ij})$ is a scalar pair-density function.
\revision{The present potential uses two EAM-type terms.
The first, as in Ref.~\cite{SI:byggmastar2022simple}, uses a Finnis-Sinclair-like pair density with a range of 5.0~\r A.
The second is a short-range coordination term, $\mathbf{q}_{i,\mathrm{CN}} = \sum_{j \ne i} \kappa_{ij} f_\mathrm{sw}(r_{ij})$, where $f_\mathrm{sw}$ is a quintic ($C^2$) switching function (implemented as \texttt{polycutoff} in the code~\cite{SI:byggmastar2022simple}) that equals 1 below 2.40~\r A and 0 above 3.00~\r A, so that $\mathbf{q}_\mathrm{CN}$ counts the neighbors within the first shell of the tetrahedral structures. Two separate Gaussian-process terms act on it, one resolving the whole coordination range and one resolving coordinations up to 8 in detail.
The total energy of $N$ atoms is therefore
\begin{align} \label{Eq:S4}
E_{\mathrm{tot.}} =\ & \sum_i^N e_0(Z_i) + \sum_{i,j>i}^{N} V_{\mathrm{pair}}(r_{ij})
  + \delta^2_{\mathrm{2b}} \sum_{i}^{N_{\mathrm{pairs}}} \sum_{s}^{M^{\mathrm{2b}}_{\mathrm{sparse}}} \alpha_{s,\mathrm{2b}} K_{\mathrm{2b}}(\mathbf{q}_{i,\mathrm{2b}}, \mathbf{q}_{s,\mathrm{2b}}) \nonumber \\
& + \delta^2_{\mathrm{3b}} \sum_{i}^{N_{\mathrm{triplets}}} \sum_{s}^{M^{\mathrm{3b}}_{\mathrm{sparse}}} \alpha_{s,\mathrm{3b}} K_{\mathrm{3b}}(\mathbf{q}_{i,\mathrm{3b}}, \mathbf{q}_{s,\mathrm{3b}}) \nonumber \\
& + \delta^2_{\mathrm{EAM}} \sum_{i}^{N} \sum_{s}^{M^{\mathrm{EAM}}_{\mathrm{sparse}}} \alpha_{s,\mathrm{EAM}} K_{\mathrm{EAM}}(\mathbf{q}_{i,\mathrm{EAM}}, \mathbf{q}_{s,\mathrm{EAM}}) \nonumber \\
& + \sum_{c\,\in\,\{\mathrm{wide},\,\mathrm{local}\}} \delta^2_{\mathrm{CN},c} \sum_{i}^{N} \sum_{s}^{M^{\mathrm{CN},c}_{\mathrm{sparse}}} \alpha_{s,\mathrm{CN},c} K_{\mathrm{CN},c}(\mathbf{q}_{i,\mathrm{CN}}, \mathbf{q}_{s,\mathrm{CN}}),
\end{align}
where $e_0$ are the isolated-atom energies of each element (which are subtracted from the target energies, as is standard for GAP) and $V_\mathrm{pair}$ is the repulsive baseline potential (also subtracted from the DFT targets), see Supplementary Note~3.}
To generate the tabulated version (tabGAP), the 2b and 3b energy terms of Eq.~\ref{Eq:S4} are mapped onto 1D and 3D grids, respectively, and evaluated by cubic-spline interpolation, which improves the computational speed by two orders of magnitude over the GAP.
The EAM terms follow the conventional format of tabulated Finnis-Sinclair EAM files (\verb|pair_style eam/fs|) with 1D cubic splines for the pair densities and embedding energies~\cite{SI:finnis1984simple}; \revision{the structure of the three potential files is described in Supplementary Note~4.}
The general tabGAP tabulation tools and the \texttt{LAMMPS} pair style are available in the open-access tabGAP package~\cite{SI:tabGAP}.

\revision{The database contains 5393 structures and 339,621 atomic environments, covering crystalline SiC, point defects, disordered SiC, elemental Si and C, and strongly repulsive configurations (main-text Fig.~1).
Each class has its own energy and force regularization (Table~\ref{tab:rmse_structures}).
Smaller regularization values give near-equilibrium crystals and point defects greater weight, while larger values accommodate the energy and force scales of disordered and strongly repulsive structures.
The energy regularization is 0.1--0.5~meV/atom for near-equilibrium crystals, strained cells and equations of state, 1--2~meV/atom for crystalline AIMD, point defects and recoil paths, 3--10~meV/atom for random, dense, molten, amorphous and elemental structures, and 10--30~meV/atom for dimers, trimers and dense high-energy structures.
Force regularization ranges from 0.02 to 2.0~eV/\r A.} 

\begin{figure}[ht!]
\sifig{8cm}{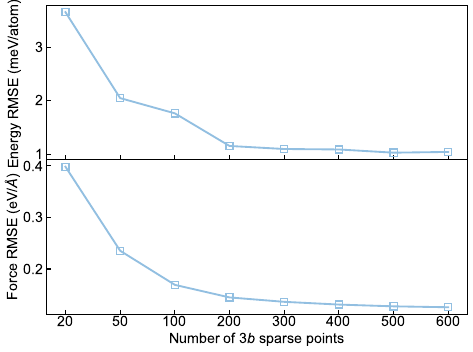}
\caption{Decrease of test errors for a set of crystalline SiC structures as a function of the number of sparse points for the 3b descriptor in the sparse Gaussian process regression.}
\label{fig:3b_sparse}
\end{figure}

A sufficient number of sparse points for the 3b descriptor was chosen based on a convergence test. Fig.~\ref{fig:3b_sparse} shows the decrease of the test errors with the number of sparse points.
Based on this, we use $M^{\mathrm{3b}}_{\mathrm{sparse}} = 300$ per triplet type.
All hyperparameters of the descriptors, kernels and tabulation are summarized in Table~\ref{Tab:S1}.

\begin{table}[ht!]
\caption{Hyperparameters of the low-dimensional GAP and of its tabulation. \revision{$\theta$ is the kernel length scale in the descriptor coordinates used by \texttt{QUIP} and $\delta$ the energy scale of each term.} The Finnis-Sinclair-like pair density of the 5~\r A EAM term needs no cutoff function, so no $\Delta r^{\mathrm{EAM}}$ is defined.}
\label{Tab:S1}
\begin{center}
\begin{tabular}{m{7cm} m{7.5cm}}
\hline
\hline
Hyperparameters & Setup \\
\hline
\multicolumn{2}{l}{\textbf{Cutoffs}} \\
$r_{\mathrm{cut}}^{\mathrm{2b}}$ \& $r_{\mathrm{cut}}^{\mathrm{EAM}}$ (\r A) & 5.0 \\
$\Delta r^{\mathrm{2b}}$ (\r A) & 1.0 \\
$r_{\mathrm{cut}}^{\mathrm{3b}}$ (\r A) & 4.1 \\
$\Delta r^{\mathrm{3b}}$ (\r A) & 0.7 \\
\revision{CN switching window (\r A)} & \revision{2.40--3.00 (quintic, $C^2$)} \\
\hline
\multicolumn{2}{l}{\textbf{2b descriptor}} \\
$M_{\mathrm{sparse}}^{\mathrm{2b}}$ & 20 per pair type \\
\revision{$\theta^{\mathrm{2b}}$, $\delta^{\mathrm{2b}}$} & \revision{0.5, 2.0~eV} \\
sparse method, kernel & ``\texttt{uniform}'', ``\texttt{ard\_se}'' \\
\hline
\multicolumn{2}{l}{\textbf{3b descriptor}} \\
$M_{\mathrm{sparse}}^{\mathrm{3b}}$ & 300 per triplet type \\
\revision{$\theta^{\mathrm{3b}}$, $\delta^{\mathrm{3b}}$} & \revision{0.5, 0.1~eV} \\
sparse method, kernel & ``\texttt{uniform}'', ``\texttt{ard\_se}'' \\
\hline
\multicolumn{2}{l}{\textbf{EAM descriptor (5~\r A)}} \\
\revision{$M_{\mathrm{sparse}}^{\mathrm{EAM}}$} & \revision{22 (C), 23 (Si), on a grid covering the sampled densities} \\
\revision{$\theta^{\mathrm{EAM}}$, $\delta^{\mathrm{EAM}}$} & \revision{1.0, 0.5~eV} \\
kernel, pair function & ``\texttt{ard\_se}'', ``\texttt{FSgen}'' of order 3 \\
\hline
\multicolumn{2}{l}{\revision{\textbf{Coordination (CN) descriptor}}} \\
\revision{wide term: $M_{\mathrm{sparse}}$, $\theta$, $\delta$} & \revision{23 per element, 1.5, 0.5~eV} \\
\revision{local term: $M_{\mathrm{sparse}}$, $\theta$, $\delta$} & \revision{13 per element on CN 0--8, 0.75, 0.25~eV} \\
\hline
\multicolumn{2}{l}{\revision{\textbf{Tabulation}}} \\
\revision{2b and EAM grids} & \revision{5000 points; pair terms on 0--5.0~\r A} \\
\revision{EAM density ranges} & \revision{0--40 (5~\r A term), 0--80 (CN term)} \\
\revision{3b grid} & \revision{$84\times84\times80$ on 0.1--4.30~\r A} \\
\hline
\hline
\end{tabular}
\end{center}
\end{table}

For more details about Gaussian approximation potentials, we refer the reader to Supplementary Refs.~\cite{SI:soap2013,SI:deringer2021gaussian,SI:bartok2010gaussian}.
The training uses the \texttt{QUIP} package~\cite{SI:quip,SI:bartok2010gaussian}, and          \texttt{LAMMPS}~\cite{SI:plimpton1995fast,SI:thompson2022lammps} was used for all MD simulations.

\revision{Training errors were evaluated as single-point energies and forces from the tabulated potential on the DFT geometries (Table~\ref{tab:rmse_structures}).
For each structure, the energy residual is $(E_\mathrm{tabGAP}-E_\mathrm{DFT})/N_\mathrm{atoms}$, with no fitted offset.
The cohesive-energy plots subtract the same DFT isolated-atom reference from both methods.
Energy RMSE weights structures equally, and force RMSE weights all Cartesian components equally (i.e. the averages do not use the fitting regularization).
The full training-set RMSEs are 268~meV/atom and 1.99~eV/\r A.
The 24 dense high-energy elemental cells contribute 74\% of the energy squared error and 64\% of the force squared error, illustrating the different scales of the classes. All classes are included in the aggregate.}

\begin{table*}[ht!]
\caption{\revision{Training errors of tabGAP against DFT per class of the fit: number of structures $N$, regularization of the energy per atom ($\sigma_E$) and of the force components ($\sigma_F$), and root-mean-square errors of the energy per atom ($E$) and of the force components ($F$), evaluated with the final tabulated model. The 21 structural classes are ordered from the most to the least tightly regularized, followed by the isolated-atom references and the full-set aggregate. Held-out errors are given in Fig.~\ref{sfig:heldout_valid}.}}
\label{tab:rmse_structures}
\begin{ruledtabular}
\begin{tabular}{@{}l r c c c c@{}}
Class & $N$ & $\sigma_E$ (meV/atom) & $\sigma_F$ (eV/\r A) & $E$ (meV/atom) & $F$ (eV/\r A) \\ \hline
\revision{Polytypes at equilibrium\footnote{2H, 3C, 4H, 6H, RS, 8H, 15R and 21R within 3~meV/atom of their minima and $|P|\le2$~GPa}} & \revision{88} & \revision{0.1} & \revision{0.10} & \revision{0.8} & \revision{0.02} \\
\revision{Strained cells (elastic)} & \revision{354} & \revision{0.1} & \revision{0.02} & \revision{0.5} & \revision{0.05} \\
\revision{Equations of state} & \revision{508} & \revision{0.5} & \revision{0.10} & \revision{8.7} & \revision{0.08} \\
\revision{AIMD of crystalline SiC} & \revision{122} & \revision{1} & \revision{0.10} & \revision{2.7} & \revision{0.26} \\
\revision{Vacancies and interstitials} & \revision{610} & \revision{1} & \revision{0.10} & \revision{8.7} & \revision{0.30} \\
\revision{Vacancy relaxation paths} & \revision{60} & \revision{1} & \revision{0.10} & \revision{10.6} & \revision{0.21} \\
\revision{Lattice distortions} & \revision{839} & \revision{2} & \revision{0.15} & \revision{8.9} & \revision{0.29} \\
\revision{Quenched SiC (AIMD)} & \revision{497} & \revision{2} & \revision{0.15} & \revision{8.8} & \revision{0.85} \\
\revision{Frenkel pairs} & \revision{81} & \revision{2} & \revision{0.20} & \revision{13.2} & \revision{1.00} \\
\revision{Quasi-static recoil paths} & \revision{242} & \revision{2} & \revision{0.20} & \revision{10.5} & \revision{0.43} \\
\revision{Compressed cells} & \revision{6} & \revision{3} & \revision{0.20} & \revision{9.3} & \revision{0.56} \\
\revision{Random structure search} & \revision{48} & \revision{3} & \revision{0.30} & \revision{127} & \revision{1.96} \\
\revision{Dense elemental structures} & \revision{39} & \revision{5} & \revision{0.50} & \revision{71.2} & \revision{2.55} \\
\revision{Molten and amorphous SiC} & \revision{429} & \revision{10} & \revision{0.50} & \revision{22.1} & \revision{0.91} \\
\revision{Si and C, crystalline and liquid} & \revision{245} & \revision{10} & \revision{0.50} & \revision{74.9} & \revision{0.48} \\
\revision{Si and C, amorphous} & \revision{132} & \revision{10} & \revision{0.50} & \revision{54.4} & \revision{1.00} \\
\revision{Dense SiC support structures} & \revision{279} & \revision{10} & \revision{1.0} & \revision{158} & \revision{2.93} \\
\revision{Dimers} & \revision{103} & \revision{10} & \revision{0.50} & \revision{384} & \revision{1.95} \\
\revision{Trimers} & \revision{556} & \revision{30} & \revision{1.5} & \revision{328} & \revision{2.75} \\
\revision{Dense high-energy SiC} & \revision{129} & \revision{30} & \revision{2.0} & \revision{337} & \revision{9.72} \\
\revision{Dense high-energy Si and C} & \revision{24} & \revision{30} & \revision{2.0} & \revision{3466} & \revision{55.62} \\
\hline
\revision{Isolated reference atoms} & \revision{2} & \revision{0.1} & \revision{0.04} & \revision{1.8} & \revision{0.00} \\
\revision{All training structures} & \revision{5393} & --- & --- & \revision{268} & \revision{1.99} \\
\end{tabular}
\end{ruledtabular}
\end{table*}

\begin{figure*}[ht!]
\sifig{0.98\linewidth}{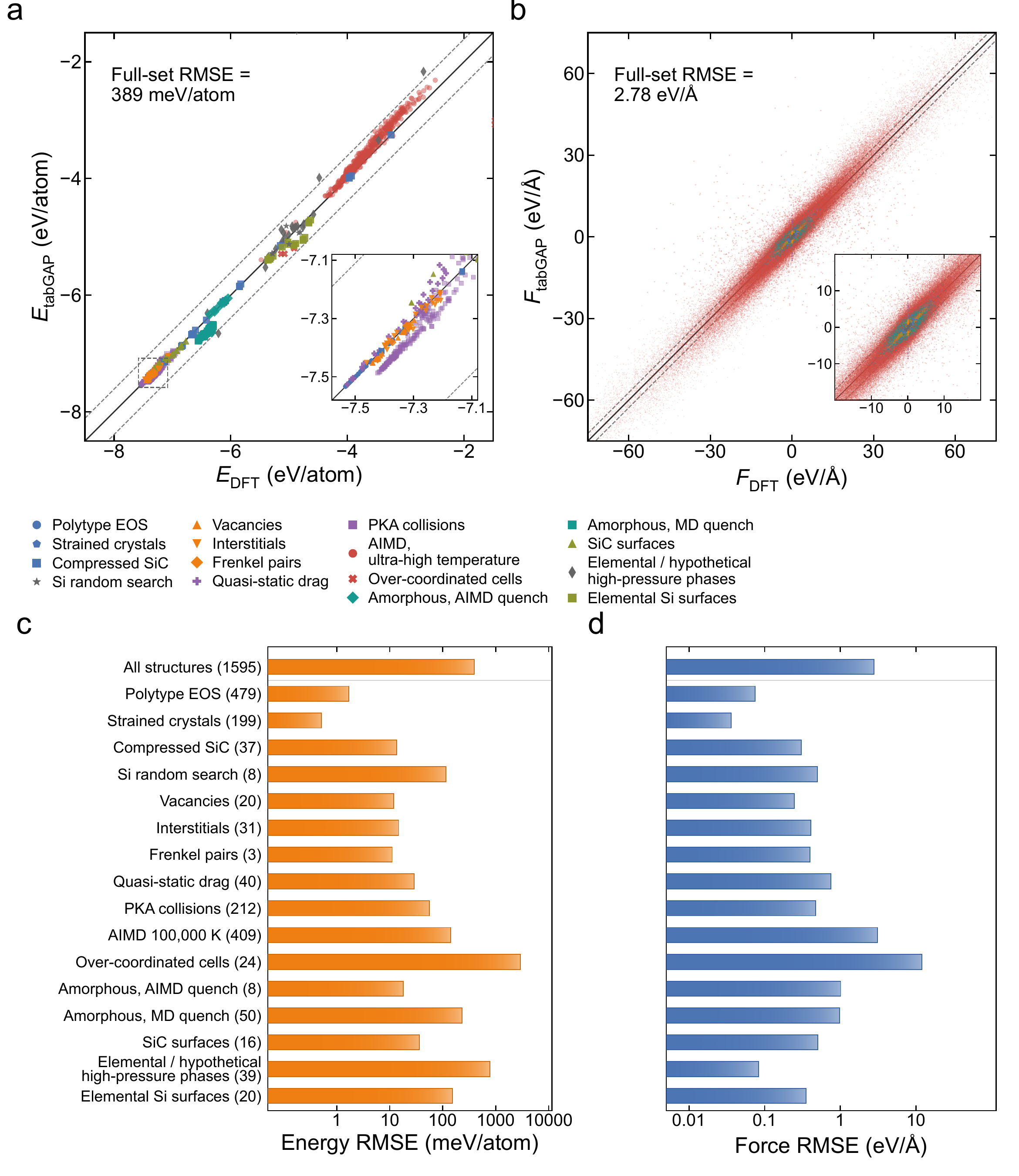}
\caption{\revision{Held-out accuracy on 1595 DFT structures.
(a) Cohesive energy per atom and (b) force components from tabGAP against DFT, displayed over $-8.5$ to $-1.5$~eV/atom and $-75$ to $75$~eV/\r A; insets magnify the low-energy and low-force regions. Colors and marker shapes distinguish the configuration types.
Solid lines show equality and dashed lines $\pm$ the full-set RMSE in both main panels and insets.
(c,d) Complete tabGAP energy and force RMSEs by all 16 classes and pooled over the full set; $N$ gives structure counts.
All RMSEs include all 1595 structures and 388,233 force components, including values outside the display windows and the five isolated energy points omitted from the scatter display.}}
\label{sfig:heldout_valid}
\end{figure*}

\revision{The test set was assembled from 2261 DFT structures and compared with the fitting and validation structures using exact geometry hashes and a permutation- and periodicity-invariant root-mean-square displacement tolerance of 0.05~\r A.
Removing 653 overlapping structures and 13 repeated geometries leaves 1595 structures and 388,233 Cartesian force components.
This set includes all 16 test classes: near-equilibrium crystals, defects, disordered structures, elemental and high-pressure cells, and 100,000~K AIMD snapshots.
The potential was evaluated without retraining on their DFT cells and atomic positions, using the same error definitions as for the training set.

Figure~\ref{sfig:heldout_valid} gives the complete errors by class and the energy and force comparisons in the populated ranges.
The full-set RMSEs are 389~meV/atom and 2.78~eV/\r A.
The relaxed-ion equation-of-state and strained-crystal energy RMSEs are 1.7 and 0.5~meV/atom; the vacancy, interstitial and Frenkel-pair values are 11.8, 14.7 and 11.1~meV/atom, with force RMSEs of 0.25--0.41~eV/\r A.
The class-resolved values distinguish these low-energy environments from the strongly repulsive structures contributing to the pooled errors.}

\clearpage

\subsection*{Supplementary Note 3: High-energy repulsive potential} \label{SuppNote_3}

The short-range repulsion is described by a screened-Coulomb pair potential refitted to all-electron DFT~\cite{SI:ziegler1985stopping,SI:nordlund1997repulsive}:
\begin{equation}\label{Eq:S5}
    E_\mathrm{rep.} = \sum_{i<j}^{N} \frac{1}{4\pi \varepsilon_{0}} \frac{Z_{i} Z_{j} e^{2}}{r_{ij}} \phi_{ij} (r_{ij}/a) f_\mathrm{cut} (r_{ij}),
\end{equation}
where $a = 0.46848~\text{\AA} / (Z_{i}^{0.23} + Z_{j}^{0.23})$ as in the universal ZBL potential, and $f_\mathrm{cut}$ is a smooth cutoff function (see Ref.~\cite{SI:byggmastar2019machine}) that drives the interaction to zero between $r_{1}$ and $r_{2}$.
\revision{The refitted screening function is a sum of three exponentials:}
\begin{equation}\label{Eq:S6}
    \phi(x) = A_1 \exp(-B_1 x) + A_2 \exp(-B_2 x) + A_3 \exp(-B_3 x),
\end{equation}
where the parameters are refitted to all-electron \textit{ab initio} data~\cite{SI:nordlund1997repulsive}.
\revision{The fitted parameters are given in Table~\ref{Tab:S2}.}
The all-electron data cover the repulsive region up to $10^{6}$~eV.
The GAP is fitted to the energy and force differences between this pair potential and the GGA-DFT data, joining the repulsive core to the interactions at larger distances.
High-energy training structures sample the transition from near-equilibrium bonding to strong repulsion.

\begin{figure}[ht!]
\sifig{14cm}{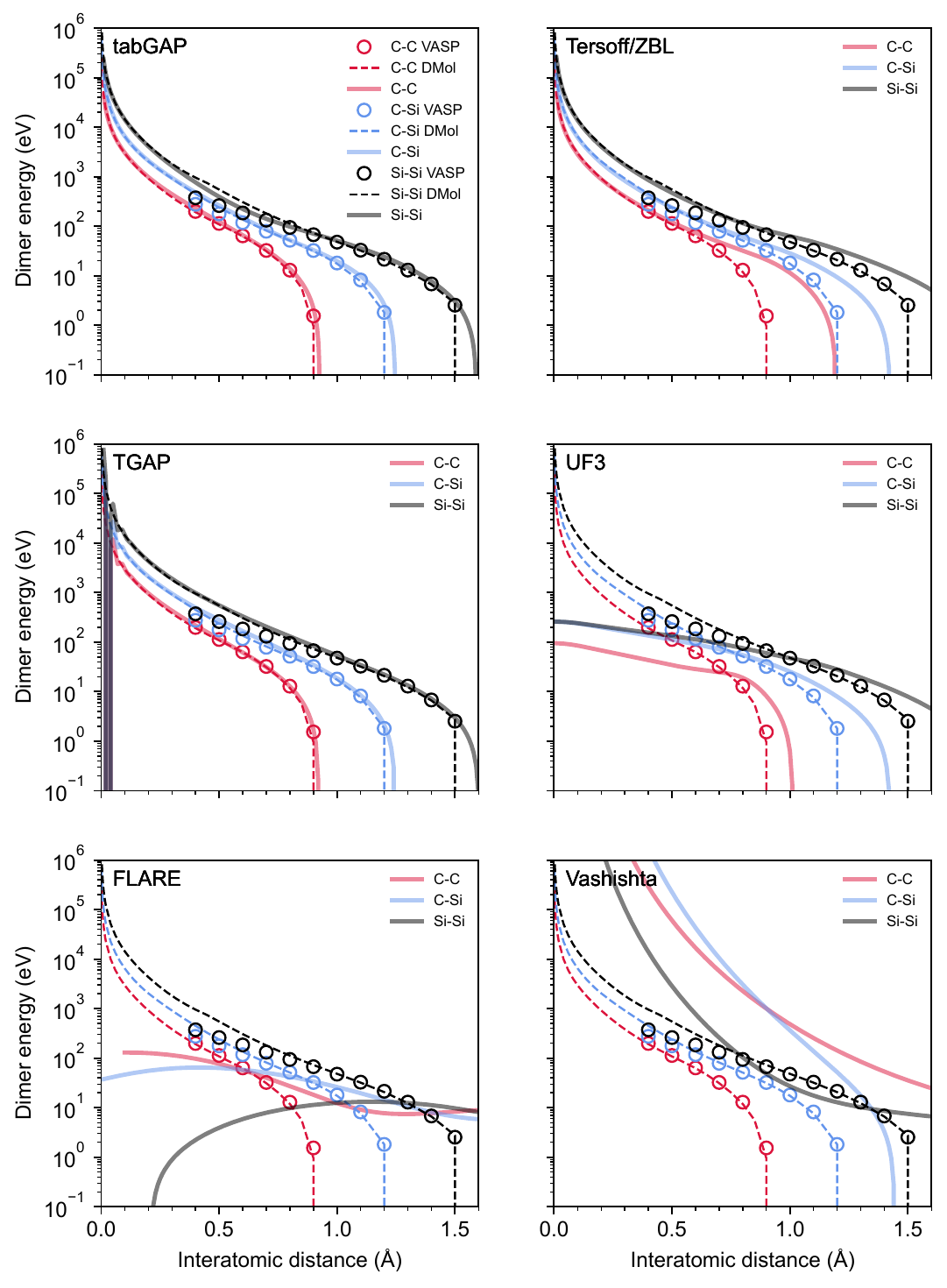}
\caption{Repulsive parts of the dimer curves in different interatomic potentials compared to DFT (\texttt{VASP}) and all-electron DFT (\texttt{DMol}) data. \revision{Energies are total dimer energies with the isolated atom energies subtracted.}}
\label{dimer_plot}
\end{figure}

The repulsive parts of the dimer curves are shown in Fig.~\ref{dimer_plot} for six interatomic potentials.
The tabGAP, Tersoff/ZBL and TGAP curves follow the all-electron repulsion, whereas UF$^{3}$, FLARE and Vashishta depart strongly from it at short distances.
For TGAP, the SOAP implementation becomes numerically unstable at extremely short distances, limiting the collision energies that can be simulated~\cite{SI:byggmastar2019machine}.

\begin{table}[ht!]
\caption{Fitted parameters of the screening function in the form of Eqs.~\ref{Eq:S5} and \ref{Eq:S6}.}  \label{Tab:S2}
\begin{ruledtabular}
\begin{tabular}{l l l l}
Parameters & Si-Si & Si-C & C-C \\
\hline
$A_1$ & 0.44688825 & \num{5.69601876e-03} & 0.0403246631 \\
$A_2$ & 0.01087316 & 0.413221372 & 0.905611306 \\
$A_3$ & 0.48300975 & 0.753530543 & \num{1.93801655e-11} \\
$B_1$ & 0.41929332 & \num{5.61247429e-25} & \num{1.58575788e-10} \\
$B_2$ & 0.41927045 & 3.65112945 & 0.823278588 \\
$B_3$ & 1.23854965 & 0.579096031 & 3.02373844 \\
$r_1$ (\r A) & 1.1 & 0.8 & 0.6 \\
$r_2$ (\r A) & 1.8 & 1.5 & 1.3 \\
\end{tabular}
\end{ruledtabular}
\end{table}

\clearpage

\subsection*{Supplementary Note 4: \revision{Energy terms and descriptor coverage}} \label{SuppNote_4}

\begin{figure*}[ht!]
\sifig{\linewidth}{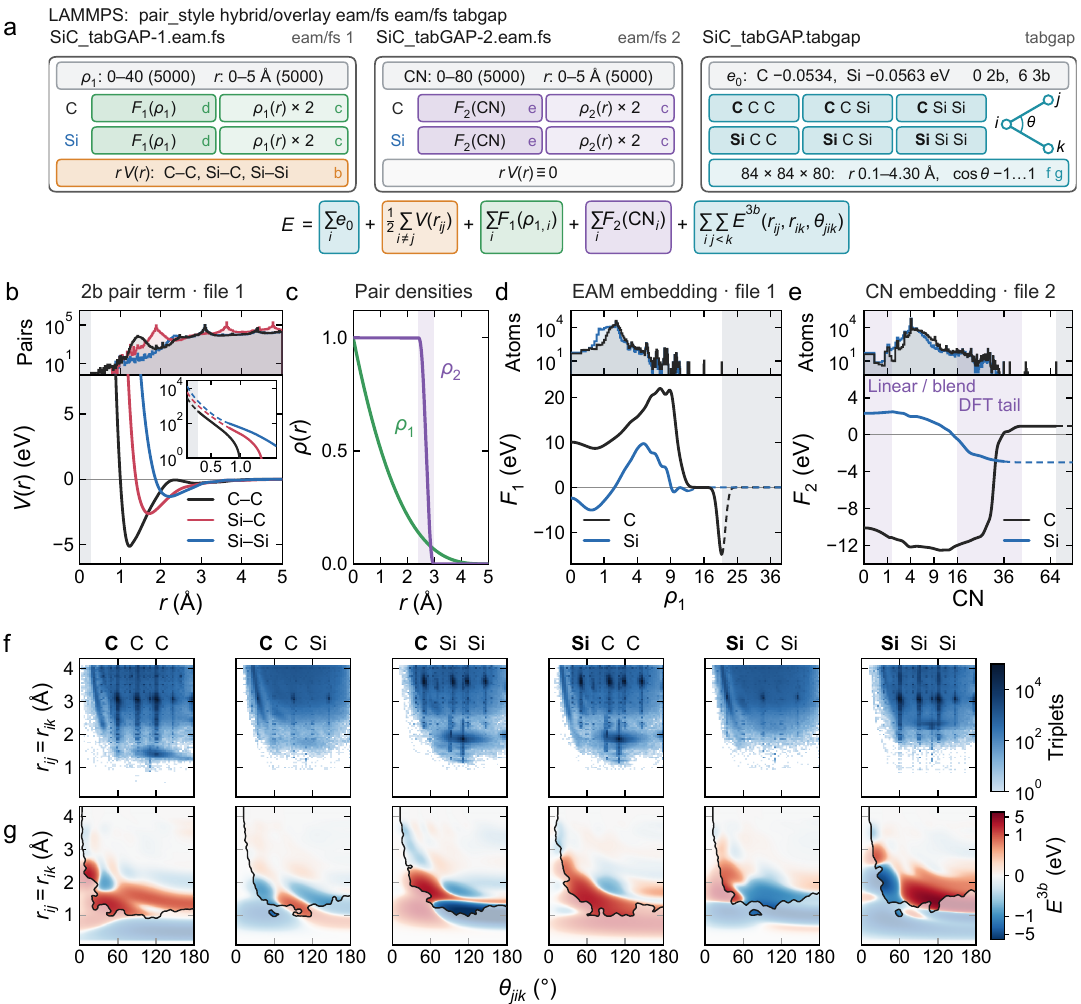}
\caption{\revision{Energy terms of tabGAP and the training-data coverage of their descriptors. (a)~The three \texttt{LAMMPS} files and the total-energy expression; matching colors identify the file supplying each term, and letters link the blocks to their panels. The central atom of each three-body block is in bold. (b)~$V(r)$ for the three pairs; top: histogram of all training pair distances below 5~\r A (0.02~\r A bins); inset: the repulsive core on a logarithmic scale. (c)~Pair densities $\rho_1(r) = (1 - r/5~\text{\AA})^3$ of file~1 and $\rho_2(r)$ of file~2, which switches from 1 to 0 between 2.40 and 3.00~\r A (shaded). (d)~$F_1(\rho_1)$ and (e)~$F_2(\mathrm{CN})$, each with the per-element distribution of the 339,621 training atoms above; square-root horizontal axes. In (e), the low-CN shaded region marks the linear continuation and quintic join of Eqs.~\eqref{eq:cn_lowend} and~\eqref{eq:cn_join}; the DFT-fitted tail starts at CN $=16$. In (b), (d) and (e), solid lines cover the range populated by the training data and dashed lines extend beyond it; grey marks the range beyond every training value. (f)~Training triplets per 0.05~\r A $\times$ 3$^\circ$ voxel on the slice $r_{ij} = r_{ik}$ for the six three-body blocks ($1.06\times10^8$ triplets from 5393 structures, no subsampling). (g)~Three-body energy $E^{3b}(r, r, \theta)$ of each block from the tabulated coefficients (symmetric logarithmic color scale); outside the black contour, which bounds the sampled voxels of (f), the map is veiled.}}
\label{sfig:terms}
\end{figure*}

\revision{The tabulated energy separates pair, three-body and density-dependent interactions (Fig.~\ref{sfig:terms}).
The first embedding term uses a local density with a 5~\r A range; the second uses the number of neighbours in the first shell.
This coordination density distinguishes the loss of a bonded neighbour around a vacancy from changes in the density of more distant shells.
The total energy is
\begin{equation}
\begin{split}
E = {} & \sum_i e_0(Z_i) + \frac{1}{2}\sum_{i\neq j} V_{Z_iZ_j}(r_{ij}) + \sum_i F_{1,Z_i}\!\left(\rho_{1,i}\right) \\
 & + \sum_i F_{2,Z_i}\!\left(\mathrm{CN}_i\right) + \sum_i\sum_{j<k} E^{3b}_{Z_iZ_jZ_k}(r_{ij}, r_{ik}, \cos\theta_{jik}),
\end{split}
\label{eq:tabgap_total}
\end{equation}
where $\rho_{1,i} = \sum_j \rho_1(r_{ij})$ and $\mathrm{CN}_i = \sum_j \rho_2(r_{ij})$.
The functions $F_1$ and $F_2$ depend on the species of the central atom. The atomic offsets $e_0$ retain the reference of the fitted energy.}

{\color{revisionred}The functions are stored in two Finnis-Sinclair EAM files and one three-body file (Table~\ref{tab:files}), used in \texttt{LAMMPS} as follows:
\begin{verbatim}
pair_style  hybrid/overlay eam/fs eam/fs tabgap
pair_coeff  * * eam/fs 1 SiC_tabGAP-1.eam.fs Si C
pair_coeff  * * eam/fs 2 SiC_tabGAP-2.eam.fs Si C
pair_coeff  * * tabgap   SiC_tabGAP.tabgap   Si C
pair_modify neigh/trim no
\end{verbatim}
The first \texttt{eam/fs} file supplies the pair energy and $F_1$, the second supplies $F_2$, and \texttt{tabgap} supplies $E^{3b}$.
\verb|neigh/trim no| supplies every sub-style with the full neighbour list.}

\begin{table*}[ht!]
\caption{\revision{Content of the three potential files. The \texttt{eam/fs} files follow the standard \texttt{setfl} (Finnis-Sinclair) layout, with the elements in the order C, Si; the \texttt{tabgap} file follows the format of the tabGAP package~\cite{SI:tabGAP}.}}
\label{tab:files}
\begin{ruledtabular}
\begin{tabular}{l l p{0.60\textwidth}}
File & Block & Content \\ \hline
\revision{\texttt{-1.eam.fs}} & \revision{$F_1(\rho_1)$, C and Si} & \revision{embedding of the 5~\r A EAM term, 5000 points on $\rho_1 = 0$--40} \\
 & \revision{$\rho_1(r)$} & \revision{pair density $(1 - r/5~\text{\AA})^3$, identical for all pairs, 5000 points on $r = 0$--5~\r A} \\
 & \revision{$r\,V(r)$, C-C, Si-C, Si-Si} & \revision{two-body Gaussian process GP plus the screened-Coulomb core of Supplementary Note~3} \\
\revision{\texttt{-2.eam.fs}} & \revision{$F_2(\mathrm{CN})$, C and Si} & \revision{coordination embedding, 5000 points on CN $= 0$--80} \\
 & \revision{$\rho_2(r)$} & \revision{neighbor counter: 1 below 2.40~\r A, quintic switch to 0 at 3.00~\r A} \\
 & \revision{$r\,V(r)$} & \revision{zero} \\
\revision{\texttt{.tabgap}} & \revision{header} & \revision{atomic reference offsets $e_0$ (C $-0.0534$, Si $-0.0563$~eV); 0 two-body and 6 three-body blocks} \\
 & \revision{$E^{3b}$, six blocks} & \revision{C-C-C, C-C-Si, C-Si-Si, Si-C-C, Si-C-Si, Si-Si-Si (central atom first); cubic B-spline coefficients on an $84\times84\times80$ grid in $r_{ij}$, $r_{ik}$ (0.1--4.30~\r A) and $\cos\theta_{jik}$ ($-1$ to 1)} \\
\end{tabular}
\end{ruledtabular}
\end{table*}

\revision{\textbf{Pair interaction and local-density embedding.}
The pair term $V(r)$ is the two-body GP (cutoff 5.0~\r A, transition width 1.0~\r A) plus the screened-Coulomb core of Supplementary Note~3, which reaches zero at 1.3~\r A (C-C), 1.5~\r A (Si-C) and 1.8~\r A (Si-Si).
Its minima are $-5.15$~eV at 1.23~\r A (C-C), $-2.64$~eV at 1.71~\r A (Si-C) and $-1.33$~eV at 2.22~\r A (Si-Si) (Fig.~\ref{sfig:terms}b).
The density function $\rho_1(r)=(1-r/5~\text{\AA})^3$ is identical for every pair (Fig.~\ref{sfig:terms}c); the species dependence enters the embedding functions $F_{1,Z}$.}

\revision{\textbf{First-shell coordination embedding.}
The coordination density uses a switch between the first (1.89~\r A) and second (3.08~\r A) neighbour shells of SiC,
\begin{equation}
\rho_2(r) = \begin{cases} 1, & r < 2.40~\text{\AA}, \\ 1 - \left(10x^3 - 15x^4 + 6x^5\right), \quad x = \dfrac{r - 2.40~\text{\AA}}{0.60~\text{\AA}}, & 2.40 \le r < 3.00~\text{\AA}, \\ 0, & r \ge 3.00~\text{\AA}, \end{cases}
\end{equation}
whose value and first two derivatives match at the endpoints.
The second file contains no pair energy.
Its embedding $F_2$ combines the wide and local Gaussian-process terms of Table~\ref{Tab:S1}: both use the same coordination density, with the local term resolving CN up to 8 (Fig.~\ref{sfig:terms}e).

At low coordination the embedding is continued linearly and joined smoothly to the fitted function.
Let $q=\mathrm{CN}$ and let $G(q)$ be the embedding including the high-CN continuation described below.
The joining points $a$ and $b$ are the grid points nearest to the nominal coordinations 1 and 1.5.
The tabulated embedding follows
\begin{equation}
F_2(q)=\begin{cases}
L(q), & 0\le q\le a,\\
P_5(q), & a<q<b,\\
G(q), & q\ge b,
\end{cases}
\qquad L(q)=G(0)+\frac{G(a)-G(0)}{a}q,
\label{eq:cn_lowend}
\end{equation}
where $P_5$ is a quintic polynomial fixed by
\begin{equation}
\begin{aligned}
P_5(a)&=G(a), & P_5'(a)&=\frac{G(a)-G(0)}{a}, & P_5''(a)&=0,\\
(P_5,P_5',P_5'')(b)&=(G,G',G'')(b).
\end{aligned}
\label{eq:cn_join}
\end{equation}
The conditions match the energy, slope and curvature at both ends; the derivatives at $b$ are evaluated by centred differences on the original grid.
For the 5000-point CN grid, $a=0.9922$ and $b=1.5043$; the function at $q\ge b$ is unchanged by this low-CN continuation.
The shaded low-CN region in Fig.~\ref{sfig:terms}e covers the linear part and its quintic join.
Above CN $=16$, the smooth tail of $G$ is fitted to DFT with the other energy terms fixed, and becomes constant beyond CN $=46$.
At zero density an isolated atom has energy $e_0+F_1(0)+F_2(0)$; the embedding sum is $-1.0$~meV for C and $-2.3$~meV for Si.}

\revision{\textbf{Three-body interaction.}
Each element-specific three-body interaction is evaluated using the stored coefficients $c_{abc}$ in a uniform cubic B-spline
\begin{equation}
E^{3b}(r_{ij}, r_{ik}, \cos\theta_{jik}) = \sum_{a,b,c=0}^{3} c_{i_0+a,\, i_1+b,\, i_2+c}\, B_a(t_0)\, B_b(t_1)\, B_c(t_2),
\end{equation}
where $i_n$ is the index of the grid interval and $t_n$ the fractional position in it, and $B_0 = (1-t)^3/6$, $B_1 = (3t^3 - 6t^2 + 4)/6$, $B_2 = (-3t^3 + 3t^2 + 3t + 1)/6$ and $B_3 = t^3/6$. The interaction is symmetric to $j$ and $k$ exchange (for rotation and permutation invariance).
The grid spacing in $r$ is 4.0/79~\r A, so that node 79 lies exactly at the 4.1~\r A cutoff of the three-body descriptor, and every node at $r \geq 4.1$~\r A is zero.
Below 0.3~\r A, the node energies include an inner cutoff $f(r_{ij})f(r_{ik})$, with $f=\chi^3(6\chi^2-15\chi+10)$ and $\chi=(r-0.2~\text{\AA})/0.1~\text{\AA}$ on 0.2--0.3~\r A.
This factor is applied once to each node, preserving symmetry under exchange of the two neighbours~\cite{SI:tabGAP}.}

The individual energy terms are not standalone interaction potentials: even a dimer receives both pair and embedding contributions.
The energy minima of a structure result from the sum over all pairs and triplets within the interaction range, together with the embedding terms.
The three-body maps therefore need not have a separate minimum at the bond length and angle of a stable crystal.

\revision{\textbf{Tabulation and sampled descriptor space.}
The tabulated potential gives mean absolute energy errors of 0.81~meV/atom for 60 relaxed bulk cells and 0.38~meV/atom for 60 strained cells, compared with 0.71 and 0.37~meV/atom for the untabulated GAP.
The distributions above panels b, d and e of Fig.~\ref{sfig:terms} and the maps of panel f cover all 5393 training structures (339,621 atoms and $1.06\times10^8$ triplets within 4.1~\r A) without subsampling.
The shortest pairs in the training data are 0.28~\r A (C-C), 0.80~\r A (Si-C) and 0.81~\r A (Si-Si); below these the pair term is the repulsive core of Supplementary Note~3.
$\rho_1$ reaches 20.6 (C) and 12.3 (Si), and CN reaches 68 (C) and 36 (Si), well inside the tabulated ranges.}

\clearpage

\subsection*{Supplementary Note 5: Fundamental physical properties} \label{SuppNote_5}

We compare the crystalline properties of tabGAP with Tersoff/ZBL~\cite{SI:devanathan1998displacement}, Vashishta~\cite{SI:vashishta2007interaction}, FLARE~\cite{SI:xie_incongruent_2026}, TGAP~\cite{SI:hamedani2025sic} and UF$^{3}$~\cite{SI:macisaac2024genetic}.
Tersoff/ZBL is widely used for collision cascades, and Vashishta reproduces the trend of the high-pressure 3C-to-RS transition~\cite{SI:xie2023uncertainty}.
FLARE was developed for bulk phases and incongruent melting; TGAP uses a SOAP descriptor and was trained on 3C-SiC for radiation damage; UF$^{3}$ uses a compact three-body representation.
The comparison therefore covers different representations and training scopes.

\begin{figure*}[ht!]
\sifig{.9\linewidth}{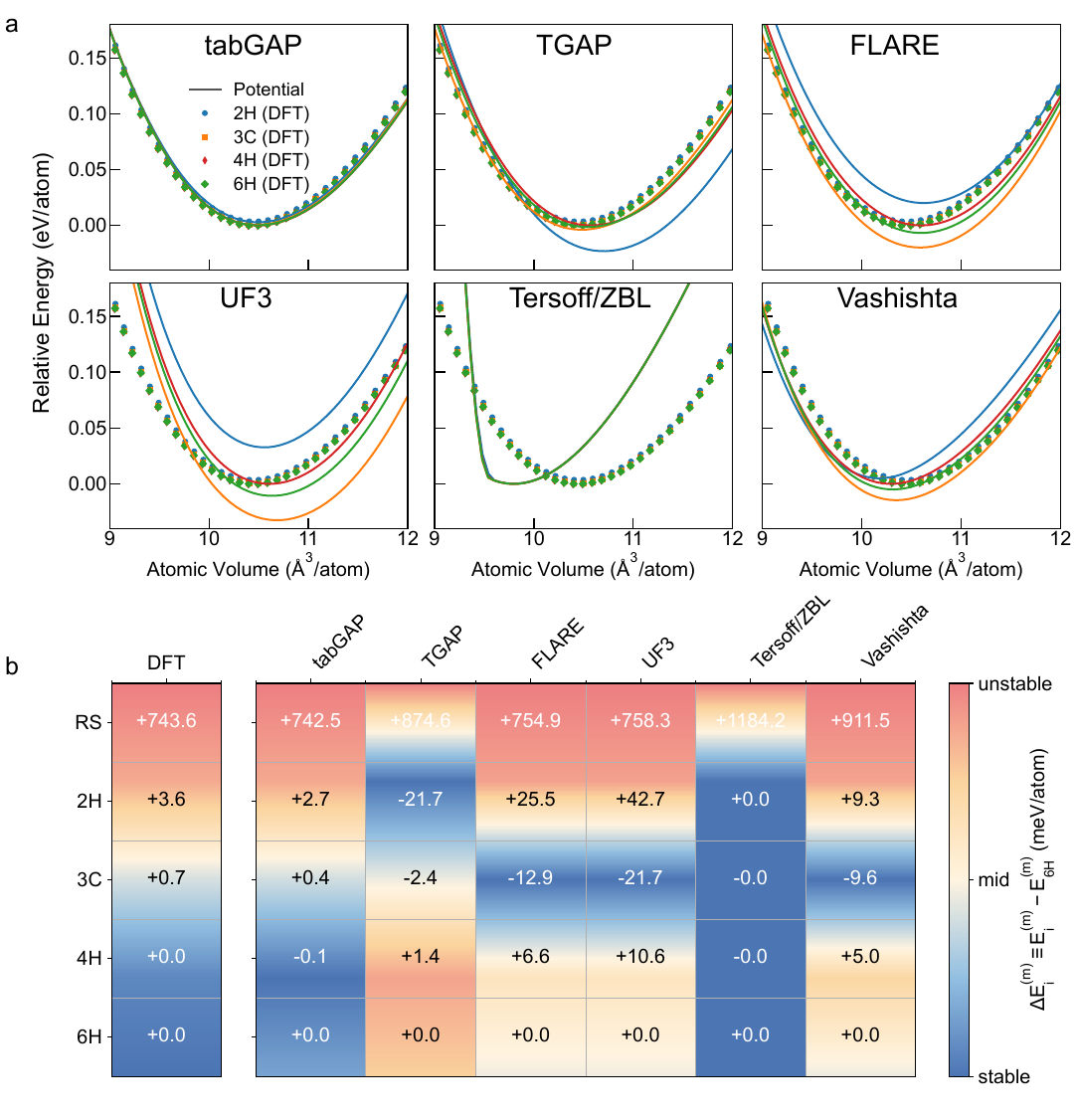}
\caption{(a) Equations of state of the SiC polytypes from different interatomic potentials (lines) compared with DFT (symbols). \revision{Each potential is referenced to the minimum of its own 4H-SiC curve and DFT to its global minimum.}
(b) Phase ordering: $E_i - E_{6\mathrm{H}}$ (meV/atom) of the fully relaxed cells, with rows in the DFT order (left). The color scale is referenced to 6H and normalized within each column; the near-degenerate 4H-6H pair \revision{(0.02~meV/atom in DFT, 0.08~meV/atom in tabGAP)} is not assigned a physical ordering.}
\label{eos}
\end{figure*}

Fig.~\ref{eos} compares the equations of state of the SiC polymorphs from the different potentials with DFT.
For the equation-of-state curves, each potential is referenced to its own 4H minimum and DFT to its global minimum.
The fully relaxed energies in panel (b) are referenced to 6H.
In DFT, 4H and 6H differ by 0.02~meV/atom, below the numerical resolution, so they are treated as degenerate.
\revision{tabGAP reproduces the grouping of the polytypes, with nearly degenerate 4H and 6H below 3C, followed by 2H and rock salt: relative to 3C, it gives $+2.34$, $-0.48$, $-0.40$ and $+742.1$~meV/atom for 2H, 4H, 6H and RS, against $+2.92$, $-0.64$, $-0.66$ and $+743.0$~meV/atom in DFT.
Its 4H and 6H are 0.08~meV/atom apart, also below the DFT resolution.
All potential curves in panel (a) were computed with the ions relaxed at fixed cell, and panel (b) uses fully relaxed cells.}

\begin{figure*}[ht!]
\sifig{\linewidth}{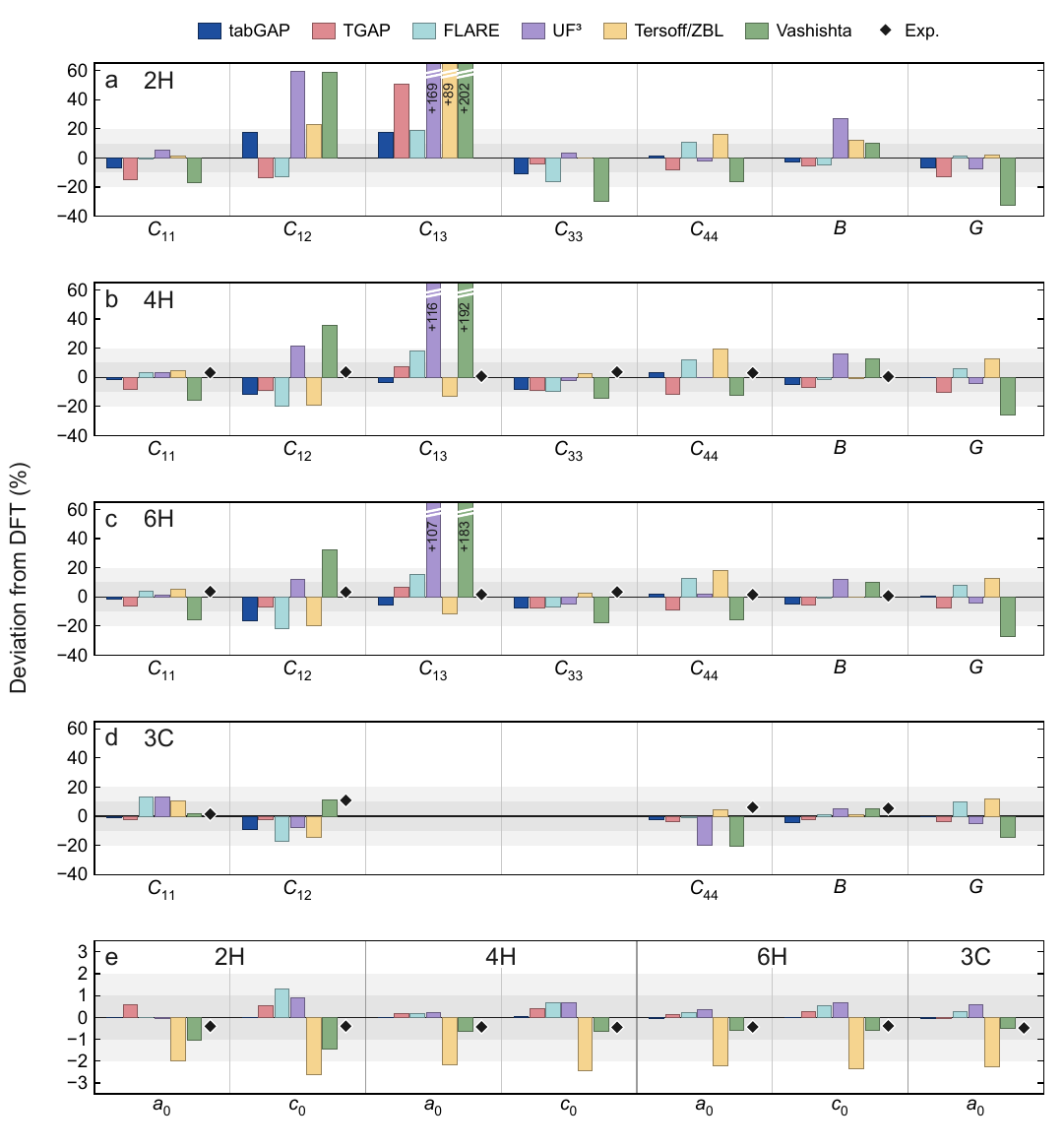}
\caption{\revision{Elastic and lattice constants of the SiC polytypes relative to DFT. (a--d) Signed deviation, $100(X/X_\mathrm{DFT}-1)$, of the elastic constants $C_{ij}$ and of the Voigt-Reuss-Hill bulk ($B$) and shear ($G$) moduli of 2H-, 4H-, 6H- and 3C-SiC for tabGAP (this work), TGAP, FLARE, UF$^{3}$, Tersoff/ZBL and Vashishta; black diamonds, experiment. (e) The same for the lattice constants $a_0$ and $c_0$. DFT elastic constants are relaxed-ion, and DFT $B$ and $G$ are the Voigt-Reuss-Hill averages of the DFT $C_{ij}$. Shaded bands: $\pm10\%$ and $\pm20\%$ (a--d), $\pm1\%$ and $\pm2\%$ (e). Bars that exceed the axis are broken and labelled with their value. No experimental elastic data exist for 2H; 4H and 6H share one measured set; the experimental $G$ is omitted (see text).}}
\label{sfig:elastic}
\end{figure*}

\revision{Fig.~\ref{sfig:elastic} compares the lattice constants, elastic constants and the Voigt-Reuss-Hill bulk and shear moduli of the four polytypes with our DFT (relaxed ions) for all six potentials. The absolute values are listed in Table~\ref{tab:elastic_abs}.
Averaged over the 26 elastic values (18 independent $C_{ij}$ and the moduli $B$ and $G$ of each polytype), the mean absolute deviation from DFT is 5.8\% for tabGAP (7.0\% over the $C_{ij}$ alone), with all values within 20\% and 21 within 10\%; its largest deviation is $C_{13}$ of 2H ($+17.3\%$), and its lattice constants are within 0.1\% (mean 0.02\%).
For the other potentials the mean absolute deviations are 9.5\% (FLARE), 9.1\% (TGAP), 12.6\% (Tersoff/ZBL), 24.1\% (UF$^{3}$) and 39.6\% (Vashishta).}
Experimental values are shown where they exist; no elastic constants have been measured for 2H, and 4H and 6H share one measured set.
\revision{The experimental shear moduli are omitted because the reported values are inconsistent with those derived from the measured $C_{ij}$ of the same sources.
All constants are relaxed-ion for both methods. The 2H $C_{13}$, the constant most sensitive to the internal relaxation, depends on the strain route for tabGAP: 58.5--60.0~GPa from stresses or energies at 0.5--1\% strain, and 64.9~GPa on the DFT strain configurations, against 49.9~GPa in DFT.
}

\begin{table*}[htbp]
\caption{Elastic constants and the Voigt-Reuss-Hill bulk and shear moduli (GPa) and lattice constants of the SiC polytypes
shown in Fig.~\ref{sfig:elastic} as deviations from DFT. \revision{DFT: this work;} Exp.: measured values~\cite{SI:yu2024comparison,SI:pizzagalli2021accurate,SI:schulz1979structure}. --: not available.}
\label{tab:elastic_abs}
\small
\renewcommand{\arraystretch}{1.08}
\begin{ruledtabular}
\begin{tabular}{l c c c c c c c c}
 & DFT & Exp. & tabGAP & TGAP & FLARE & UF$^{3}$ & Tersoff/ZBL & Vashishta \\ \hline
\multicolumn{9}{l}{\textbf{2H-SiC}} \\
$C_{11}$ & \revision{499.1} & -- & \revision{465.8} & 425.4 & \revision{494.4} & 524.6 & 506.1 & 415.2 \\
$C_{12}$ & \revision{99.6} & -- & \revision{116.7} & 86.3 & \revision{86.7} & 158.5 & 122.6 & 158.1 \\
$C_{13}$ & \revision{49.9} & -- & \revision{58.5} & 75.3 & \revision{59.2} & 134.2 & 94.3 & 150.9 \\
$C_{33}$ & \revision{534.8} & -- & \revision{477.7} & 511.9 & \revision{446.0} & 551.2 & 534.5 & 376.5 \\
$C_{44}$ & \revision{151.7} & -- & \revision{153.3} & 139.0 & \revision{168.1} & 148.8 & 176.1 & 126.7 \\
$B$ & \revision{214.6} & -- & \revision{208.3} & 203.5 & \revision{204.5} & 272.7 & 241.0 & 236.0 \\
$G$ & \revision{186.4} & -- & \revision{173.1} & 162.8 & \revision{189.0} & 172.9 & 190.5 & 126.1 \\
$a_0$ (\r A) & \revision{3.091} & 3.079 & \revision{3.091} & 3.110 & \revision{3.092} & 3.090 & 3.030 & 3.060 \\
$c_0$ (\r A) & \revision{5.073} & 5.053 & \revision{5.074} & 5.100 & \revision{5.139} & 5.120 & 4.940 & 5.000 \\
\hline
\multicolumn{9}{l}{\textbf{4H-SiC}} \\
$C_{11}$ & \revision{485.5} & 501.0 & \revision{477.0} & 445.8 & \revision{500.4} & 501.3 & 506.6 & 409.9 \\
$C_{12}$ & \revision{107.1} & 111.0 & \revision{94.6} & 97.9 & \revision{85.9} & 130.0 & 86.6 & 145.1 \\
$C_{13}$ & \revision{51.6} & 52.0 & \revision{49.6} & 55.5 & \revision{60.8} & 111.4 & 44.9 & 150.8 \\
$C_{33}$ & \revision{533.6} & 553.0 & \revision{488.2} & 485.2 & \revision{481.8} & 523.1 & 547.7 & 455.5 \\
$C_{44}$ & \revision{158.3} & 163.0 & \revision{162.9} & 139.5 & \revision{177.7} & 158.5 & 189.1 & 138.8 \\
$B$ & \revision{213.9} & 215.0 & \revision{203.2} & 199.4 & \revision{210.7} & 247.9 & 212.6 & 240.6 \\
$G$ & \revision{185.2} & \revision{--} & \revision{185.3} & 166.3 & \revision{196.8} & 177.8 & 208.8 & 137.1 \\
$a_0$ (\r A) & \revision{3.093} & 3.080 & \revision{3.093} & 3.099 & \revision{3.099} & 3.100 & 3.026 & 3.074 \\
$c_0$ (\r A) & \revision{10.128} & 10.082 & \revision{10.131} & 10.169 & \revision{10.197} & 10.198 & 9.883 & 10.063 \\
\hline
\multicolumn{9}{l}{\textbf{6H-SiC}} \\
$C_{11}$ & \revision{483.5} & 501.0 & \revision{477.3} & 452.3 & \revision{503.2} & 488.6 & 507.2 & 407.4 \\
$C_{12}$ & \revision{108.1} & 111.5 & \revision{90.7} & 100.3 & \revision{84.5} & 121.2 & 87.0 & 142.7 \\
$C_{13}$ & \revision{51.2} & 52.0 & \revision{48.4} & 54.5 & \revision{59.1} & 105.8 & 45.3 & 145.1 \\
$C_{33}$ & \revision{535.2} & 553.0 & \revision{495.9} & 495.9 & \revision{499.6} & 508.6 & 548.6 & 439.1 \\
$C_{44}$ & \revision{160.6} & 163.0 & \revision{163.0} & 146.1 & \revision{181.4} & 163.1 & 189.1 & 135.2 \\
$B$ & \revision{213.7} & 215.0 & \revision{202.8} & 202.1 & \revision{212.3} & 239.0 & 213.1 & 235.4 \\
$G$ & \revision{185.8} & \revision{--} & \revision{186.6} & 171.1 & \revision{200.6} & 178.3 & 208.9 & 135.2 \\
$a_0$ (\r A) & \revision{3.094} & 3.081 & \revision{3.094} & 3.098 & \revision{3.101} & 3.105 & 3.026 & 3.077 \\
$c_0$ (\r A) & \revision{15.184} & 15.125 & \revision{15.185} & 15.222 & \revision{15.268} & 15.286 & 14.825 & 15.096 \\
\hline
\multicolumn{9}{l}{\textbf{3C-SiC}} \\
$C_{11}$ & \revision{384.3} & 390.0 & \revision{379.7} & 374.5 & \revision{434.1} & 435.1 & 426.0 & 390.1 \\
$C_{12}$ & \revision{128.1} & 142.0 & \revision{116.6} & 124.9 & \revision{106.4} & 118.6 & 109.9 & 142.6 \\
$C_{44}$ & \revision{241.4} & 256.0 & \revision{236.1} & 231.8 & \revision{238.8} & 193.3 & 252.7 & 191.0 \\
$B$ & \revision{213.5} & 225.0 & \revision{204.3} & 208.1 & \revision{215.7} & 224.1 & 215.3 & 225.1 \\
$G$ & \revision{187.2} & \revision{--} & \revision{186.7} & 180.8 & \revision{205.3} & 178.4 & 209.4 & 160.5 \\
$a_0$ (\r A) & \revision{4.379} & 4.358 & \revision{4.376} & 4.377 & \revision{4.392} & 4.404 & 4.280 & 4.358 \\
\end{tabular}
\end{ruledtabular}
\end{table*}


\clearpage

\subsection*{Supplementary Note 6: \revision{Phonons and the non-analytical correction for a short-ranged potential}} \label{SuppNote_6}

\begin{figure*}[ht!]
\sifig{\linewidth}{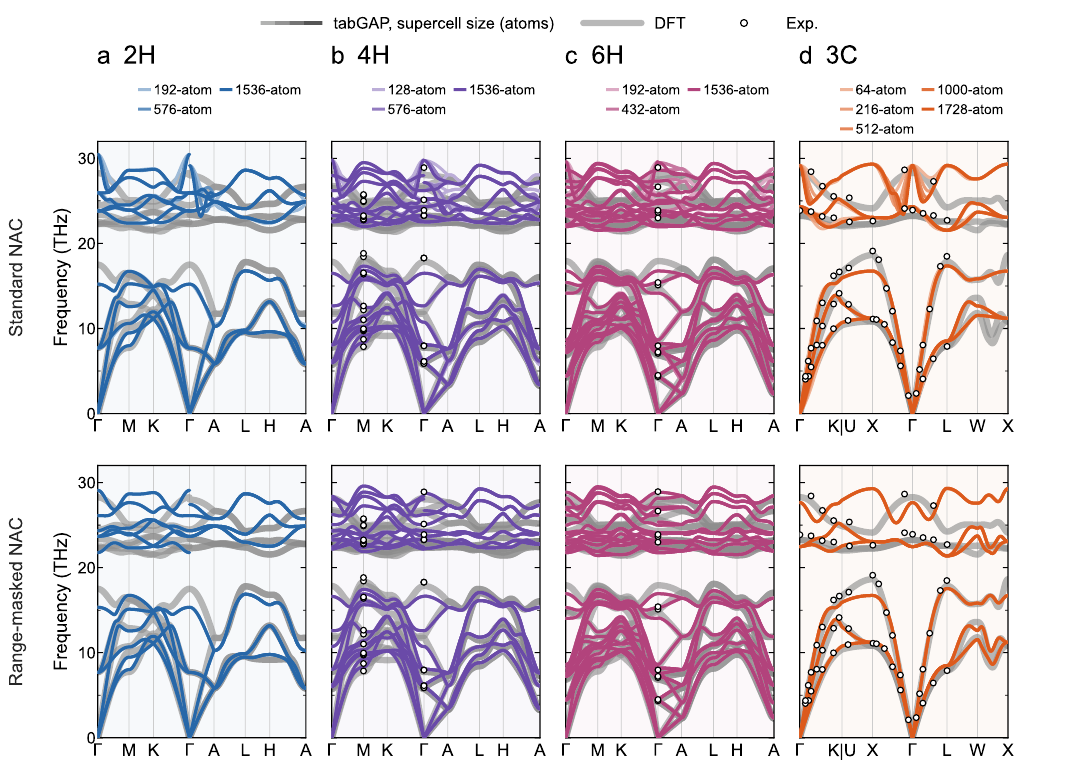}
\caption{\revision{Dipole correction for tabGAP force constants of 2H-, 4H-, 6H- and 3C-SiC, all evaluated at a fixed displacement of 0.01~\r A for this supercell comparison. Top: the standard finite-supercell correction applied to the tabGAP force constants (light to dark: small to large cells, with atom counts in the color keys), illustrating the cell-dependent interpolation near $\Gamma$. Bottom: the direct primitive-lattice near-field subtraction at $r_c=7$~\r A combined with the converged full Ewald sum. Grey: DFT with the same full dipole kernel subtracted on its actual force supercell and added back after short-range interpolation. At $\Gamma$, each segment retains its directional limit. Circles: inelastic x-ray scattering for 3C~\cite{SI:serrano2002determination}, Raman $\Gamma$ modes for 4H~\cite{SI:feldman1968phonon,SI:rao2023thermooptic} and 6H~\cite{SI:Bacherikov2020}, and $M$-point phonon energies for 4H~\cite{SI:klahold2020band}. The room-temperature Raman data of Rao et al. were measured along the $c$ axis. The $M$-point data were measured at 1.4--2~K, with reported energy uncertainties of approximately $\pm0.6$~meV.}}
\label{sfig:nac}
\end{figure*}

\revision{The phonon dispersions of the main text (Fig.~2) include the long-range dipole-dipole interaction through the non-analytical correction (NAC), which produces the splitting of the longitudinal and transverse optical (LO-TO) modes at $\Gamma$. For the 3C DFT reference, the primitive-cell relaxation and DFPT used a $16\times16\times16$ mesh, and the 64- and 216-atom finite-displacement calculations used $4\times4\times4$ and $3\times3\times3$ meshes, respectively, with an electronic convergence threshold of $10^{-8}$~eV.}
\revision{Figure~\ref{sfig:nac} compares the standard finite-supercell correction applied to tabGAP with the range-separated construction used in the main text.}

\revision{\textbf{Dipole correction and range separation.}
The standard interpolation of polar DFT force constants subtracts the dipole contribution folded on the actual force-constant supercell, interpolates the remaining short-range part and adds the long-range dynamical matrix~\cite{SI:gonze1997dynamical,SI:togo2023first}:
\begin{equation}
D(\mathbf q)=\mathrm{FT}[\Phi_{\mathrm{DFT}}-\Phi_{\mathrm{dd}}^{\mathrm{SC}}](\mathbf q)+D_{\mathrm{dd}}(\mathbf q).
\end{equation}
Here $\Phi_{\mathrm{dd}}^{\mathrm{SC}}$ is obtained from $D_{\mathrm{dd}}$ at all wave vectors commensurate with that supercell.
A local potential has no explicit dipole tail. Applying this subtraction to tabGAP therefore introduces a compensating tail whose cancellation depends on the wave vector and supercell (top row of Fig.~\ref{sfig:nac}).
For tabGAP we instead use a fixed real-space matching radius $r_c=7$~\r A:
\begin{equation}
D(\mathbf q)=\mathrm{FT}[\Phi_{\mathrm{tabGAP}}-\Phi_{\mathrm{dd}}^{r<r_c}](\mathbf q)+D_{\mathrm{dd}}(\mathbf q).
\end{equation}
The off-diagonal blocks of $\Phi_{\mathrm{dd}}^{r<r_c}$ are evaluated directly from the anisotropically screened Coulomb Hessian for primitive-lattice pairs inside $r_c$; each self block is the negative sum of its off-diagonal row.
No auxiliary supercell is used to construct these near-field blocks.
The full $D_{\mathrm{dd}}$ contains both real- and reciprocal-space Ewald sums, with the same acoustic-sum convention.
The fixed matching radius specifies how the dipole correction is joined to the local model; it was not fitted to the phonon dispersion.}

\revision{\textbf{Symmetry and numerical checks.}
We use the dielectric tensor and site-resolved Born tensors from our DFPT calculations, with the Born tensors mapped to the corresponding primitive sites.
Only symmetry-equivalent sites are averaged, and charge neutrality is imposed on the Born tensors.
The force-constant representation is projected onto primitive-cell space-group symmetry, permutation symmetry and the acoustic sum rule; for DFT this projection is applied after subtracting the folded dipole contribution.
The same matrices are used for dispersions and densities of states.
For all four polytypes, halving and doubling the Ewald separation parameter and independently extending the real- and reciprocal-space sums change the optical frequencies along the plotted paths by less than $10^{-4}$~THz.
Changing an auxiliary zero-force-constant storage cell leaves the dipole correction unchanged within numerical precision.
The supercell comparisons in Fig.~\ref{sfig:nac} use a fixed displacement of 0.01~\r A for every polytype.
For the two largest tested tabGAP force-constant supercells, the maximum frequency difference along the full paths is 0.0035, 0.0030, 0.0028 and 0.0019~THz for 2H, 4H, 6H and 3C, respectively (Fig.~\ref{sfig:nac}).
For 2H, reducing the displacement from $1.953125\times10^{-5}$ to $9.765625\times10^{-6}$~\r A changes full-path frequencies by at most 0.00081~THz. The smaller displacement was used in the main-text dispersion and density of states.
The other polytypes were not refined in displacement size; rotational-sum-rule convergence was not established.}

\revision{\textbf{Directional limits and comparison with DFT.}
At $\Gamma$, the non-analytical term is evaluated with the direction of the adjacent path segment.
The basal-plane $E_1$(LO) and axial $A_1$(LO) modes of a hexagonal polytype can therefore approach different frequencies at the same plotted $\Gamma$ point.
For 2H, the corrected tabGAP limits are 29.14 and 27.48~THz, compared with 28.26 and 28.00~THz in DFT.
The larger 2H splitting persists after convergence of the dipole sums and displacement-size check. The corrected model also differs from DFT in the optical branches and has a softer 3C longitudinal acoustic branch near $X$.
Small negative acoustic frequencies persist in narrow regions near $\Gamma$ in the hexagonal DFT paths (up to 0.25~THz in magnitude).
An explicit 2H check traces this behaviour to a linear-in-wave-vector dynamical-matrix residual already present in the input force constants; changing the Ewald parameters does not remove it.
All signed frequencies are retained in the numerical data; no negative mode is replaced by its absolute value.}

\clearpage

\subsection*{Supplementary Note 7: Robustness of tabGAP in the ultrahigh-temperature regime} \label{SuppNote_7}

\begin{figure*}[ht!]
\sifig{\linewidth}{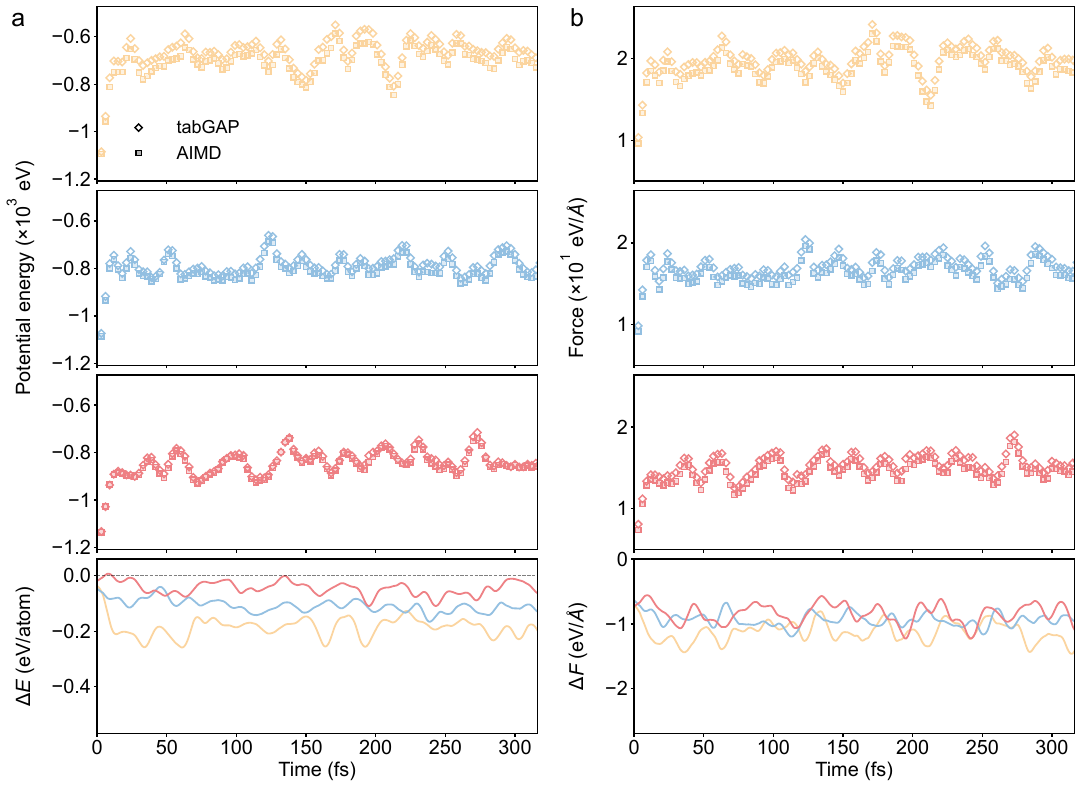}
\caption{Comparison of (a) potential energy and (b) \revision{mean atomic force magnitude $\langle|F|\rangle = N^{-1}\sum_i |\mathbf{F}_i|$ from tabGAP and AIMD at about 100,000~K. The three upper rows correspond to 216-atom cells whose lengths are changed isotropically by $-5\%$, $0\%$ and $+5\%$ from the reference cell; tabGAP (diamonds) is evaluated as single points on every AIMD frame (squares), and the curves are 8~fs running averages. The bottom panels show the deviations $\Delta E = (E_{\mathrm{AIMD}} - E_{\mathrm{tabGAP}})/N$ and $\Delta F = \langle|F|\rangle_{\mathrm{AIMD}} - \langle|F|\rangle_{\mathrm{tabGAP}}$. Yellow, blue and red denote the $-5\%$, $0\%$ and $+5\%$ cells.}}
\label{10w}
\end{figure*}

We evaluated tabGAP on AIMD trajectories at about 100,000~K for the three reference volumes used in the amorphous calculations: $L/L_0=0.95$, 1.00 and 1.05 (Fig.~\ref{10w}).
The length ratios describe the imposed volume relative to the room-temperature crystal; all three cells have high positive pressures, from about 45~GPa in the expanded cell to 150~GPa in the compressed cell.
\revision{Single-point tabGAP energies and forces were computed on every AIMD frame.
Over the displayed 0--316~fs interval, the mean energy offsets are 0.045, 0.11 and 0.19~eV/atom for the expanded, unstrained and compressed cells.
The corresponding force-component RMSEs are 2.3, 2.6 and 3.0~eV/\r A, or 17--19\% of the root-mean-square AIMD force.
The deviations increase with compression.}

\clearpage

\subsection*{Supplementary Note 8: Phase diagram calculation} \label{SuppNote_8}

\begin{figure*}[ht!]
\sifig{.85\linewidth}{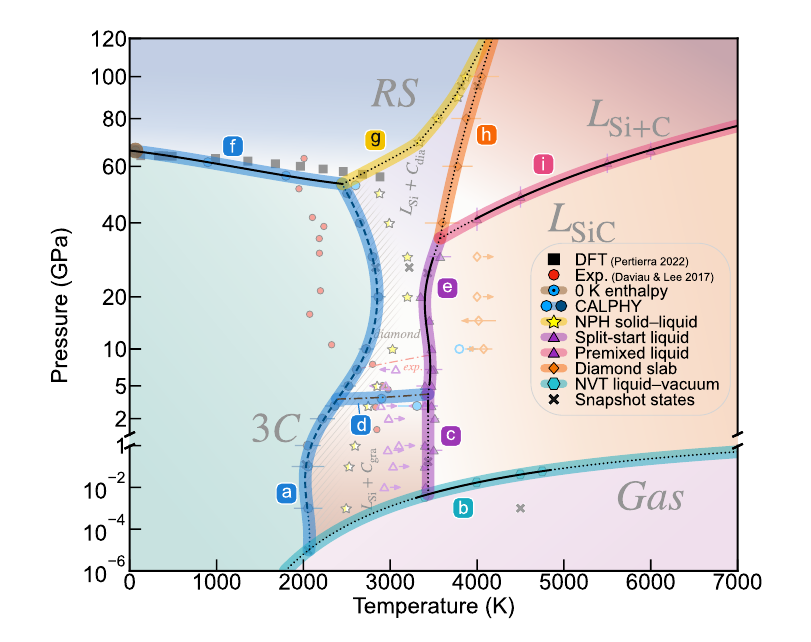}
\caption{\revision{Methods used to construct the $P$-$T$ diagram of SiC (Fig.~4a of the main text). Colored bands identify the method (legend); the letters progress from the low-pressure region to the high-pressure region: (a) 3C | $L_\mathrm{Si}+\mathrm{C}$, \texttt{CALPHY} common tangent; (b) evaporation estimate from $NVT$ liquid-vacuum slabs; (c) $L_\mathrm{Si}+C_\mathrm{gra}$ | $L_\mathrm{SiC}$, graphite dissolution from phase-separated starts; (d) graphite | diamond, \texttt{CALPHY}; (e) $L_\mathrm{Si}+C_\mathrm{dia}$ | $L_\mathrm{SiC}$, diamond dissolution from phase-separated starts; (f) 3C | RS, fixed-temperature Frenkel-Ladd integration and the static 0~K enthalpy crossing (brown disc); (g) RS | $L_\mathrm{Si}+C_\mathrm{dia}$, two-phase $NPH$; (h) $L_\mathrm{Si}+C_\mathrm{dia}$ | $L_\mathrm{Si+C}$, diamond slabs in liquid Si; (i) $L_\mathrm{SiC}$ | $L_\mathrm{Si+C}$, demixing of premixed liquids. Dissolution estimates (c, e) use phase-separated runs and liquid compositions. Dotted pieces are not located. The red dash-dot line is the experimental graphite-diamond line; the hatched field is metastable 3C up to the two-phase $NPH$ superheating limit (stars).}}
\label{ptindicator}
\end{figure*}

\revision{Figure~\ref{ptindicator} identifies the Gibbs-energy, interface, dissolution and demixing calculations used for segments (a)--(i).
We used \texttt{LAMMPS} and \texttt{CALPHY} 2.1.2~\cite{SI:menon2021automated}; Gibbs energies are per atom.
Filled symbols mark estimates, open symbols with arrows mark bounds, and dotted lines show unresolved segments.}

\subsubsection*{Simulation settings}

Unless otherwise stated, molecular-dynamics simulations were carried out in \texttt{LAMMPS} with a time step of 0.5~fs, Nos\'e-Hoover thermostats with a damping constant of 0.1~ps and, where the pressure was controlled, Nos\'e-Hoover (MTK) barostats with a damping constant of 1.0~ps.
\revision{The direct simulations used cells of 1728 atoms (Si$_{1-x}$C$_x$ liquids, $NPT$, isotropic coupling), 2800 atoms (a diamond slab in liquid Si, $NPT$, anisotropic coupling, followed in the 800~ps runs by coupling along the slab normal only), 5184 or 12,288 atoms (two-phase 3C/liquid and rock-salt/liquid cells, $NPH$ with the barostat acting along the interface normal only) and 6144 atoms (liquid slabs with a vacuum gap of three slab lengths, $NVT$).
Production runs lasted 80--100~ps for the demixing of the premixed liquid, 150--800~ps for the diamond slabs, 500~ps for the liquid-vacuum slabs and the rock-salt/liquid cells, 600--900~ps for the split-start liquids and 500--2000~ps for the 3C/liquid cells.
Free energies were computed with \texttt{CALPHY} in cells of 1000 atoms (3C, rock salt, diamond, liquids) and 960 atoms (graphite), with Nos\'e-Hoover thermostat and barostat damping constants of 0.1~ps for equilibration and reversible scaling, and a Langevin thermostat with a damping constant of 0.1~ps during the Frenkel-Ladd and Uhlenbeck-Ford ($p = 50$, $\sigma = 1.5$~\r A) switching; the pressure was coupled anisotropically for graphite and isotropically otherwise.
The reversible-scaling calculations used a 0.5~fs time step, 20,000 equilibration steps, 60,000 (solids) or 100,000 (liquids) switching steps and 200,000--370,000 sweep steps in each direction.
The fixed-temperature Frenkel-Ladd calculations used 0.5~fs and 60,000--150,000 switching steps for graphite and diamond, and 1~fs with 100,000 switching steps for the 3C-RS boundary.}

\subsubsection*{\revision{(a) Decomposition of 3C into liquid silicon and carbon}}

\revision{The Gibbs energies $G(T)$ of 3C, of the carbon phase (graphite or diamond) and of Si$_{1-x}$C$_x$ liquids with $x = 0$, 0.05, 0.1, 0.2 and 0.3 were computed by Frenkel-Ladd (solids) or Uhlenbeck-Ford (liquids) integration at a reference temperature, followed by a reversible-scaling sweep in temperature.
We fitted $g_L(x)=h(x)+k_\mathrm{B}T[x\ln x+(1-x)\ln(1-x)]$, with $h(x)$ a cubic fit after subtracting ideal mixing from the calculated Gibbs energies.
At overall composition $x_\mathrm{C}=0.5$, the lever rule gives a carbon-phase fraction $(0.5-x_L)/(1-x_L)$. The common-tangent mixture has Gibbs energy
\begin{equation}
G_\mathrm{mix}(0.5) = \min_{x_L}\left[g_L(x_L) + \left(g_\mathrm{C} - g_L(x_L)\right)\frac{0.5 - x_L}{1 - x_L}\right], \qquad g_\mathrm{C} = \min(g_\mathrm{graphite}, g_\mathrm{diamond}),
\end{equation}
and the decomposition temperature $T_\mathrm{dec}$ is the zero of $G_\mathrm{mix}(0.5) - G_\mathrm{3C}$.
Above about 1~GPa, the reversible-scaling barostat used $\lambda P$, with switching parameter $\lambda$ and target pressure $P$. At 30~GPa, the resulting liquid-Si free energy agrees with an independent fixed-temperature integration within 0.1~meV/atom.
The decomposition temperatures (Table~\ref{tab:pt_b}) rise from 2055~K at 10~bar to 2863~K at 20~GPa, then fall to 2504~K at 50~GPa. The plotted spline differs from these points by at most 34~K.
Two-phase $NPH$ simulations of 3C in contact with a homogeneous melt (5184 or 12,288 atoms) give the superheating edge of 3C, 300--550~K higher (2453--2532~K at 1~MPa, 3024--3029~K at 10~GPa, 3198~K at 20 and 30~GPa); they delimit the hatched superheated 3C field in Fig.~4a.}

\begin{table}[ht!]
\caption{\revision{Located points of the decomposition tie line (a). The uncertainty is the summed dissipation bound of the free-energy integrations; $x_\mathrm{C}$ is the carbon fraction of the tangent liquid.}}
\label{tab:pt_b}
\begin{ruledtabular}
\begin{tabular}{c c c c c}
$P$ (GPa) & $T_\mathrm{dec}$ (K) & $\pm$ (K) & carbon phase & $x_\mathrm{C}$ \\ \hline
\revision{0.001} & \revision{2055} & \revision{163} & \revision{graphite} & \revision{0.06} \\
\revision{0.1} & \revision{2054} & \revision{188} & \revision{graphite} & \revision{0.04} \\
\revision{1} & \revision{2051} & \revision{148} & \revision{graphite} & \revision{0.04} \\
\revision{2} & \revision{2219} & \revision{145} & \revision{graphite} & \revision{0.05} \\
\revision{3} & \revision{2358} & \revision{109} & \revision{graphite} & \revision{0.08} \\
\revision{5} & \revision{2517} & \revision{89} & \revision{diamond} & \revision{0.14} \\
\revision{7} & \revision{2615} & \revision{58} & \revision{diamond} & \revision{0.17} \\
\revision{10} & \revision{2723} & \revision{87} & \revision{diamond} & \revision{0.19} \\
\revision{20} & \revision{2863} & \revision{83} & \revision{diamond} & \revision{0.25} \\
\revision{30} & \revision{2796} & \revision{54} & \revision{diamond} & \revision{0.25} \\
\revision{40} & \revision{2650} & \revision{52} & \revision{diamond} & \revision{0.23} \\
\revision{50} & \revision{2504} & \revision{49} & \revision{diamond} & \revision{0.21} \\
\end{tabular}
\end{ruledtabular}
\end{table}

\subsubsection*{(b) Evaporation estimate: liquid-vacuum simulations}

\revision{A liquid SiC slab was placed in an $NVT$ cell with a vacuum of three slab lengths along $x$ and run for 500~ps; the last half of each run was analyzed.
Atoms in clusters of fewer than 20 atoms (connectivity cutoff 3.0~\r A) outside the slab were counted as vapor, $N_\mathrm{vap}$, and the vacuum volume was taken from the instantaneous box, $V_\mathrm{vac} = (L_x - t)\,L_yL_z$, where $t$ is the slab thickness between the two Gibbs dividing surfaces at half the liquid density.}
From these counts we calculated \begin{equation} P_\mathrm{atom} = \frac{\langle N_\mathrm{vap}\rangle k_\mathrm{B}T}{V_\mathrm{vac}}, \end{equation}
\revision{with block-mean standard errors. The atom count is roughly 7--8 times the number of identified vapor clusters, so $P_\mathrm{atom}$ does not give the ideal-gas pressure of independent vapor species. Stationarity was checked from the two half-window vapor populations and the drift of the carbon network.
The resulting values are $35.3\pm6.1$, $172.8\pm10.5$, $415.1\pm12.9$ and $584.5\pm17.2$~bar at 3397, 3995, 4498 and 4751~K.}
Their temperature dependence was fitted in the Clausius-Clapeyron form~\cite{SI:koutsoyiannis2012clausius}
\begin{equation}
\ln(P_\mathrm{atom}/\mathrm{bar}) = A + \frac{B}{T}, \qquad \revision{A = 13.05\pm0.30,\quad B = -31{,}644\pm1361~\mathrm{K}},
\end{equation}
\revision{with inverse-variance weights on $\ln P_\mathrm{atom}$ ($R^2 = 0.997$, reduced $\chi^2 = 0.83$).
The line is solid over the fitted range plus 100~K at each end; extrapolations are dotted.}

\subsubsection*{(c) Dissolution of graphite into the 1:1 liquid}

\revision{Phase-separated 1:1 Si-C cells were simulated under $NPT$ for 600--900~ps and analyzed over the last 200~ps. Carbon clusters were identified with a 1.85~\r A C--C cutoff. We measured the fraction $f_C$ of all carbon atoms in clusters of at least 30 atoms and the carbon fraction $x_\mathrm{liq}$ among atoms more than 2.6~\r A from those clusters. Runs were classified as nearly homogeneous when $\langle x_\mathrm{liq}\rangle\geq0.47$ or $\langle f_C\rangle\leq0.10$, and as two-phase when $\langle x_\mathrm{liq}\rangle\leq0.45$, $\langle f_C\rangle\geq0.15$ and the fitted decline in $f_C$ was no greater than 0.03 per 100~ps. These are classification tolerances; intermediate states were left unresolved.
Linear composition fits, anchored at $x_\mathrm{liq}=0.5$ for the first nearly homogeneous run, gave 3407, 3398, 3500, 3400 and 3514~K at 0.01, 0.1, 0.6, 1 and 2~GPa ($\pm50$--100~K).
At 3~GPa, a 400~ps series bracketed dissolution at 3400--3550~K, whereas 600~ps runs became homogeneous at 3400~K. These are finite-time estimates.
A \texttt{CALPHY} common tangent at 3~GPa gives $3308\pm220$~K. The five composition fits, the 400~ps bracket midpoint and this common-tangent result give the plotted weighted mean of 3436~K.
}

\subsubsection*{(d) Graphite | diamond inside the decomposed field}

\revision{The carbon precipitated by the decomposition is graphite below and diamond above the pure-carbon coexistence line, which was obtained from pressure-consistent reversible-scaling calculations of graphite and diamond.
The compact and expanded graphite branches were swept separately because of a first-order volume change near 2900--3100~K; the lower-free-energy, expanded branch was used at each temperature.
The crossings, 3.66, 3.73, 3.83, 4.11 and 4.60~GPa at 2500, 2800, 3000, 3400 and 4000~K, are described by
\begin{equation}
P_\mathrm{gd} = 4.7543 - 1.1206\times10^{-3}\,T + 2.7245\times10^{-7}\,T^2 \quad (\mathrm{GPa},\ \mathrm{K}),
\end{equation}
with a largest residual of 0.036~GPa.
Fixed-temperature Frenkel-Ladd calculations of both phases at five pressures (three seeds each), give $3.67\pm0.04$~GPa at 2900~K and $4.02\pm0.03$~GPa at 3400~K, within 0.13~GPa of the line; the slope of $\Delta G$ with pressure, 24~meV/atom per GPa, equals the volume difference of the two phases.
The line is drawn from its junction with (a) at 2390~K to its junction with (c) at 3436~K.}

\subsubsection*{(e) Dissolution of diamond into the 1:1 liquid}

\revision{The phase-separated protocol and phase criteria are the same as for graphite in (c).
Above the graphite-diamond line (d) the brackets (highest two-phase/lowest homogeneous temperature) are 3450/3500~K at 5, 7 and 10~GPa, 3400/3500~K at 15~GPa, 3300/3400~K at 20~GPa and 3450/3700~K at 30~GPa.
The plotted curve minimizes integrated squared curvature within these temperature brackets.
The separate diamond-slab simulations are described in (h).}

\subsubsection*{(f) 3C | rock salt: static enthalpy and fixed-temperature free energies}

\revision{At 0~K the transition pressure is the enthalpy crossing of the equations of state of 3C and rock salt, 66.3~GPa (brown disc in Fig.~\ref{ptindicator}).
At each finite temperature, Frenkel-Ladd calculations used 1000-atom cells at four pressures, with spring constants obtained from mean-square displacements.
The transition pressure is the zero of a linear fit to $G_\mathrm{RS}-G_\mathrm{3C}$. Its slope agrees with the calculated volume difference within 1\%; both are $-9.80$~meV/(atom GPa) at 1800~K.
The crossings are $61.62\pm0.2$, $56.42\pm0.2$ and $52.84\pm0.3$~GPa at 900, 1800 and 2600~K, with dissipation below 3~meV/atom.
The line is a smoothing spline through the 0, 900 and 1800~K points, ending at the decomposition line.
The DFT-PBE quasi-harmonic boundary of Ref.~\cite{SI:pertierra2022pressure} gives 64.2~GPa at 127~K and 63.2, 60.6 and 57.3~GPa at 900, 1800 and 2600~K (interpolated), within 2.1~GPa of tabGAP below 1000~K and 4.2--4.5~GPa above it at 1800--2600~K.}

\subsubsection*{(g) Decomposition of rock salt}

\revision{Rock-salt/melt interfaces simulated for 500~ps in $NPH$ reached $3335\pm8$, $3544\pm12$, $3779\pm17$ and $3868\pm11$~K at 70, 80, 90 and 100~GPa. These temperatures still drifted and are treated as upper bounds.
The dotted quadratic starts at the (f)/(a) junction, 53.4~GPa and 2454~K, and remains below all four bounds.}

\subsubsection*{(h) Diamond in liquid silicon at high pressure}

\revision{A 1400-atom diamond slab in contact with 1400 liquid-Si atoms was simulated under $NPT$. Crystalline carbon required four $q_6$-aligned carbon neighbors within 1.95~\r A.
At 40, 60, 80 and 100~GPa, diamond order persists at 3400, 3600, 3705 and 3850~K and is lost at 3798, 3950, 4047 and 4250~K. The bracket midpoints and half-widths are $3599\pm199$, $3775\pm175$, $3876\pm171$ and $4050\pm200$~K.
The line is dotted because loss of diamond order does not establish dissolution: 300--525~ps runs at 40--80~GPa retained segregated carbon up to 4400--4800~K.}

\subsubsection*{(i) Demixing of the 1:1 liquid}

\revision{A homogeneous 1:1 liquid of 1728 atoms, melted at 7000~K, was held under $NPT$ at fixed temperature at a sequence of pressures.
We identified demixing from persistent carbon-rich clusters and increased carbon-composition fluctuations in $3^3$ and $4^3$ subcells, supported by the evolution of C--C coordination. The fluctuations were normalized by the binomial random-mixing reference; homogeneous samples remained close to this reference. At 4500~K, phase-separated starts converged to the same states.
The brackets (highest homogeneous/lowest demixed pressure) give $41.25\pm3.75$, $48.75\pm3.75$, $61.25\pm3.75$ and $66.25\pm3.75$~GPa at 4000, 4500, 5500 and 6000~K, through which a smoothing spline is drawn.}

\subsubsection*{Snapshot states}

The states shown in Fig.~4b of the main text are placed on the boundaries of this diagram according to their phase content (Table~\ref{tab:pt_snap}).

\begin{table}[ht!]
\caption{\revision{Snapshot states of Fig.~4b of the main text.}}
\label{tab:pt_snap}
\begin{ruledtabular}
\begin{tabular}{c c c p{0.62\columnwidth}}
State & $T$ (K) & $P$ (GPa) & Content \\ \hline
\revision{i} & \revision{2055} & \revision{$9.5\times10^{-6}$} & \revision{Si leaves the surface, graphitic carbon remains} \\
\revision{ii} & \revision{2028} & \revision{0.127} & \revision{incongruent melting: 3C, liquid Si and graphite (on a)} \\
\revision{iii} & \revision{3436} & \revision{0.171} & \revision{graphite dissolving into the 1:1 liquid (on c)} \\
\revision{iv} & \revision{4500} & \revision{0.001} & \revision{vapor of atoms and small molecules} \\
\revision{v} & \revision{2831} & \revision{27.1} & \revision{decomposition with diamond clusters (on a)} \\
\revision{vi} & \revision{3219} & \revision{27.1} & \revision{liquid Si and diamond (superheating edge)} \\
\revision{vii} & \revision{3432} & \revision{25.7} & \revision{homogeneous 1:1 liquid (on e)} \\
\revision{viii} & \revision{3814} & \revision{95.3} & \revision{rock salt, liquid Si and diamond (on g)} \\
\revision{ix} & \revision{4007} & \revision{95.3} & \revision{diamond, liquid carbon and liquid Si (on h)} \\
\end{tabular}
\end{ruledtabular}
\end{table}

\clearpage

\subsection*{Supplementary Note 9: \revision{Point defects and threshold displacement energies}} \label{SuppNote_9}

This note gives the point-defect formation energies of the SiC polytypes, vacancy migration in 3C-SiC and the low-index threshold displacement energies (TDEs). DFT provides the energetic references, and AIMD and experiment provide additional comparisons for the TDEs.

\begin{figure}[ht!]
\sifig{\linewidth}{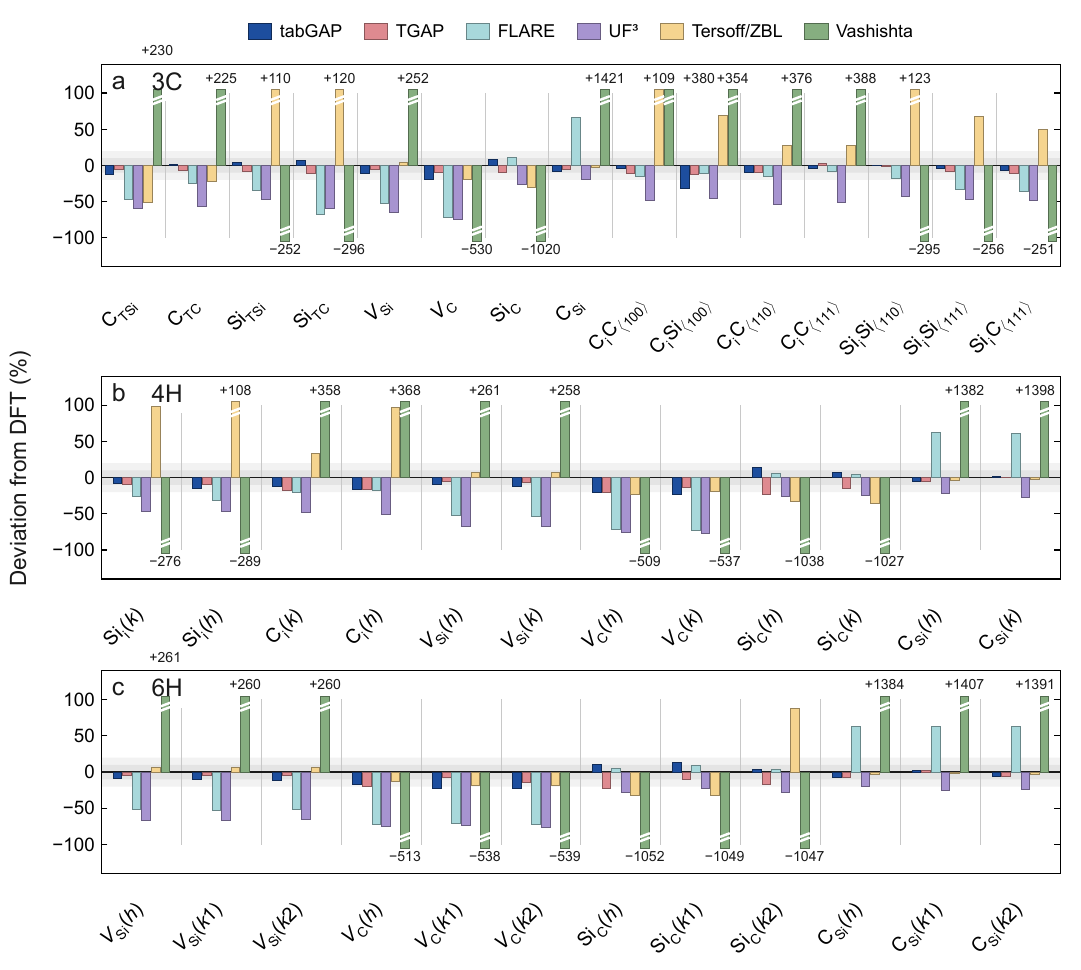}
\caption{\revision{Point-defect formation energies in (a) 3C-, (b) 4H- and (c) 6H-SiC, shown as the signed deviation from our DFT results, $100\,(E_\mathrm{f}/E_\mathrm{f}^\mathrm{DFT}-1)$. Grey bands mark $\pm10\%$ and $\pm20\%$. All defects are neutral and use the Si-rich chemical potentials of the method being evaluated. Labels identify the starting configurations, with $h$ and $k$ denoting the sites in 4H and 6H. Bars beyond $\pm100\%$ are broken and labelled with their values. Absolute energies and the separate DFT sources are listed in Table~\ref{tab:Ef_abs}.}}
\label{fig:Ef}
\end{figure}

\textbf{Formation energies.}
\revision{Figure~\ref{fig:Ef} compares point-defect formation energies in 3C-, 4H- and 6H-SiC from tabGAP, TGAP, UF$^{3}$, FLARE, Tersoff/ZBL and Vashishta with our DFT results. The plot shows deviations from DFT; absolute energies and literature values are listed in Table~\ref{tab:Ef_abs}.
Our DFT and IAP values are for neutral defects in the Si-rich limit, each method taking its chemical potentials from its own reservoirs ($\mu_\mathrm{Si}$ from diamond Si and $\mu_\mathrm{C} = E_\mathrm{SiC} - \mu_\mathrm{Si}$, with $E_\mathrm{SiC}$ per SiC formula unit).
The tabGAP calculations used 64-atom (3C), 72-atom (4H) and 108-atom (6H) reference supercells at its own lattice parameters.
The tetrahedral interstitials are labelled as in the literature: $\mathrm{C}_{T\mathrm{Si}}$, for example, is a C atom surrounded by four Si atoms.
In the IAP relaxations, the guest atom was fixed in the four 3C tetrahedral configurations, and 3C dumbbell atoms could move only along their initial axis, except for the freely relaxed $\mathrm{C}_{\mathrm{i}}\mathrm{C}_{\langle100\rangle}$ state. Other coordinates were relaxed at fixed cell. DFT used symmetry-enabled fixed-cell relaxation without selective-coordinate constraints. Tersoff/ZBL and Vashishta retained experimental reference cells for 22 and 24 free-relaxation cases, respectively. The labels identify starting configurations, not necessarily a common relaxed defect identity; $h$ and $k$ distinguish the 4H sites, and the 6H comparison includes vacancies and antisites. Literature DFT values are listed separately for context.
}

\revision{tabGAP follows the DFT trends of the interstitial configurations and antisites across the three polytypes.
In 3C the mean absolute deviation from our DFT results is 0.65~eV, with a median absolute relative deviation of 7.3\%.
Over the 39 defects, the mean absolute deviation from our DFT results is 0.64~eV for tabGAP and 0.58~eV for TGAP; 34 of the 39 tabGAP values lie within 20\% of DFT. The complete comparison is given in Table~\ref{tab:benchmark}.}

\revision{For the 21 vacancy and interstitial configurations in 3C and 4H, the mean absolute deviation from our DFT calculations is 0.80~eV, and the mean absolute relative deviation is 11.1\%.
The 15 corresponding vacancy--interstitial sums at infinite separation give a mean absolute deviation of 1.47~eV (10.7\%).
}

\begin{table*}[htbp]
\caption{Formation energies (eV) for the initial defect configurations of Fig.~\ref{fig:Ef}. \revision{Our DFT and IAP values are for neutral defects in the Si-rich limit,
with each method using its own chemical potentials. The deviations in Fig.~\ref{fig:Ef} use our DFT results.
Literature DFT values are PBE/Si-rich for 3C~\cite{SI:hamedani2025sic}, HSE06/C-rich for 4H~\cite{SI:kobayashi2019native}, and SCAN/Si-rich for 6H~\cite{SI:wang2025scan}. Large negative energies reflect reconstruction during relaxation. --: not available.}}
\label{tab:Ef_abs}
\footnotesize
\renewcommand{\arraystretch}{0.86}
\begin{ruledtabular}
\begin{tabular}{l c c c c c c c c}
Defect & \shortstack{DFT\\(this work)} & \shortstack{Literature\\DFT} & tabGAP & TGAP & FLARE & UF$^{3}$ & Tersoff/ZBL & Vashishta \\ \hline
\multicolumn{9}{l}{\textbf{3C-SiC}} \\
$\mathrm{C}_{\mathrm{TSi}}$ & \revision{10.01} & 9.96 & \revision{8.72} & \revision{9.52} & \revision{5.35} & \revision{4.09} & \revision{4.93} & \revision{32.97} \\
$\mathrm{C}_{\mathrm{TC}}$ & \revision{11.29} & 10.96 & \revision{11.39} & \revision{10.49} & \revision{8.46} & \revision{4.95} & \revision{8.74} & \revision{36.67} \\
$\mathrm{Si}_{\mathrm{TSi}}$ & \revision{10.85} & 10.21 & \revision{11.25} & \revision{10.01} & \revision{7.10} & \revision{5.74} & \revision{22.81} & \revision{-16.51} \\
$\mathrm{Si}_{\mathrm{TC}}$ & \revision{9.14} & 8.46 & \revision{9.75} & \revision{8.13} & \revision{2.94} & \revision{3.68} & \revision{20.07} & \revision{-17.88} \\
$V_{\mathrm{Si}}$ & \revision{7.76} & 7.72 & \revision{6.92} & \revision{7.33} & \revision{3.73} & \revision{2.74} & \revision{8.10} & \revision{27.34} \\
$V_{\mathrm{C}}$ & \revision{4.20} & 4.20 & \revision{3.38} & \revision{3.78} & \revision{1.16} & \revision{1.10} & \revision{3.38} & \revision{-18.04} \\
$\mathrm{Si}_{\mathrm{C}}$ & \revision{3.56} & 3.33 & \revision{3.85} & \revision{3.23} & \revision{3.97} & \revision{2.64} & \revision{2.50} & \revision{-32.71} \\
$\mathrm{C}_{\mathrm{Si}}$ & \revision{4.04} & 3.94 & \revision{3.71} & \revision{3.84} & \revision{6.73} & \revision{3.24} & \revision{3.95} & \revision{61.49} \\
$\mathrm{C}_{\mathrm{i}}\mathrm{C}_{\langle 100 \rangle}$ & \revision{6.86} & 6.79 & \revision{6.59} & \revision{6.11} & \revision{5.84} & \revision{3.55} & \revision{14.32} & \revision{32.97} \\
$\mathrm{C}_{\mathrm{i}}\mathrm{Si}_{\langle 100 \rangle}$ & \revision{7.84} & 7.60 & \revision{5.36} & \revision{6.84} & \revision{6.96} & \revision{4.31} & \revision{13.29} & \revision{35.64} \\
$\mathrm{C}_{\mathrm{i}}\mathrm{C}_{\langle 110 \rangle}$ & \revision{7.36} & 7.14 & \revision{6.62} & \revision{6.64} & \revision{6.29} & \revision{3.40} & \revision{9.45} & \revision{35.03} \\
$\mathrm{C}_{\mathrm{i}}\mathrm{C}_{\langle 111 \rangle}$ & \revision{7.37} & -- & \revision{7.08} & \revision{7.57} & \revision{6.76} & \revision{3.57} & \revision{9.46} & \revision{35.92} \\
$\mathrm{Si}_{\mathrm{i}}\mathrm{Si}_{\langle 110 \rangle}$ & \revision{8.65} & 8.37 & \revision{8.62} & \revision{8.53} & \revision{7.14} & \revision{4.94} & \revision{19.30} & \revision{-16.91} \\
$\mathrm{Si}_{\mathrm{i}}\mathrm{Si}_{\langle 111 \rangle}$ & \revision{10.71} & -- & \revision{10.25} & \revision{9.90} & \revision{7.10} & \revision{5.73} & \revision{18.04} & \revision{-16.73} \\
$\mathrm{Si}_{\mathrm{i}}\mathrm{C}_{\langle 111 \rangle}$ & \revision{11.06} & -- & \revision{10.25} & \revision{9.90} & \revision{7.10} & \revision{5.73} & \revision{16.55} & \revision{-16.73} \\
\hline
\multicolumn{9}{l}{\textbf{4H-SiC}} \\
$\mathrm{Si}_{\mathrm{i}}(k)$ & \revision{9.58} & 12.46 & \revision{8.84} & \revision{8.62} & \revision{7.14} & \revision{5.07} & \revision{19.04} & \revision{-16.90} \\
$\mathrm{Si}_{\mathrm{i}}(h)$ & \revision{9.00} & 10.87 & \revision{7.69} & \revision{8.15} & \revision{6.21} & \revision{4.82} & \revision{18.70} & \revision{-17.04} \\
$\mathrm{C}_{\mathrm{i}}(k)$ & \revision{7.24} & 6.32 & \revision{6.34} & \revision{5.94} & \revision{5.79} & \revision{3.74} & \revision{9.69} & \revision{33.12} \\
$\mathrm{C}_{\mathrm{i}}(h)$ & \revision{7.07} & 6.21 & \revision{5.92} & \revision{5.93} & \revision{5.85} & \revision{3.51} & \revision{13.90} & \revision{33.12} \\
$V_{\mathrm{Si}}(h)$ & \revision{7.58} & 7.10 & \revision{6.86} & \revision{7.18} & \revision{3.63} & \revision{2.48} & \revision{8.10} & \revision{27.34} \\
$V_{\mathrm{Si}}(k)$ & \revision{7.60} & 7.16 & \revision{6.68} & \revision{7.12} & \revision{3.53} & \revision{2.49} & \revision{8.10} & \revision{27.26} \\
$V_{\mathrm{C}}(h)$ & \revision{4.41} & 4.88 & \revision{3.49} & \revision{3.52} & \revision{1.25} & \revision{1.07} & \revision{3.38} & \revision{-18.03} \\
$V_{\mathrm{C}}(k)$ & \revision{4.15} & 4.84 & \revision{3.18} & \revision{3.61} & \revision{1.12} & \revision{0.95} & \revision{3.37} & \revision{-18.16} \\
$\mathrm{Si}_{\mathrm{C}}(h)$ & \revision{3.70} & 4.63 & \revision{4.23} & \revision{2.85} & \revision{3.89} & \revision{2.73} & \revision{2.47} & \revision{-34.74} \\
$\mathrm{Si}_{\mathrm{C}}(k)$ & \revision{3.66} & 4.58 & \revision{3.92} & \revision{3.10} & \revision{3.80} & \revision{2.75} & \revision{2.36} & \revision{-33.96} \\
$\mathrm{C}_{\mathrm{Si}}(h)$ & \revision{4.11} & 2.65 & \revision{3.87} & \revision{3.87} & \revision{6.65} & \revision{3.21} & \revision{3.96} & \revision{60.90} \\
$\mathrm{C}_{\mathrm{Si}}(k)$ & \revision{4.07} & 2.61 & \revision{4.11} & \revision{4.10} & \revision{6.54} & \revision{2.94} & \revision{3.96} & \revision{61.03} \\
\hline
\multicolumn{9}{l}{\textbf{6H-SiC}} \\
$V_{\mathrm{Si}}(h)$ & \revision{7.57} & 7.67 & \revision{6.87} & \revision{7.21} & \revision{3.63} & \revision{2.52} & \revision{8.10} & \revision{27.35} \\
$V_{\mathrm{Si}}(k1)$ & \revision{7.60} & 7.69 & \revision{6.85} & \revision{7.23} & \revision{3.57} & \revision{2.51} & \revision{8.10} & \revision{27.35} \\
$V_{\mathrm{Si}}(k2)$ & \revision{7.57} & 7.73 & \revision{6.72} & \revision{7.21} & \revision{3.70} & \revision{2.62} & \revision{8.10} & \revision{27.24} \\
$V_{\mathrm{C}}(h)$ & \revision{4.36} & 4.59 & \revision{3.63} & \revision{3.53} & \revision{1.22} & \revision{1.07} & \revision{3.77} & \revision{-18.03} \\
$V_{\mathrm{C}}(k1)$ & \revision{4.15} & 4.51 & \revision{3.23} & \revision{3.84} & \revision{1.20} & \revision{1.08} & \revision{3.38} & \revision{-18.18} \\
$V_{\mathrm{C}}(k2)$ & \revision{4.10} & 4.58 & \revision{3.18} & \revision{3.54} & \revision{1.12} & \revision{0.95} & \revision{3.37} & \revision{-18.02} \\
$\mathrm{Si}_{\mathrm{C}}(h)$ & \revision{3.66} & 3.76 & \revision{4.06} & \revision{2.84} & \revision{3.85} & \revision{2.62} & \revision{2.47} & \revision{-34.81} \\
$\mathrm{Si}_{\mathrm{C}}(k1)$ & \revision{3.66} & 3.74 & \revision{4.14} & \revision{3.27} & \revision{4.01} & \revision{2.83} & \revision{2.48} & \revision{-34.71} \\
$\mathrm{Si}_{\mathrm{C}}(k2)$ & \revision{3.64} & 3.76 & \revision{3.75} & \revision{3.03} & \revision{3.79} & \revision{2.61} & \revision{6.84} & \revision{-34.46} \\
$\mathrm{C}_{\mathrm{Si}}(h)$ & \revision{4.10} & 4.49 & \revision{3.80} & \revision{3.82} & \revision{6.66} & \revision{3.26} & \revision{3.96} & \revision{60.80} \\
$\mathrm{C}_{\mathrm{Si}}(k1)$ & \revision{4.06} & 4.47 & \revision{4.15} & \revision{4.13} & \revision{6.60} & \revision{3.02} & \revision{3.96} & \revision{61.19} \\
$\mathrm{C}_{\mathrm{Si}}(k2)$ & \revision{4.08} & 4.44 & \revision{3.84} & \revision{3.82} & \revision{6.66} & \revision{3.11} & \revision{3.96} & \revision{60.87} \\
\end{tabular}
\end{ruledtabular}
\end{table*}

\begin{figure}[ht!]
\centering
\sifig{\linewidth}{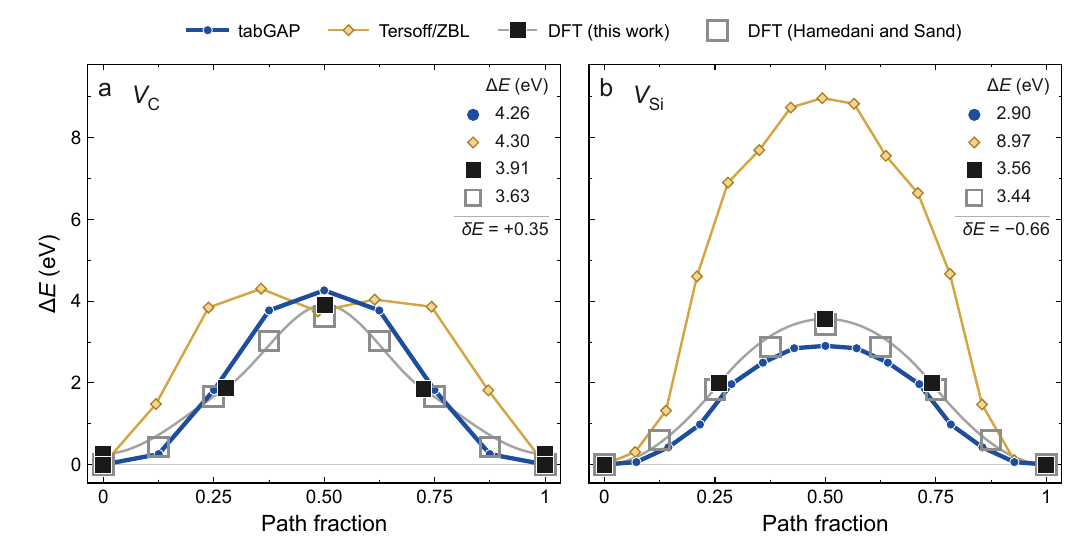}
\caption{\revision{Vacancy migration in 3C-SiC: (a)~$V_\mathrm{C}$ and (b)~$V_\mathrm{Si}$ hopping to a neighbouring site of the same sublattice. Lines and circles: tabGAP; lines and diamonds: Tersoff/ZBL~\cite{SI:devanathan1998displacement}; filled squares: our DFT, with a grey line as a guide to the eye; open squares: DFT profiles of Hamedani and Sand~\cite{SI:hamedani2025sic}.  The path fraction is the cumulative atomic displacement along each path, and energies are relative to the initial vacancy. $\Delta E$ lists the barriers of tabGAP, Tersoff/ZBL, our DFT and Ref.~\cite{SI:hamedani2025sic}, and $\delta E$ the difference between tabGAP and our DFT. In (a), our DFT barrier is the stationary midpoint relative to the re-relaxed vacancy, which lies 0.26~eV below the band endpoints. For Tersoff/ZBL in (b) the converged CI-NEB profile is shown.}}
\label{sfig:migration3c}
\end{figure}

\revision{\textbf{Vacancy migration.}
Fig.~\ref{sfig:migration3c} compares the migration of the carbon and the silicon vacancy to a neighbouring site of their own sublattice in 3C-SiC, in a 63-atom cell (2$\times$2$\times$2 conventional cell at the DFT lattice parameter).
The DFT paths were computed with the nudged elastic band method (PBE, 900~eV, 3$\times$3$\times$3 $\Gamma$-centred $k$-points, forces below 0.03~eV/\r A) and the tabGAP and Tersoff/ZBL~\cite{SI:devanathan1998displacement} paths with the climbing-image method (9 and 15 images, forces below 0.01~eV/\r A), each potential at its own lattice parameter.
tabGAP gives barriers of 4.26~eV for $V_\mathrm{C}$ and 2.90~eV for $V_\mathrm{Si}$, against 3.91 and 3.56~eV in our DFT and 3.63 and 3.44~eV in the DFT of Hamedani and Sand~\cite{SI:hamedani2025sic}, deviations of $+9.0\%$ and $-18.5\%$ from our DFT and of $+17\%$ and $-16\%$ from the published values.
Tersoff/ZBL gives 4.30 and 8.97~eV; its silicon-vacancy barrier is 2.5 times the DFT value.}

\textbf{Threshold displacement energies.}

\begin{figure}[ht!]
\centering
\sifig{0.72\linewidth}{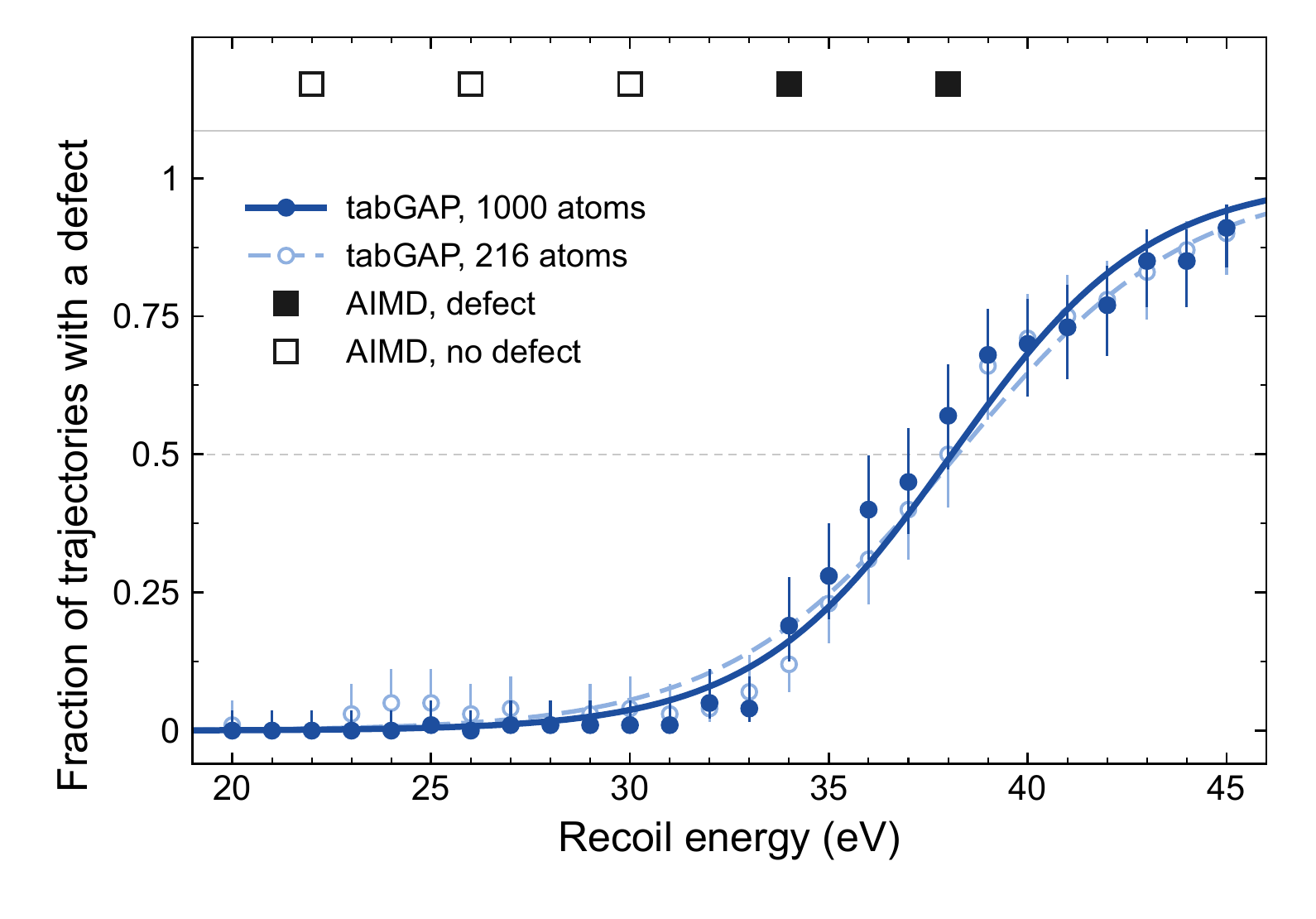}
\caption{\revision{Statistical threshold of the carbon recoil along \hkl[-1-1-1] in 3C-SiC: fraction of the tabGAP trajectories whose quenched frame after 10~ps holds a defect (any lattice site left empty), for 100 thermal start states per recoil energy in 1000-atom (filled circles) and 216-atom (open circles) cells; bars are 95\% intervals and the lines are logistic fits, crossing one half at 38.1 and 38.2~eV; the 216-atom fit uses the first rising branch. Squares: the AIMD trajectories in the 216-atom cell, one per recoil energy, evaluated at 640 fs (filled, a defect survives; open, none).}}
\label{sfig:tde_pe}
\end{figure}

Table~\ref{tab:tde_values} compares the low-index TDEs of 3C-SiC with literature DFT, experiment and other potentials.
\revision{The tabGAP column reports directional values at 30~K. Separately, statistical survival scans examine the cell-size and observation-time dependence, using a Wigner-Seitz analysis of the quenched frame after 10~ps.
For Si\hkl[111], Si\hkl[-1-1-1] and the C directions, we sampled recoil energies in 1~eV steps with 100 thermal seeds in 216-atom cells; C\hkl[100] uses 50 seeds in a 1000-atom cell.
The directional maps in main-text Fig.~6 compare tabGAP and Tersoff/ZBL under the same 30~K protocol. Both potentials give higher mean thresholds for Si than for C recoils, with Tersoff/ZBL shifted to higher energies, particularly for Si.

For C\hkl[-1-1-1], we also compared AIMD and tabGAP in identical 64- and 216-atom cells with a 640~fs observation window.
AIMD used one trajectory per recoil energy from the perfect lattice, while tabGAP used 100 thermal seeds per energy.
In both AIMD cells, a surviving defect first appears at 34~eV.
The tabGAP 50\% survival energies over the same window are 37.7, 37.6 and 37.9~eV in 64-, 216- and 1000-atom cells.
After 10~ps, the values are 38.2 and 38.1~eV in the 216- and 1000-atom cells (Fig.~\ref{sfig:tde_pe}); the 216-atom fit uses the first rising branch.

Against the literature DFT values, five of the six tabGAP entries differ by less than 12\%, including C\hkl[-1-1-1] (39.5 against 38.0~eV), and Si\hkl[-1-1-1] differs by 30\%.
For Si\hkl[100], tabGAP gives 42~eV, compared with the 46~eV DFT value of Lucas and Pizzagalli~\cite{SI:lucas2005ab}.
The mean absolute deviation over all six directions is 2.7~eV (10.9\%).}

\begin{table}[htbp]
  \centering
  \caption{TDE values (in eV) along six low-index directions in 3C-SiC. Si \hkl[111] and Si \hkl[-1-1-1] are the close and open directions; for C recoils the directions are reversed. \revision{The DFT values are from Lucas and Pizzagalli~\cite{SI:lucas2005ab}.}}
  \label{tab:tde_values} \footnotesize \setlength{\tabcolsep}{2pt}
  \begin{ruledtabular}
    \begin{tabular}{l c c c c c c c c c}
    Direction & Exp.\footnote{Refs.~\cite{SI:geiczy1971luminescence,SI:honstvet1980determination,SI:lefevre2009silicon}} & DFT\footnote{Ref.~\cite{SI:lucas2005ab}} & tabGAP & TGAP\footnote{Ref.~\cite{SI:hamedani2025sic}} & Tersoff\footnote{Tersoff/ZBL in Ref.~\cite{SI:devanathan1998displacement}} & DP\footnote{DP-ZBL in Ref.~\cite{SI:liu2024deep}} & MEAM\footnote{Ref.~\cite{SI:kang2014governing}} & EDIP\footnote{Ref.~\cite{SI:jiang2012carbon}} & GW\footnote{GW-ZBL in Ref.~\cite{SI:gao2002empirical}} \\
    \hline
    Si\hkl[100]    & 23.0, 25.0, 45.0 & 46.0  & \revision{42.0}  & 34.0  & 40.0  & 33.5  & 16.0  & 38.0  & 18.0 \\
    Si\hkl[111]    &  --              & 21.0  & \revision{20.8}  & 20.5  & 21.0  & 23.0  & 18.0  & 28.0  & 16.0 \\
    Si\hkl[-1-1-1] & 18.0             & 22.0  & \revision{28.6}  & 48.5  & 35.0  & 43.5  & 24.0  & 24.0  & 32.0 \\
    C\hkl[100]     &  --              & 18.0  & \revision{16.2}  & 16.5  & 14.0  & 15.0  & 22.0  & 18.0  & 16.0 \\
    C\hkl[111]     &   --             & 16.0  & \revision{17.9}  & 18.5  & 21.0  & 15.5  & 32.0  & 18.0  & 14.0 \\
    C\hkl[-1-1-1]  & 22.0             & 38.0  & \revision{39.5}  & 52.5  & 51.0  & 43.5  & 34.0  & 26.0  & 24.0 \\
    \end{tabular}
  \end{ruledtabular}
\end{table}

\clearpage

\subsection*{Supplementary Note 10: Computational speed} \label{SuppNote_10}

\begin{figure}[ht!]
    \centering
    \sifig{\linewidth}{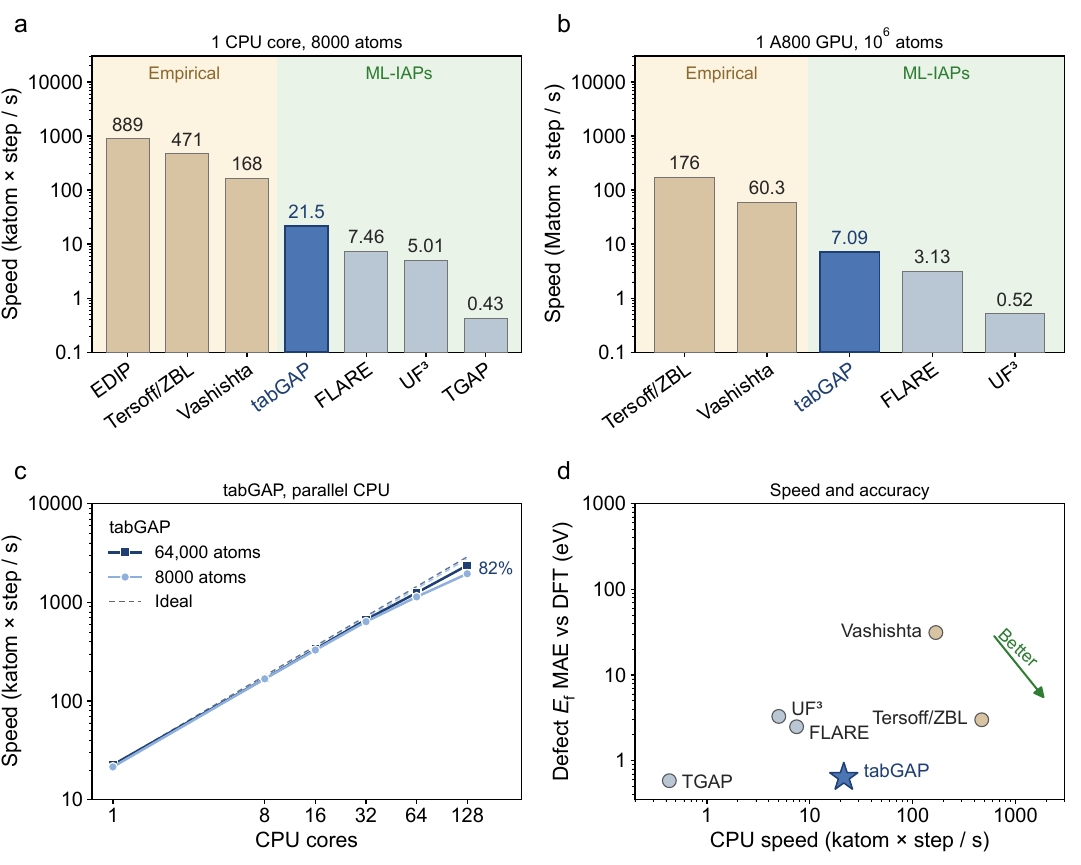}
    \caption{\revision{Computational speed of SiC potentials available in \texttt{LAMMPS}~\cite{SI:devanathan1998displacement,SI:vashishta2007interaction,SI:jiang2012carbon,SI:macisaac2024genetic,SI:hamedani2025sic,SI:xie_incongruent_2026} (logarithmic axes). (a)~One CPU core: all potentials with the same \texttt{LAMMPS} version and compiler flags (FLARE with a build that adds its plugin) on one core of an Intel Xeon Platinum 8358P (2.6~GHz), 8000-atom 3C-SiC ($NVE$, 1~fs, about 300~K); median of three runs (spread $\le 2.6\%$). (b)~One NVIDIA A800 GPU, \texttt{LAMMPS}-\texttt{KOKKOS}, one-million-atom 3C-SiC; median of three runs (spread $\le 1.2\%$); EDIP and TGAP have no GPU implementation. (c)~Parallel scaling of tabGAP on the CPU nodes of (a) for 8000- and 64,000-atom cells, with ideal scaling from one core (dashed) and the parallel efficiency at 128 cores. (d)~Single-core speed from (a) against the mean absolute deviation from our DFT results for the configurations of Fig.~\ref{fig:Ef} (39 for each IAP). tabGAP is the potential of this work..}}
    \label{fig:cost}
\end{figure}

\revision{Figure~\ref{fig:cost} compares the single-core and single-GPU speed of SiC potentials and the CPU parallel scaling of tabGAP.
All potentials were run with the same \texttt{LAMMPS} version (22 Jul 2025, GCC 11.4, \texttt{-O3}) on one core of an Intel Xeon Platinum 8358P (2.6~GHz), for an 8000-atom 3C-SiC cell ($NVE$, 1~fs time step, about 300~K); each value is the median of three runs, whose spread is at most 2.6\%.
tabGAP runs 21.5~thousand atom-steps per second, 4.3 times faster than UF$^{3}$ (5.0), 2.9 times faster than FLARE (7.5) and 50 times faster than TGAP (0.43); the empirical potentials run at 168 (Vashishta), 471 (Tersoff/ZBL) and 889 (EDIP) thousand atom-steps per second. FLARE used the authors' 2026 mapped B2 model (\texttt{lmp.flare}, model repository commit \texttt{24bd275}) for both accuracy and timing tests, with FLARE++ 1.3.0 compiled into the same \texttt{LAMMPS} version using the same compiler flags.}
The tabulated form makes the cost of tabGAP independent of the size of the training set and of the number of sparse points.
\revision{On the same nodes, tabGAP scales to 1.25 and 2.37~million atom-steps per second on 64 and 128 cores for a 64,000-atom cell, parallel efficiencies of 86\% and 82\% relative to one core, and above 92\% up to 32 cores for both the 8000- and the 64,000-atom cell (Fig.~\ref{fig:cost}c).
On one NVIDIA A800 GPU (80~GB PCIe), with the \texttt{KOKKOS} implementations (\texttt{LAMMPS} 11 Feb 2026, CUDA 12.4), tabGAP runs 7.09~million atom-steps per second for a one-million-atom 3C-SiC cell, 13.8 times faster than UF$^{3}$ (0.52), while Tersoff/ZBL and Vashishta run at 176 and 60 (Fig.~\ref{fig:cost}b); energies on the GPU agree with the CPU implementation to $10^{-11}$~eV.
The FLARE++ 1.3.0 \texttt{KOKKOS} pair style was run in a separate \texttt{LAMMPS} 22 Jul 2025 build with a full neighbour list and Newton's third law.
It gives 3.13~million atom-steps per second, with tabGAP approximately 2.3 times faster.
The GPU energies and forces were verified against the CPU implementation.
Figure~\ref{fig:cost}d relates single-core speed to the mean absolute deviation from our DFT results, using all 39 configurations for each of the six IAPs.
The empirical potentials are faster, with deviations of 3.0--31.3~eV; TGAP has a deviation below 1~eV and is 50 times slower than tabGAP.}

\subsection*{Supplementary Note 11: \revision{Comparison of accuracy and computational cost}} \label{SuppNote_11}

\revision{Figure~7 compares the accuracy of six SiC potentials across crystalline properties, point defects, recoil paths and amorphous structure, together with the cost of one atom-step.
Table~\ref{tab:benchmark} gives the mean deviations in physical units and the relative deviations plotted for each category.}

\revision{\textbf{Quantities and references.}
The comparison includes quantities with a first-principles reference and values for tabGAP and at least one other potential.
The references are our DFT calculations, AIMD melt-quench trajectories and the PBE 3C-RS crossing of Ref.~\cite{SI:pertierra2022pressure}.
Each quantity uses the same reference for every potential; experiment is not used in these deviations. Literature TDEs are compared separately in Table~\ref{tab:tde_values} because their simulation protocols differ.
The 100 quantities form the 11 categories of Fig.~7a,b: the lattice constants $a_0$ and $c_0$ of 2H, 4H, 6H and 3C (7); the energy differences $E(\mathrm{2H})-E(\mathrm{3C})$ and $E(\mathrm{RS})-E(\mathrm{3C})$ of the relaxed crystals, each potential at its own minima, which enter the category $\Delta E$ together (2); the 3C-RS transition pressure $P_\mathrm{t}$ (1; tabGAP, FLARE, Vashishta); the elastic constants $C_{ij}$, $B$ and $G$ of the four polytypes (26, DFT with relaxed ions); the formation energies $E_\mathrm{f}$ of the neutral defects (39); the ten quasi-static recoil paths of Fig.~5 (10, DFT); and the pair distribution $g(r)$, the bond-angle distribution $P(\theta)$, the ring-size distribution $R(n)$, the distribution $P(n_\mathrm{homo})$ of like-species (homonuclear) neighbours and the distribution $P(z_\mathrm{Si})$ of the Si coordination at the three densities of Fig.~3 (3 each, AIMD).
The values of the other potentials are those shown in Figs.~\ref{sfig:elastic} and~\ref{fig:Ef}, in Figs.~3 and~5 of the main text, together with their polytype energies, the 3C-RS pressures of FLARE and Vashishta, the amorphous $g(r)$ and bond angles of UF$^{3}$, Vashishta and FLARE and the recoil paths of FLARE, evaluated with the same references and analysis.
}

\revision{\textbf{Amorphous structure.}
Ring, neighbor and coordination distributions were obtained with the melt-quench protocol of Fig.~3: 216-atom cells at $L/L_0=0.95$, 1.00 and 1.05, three seeds each, melted at 5500~K, held at 3000~K and quenched to 300~K in 13.5~ps, and analyzed with the same code, cutoffs and 300 frames.
Vashishta and UF$^{3}$ melt (mean-square displacement in the 3000~K stage of 73--191 and 289--424~\r A$^2$, against about 1~\r A$^2$ for a crystal) and give the distributions of Fig.~7.
Vashishta forms an almost chemically ordered network (like-species bonds account for 2--6\% of all bonds, against 30--49\% in AIMD) and partly recrystallizes at $L/L_0=0.95$ (1.2--1.6 six-membered rings per atom, against 0.97 in AIMD); UF$^{3}$ forms networks rich in like-species bonds (36--79\% of all bonds, Si--Si more than C--C, against 30--49\% in AIMD).

}

\revision{\textbf{Cost.}
Cost is the time per atom-step under the CPU and GPU protocols of Note~10 (Fig.~\ref{fig:cost}). FLARE uses FLARE++ and its \texttt{KOKKOS} implementation; TGAP has no GPU implementation.}

\revision{\textbf{Error measures.}
Scalar errors are $\delta=|x-x_\mathrm{ref}|/|x_\mathrm{ref}|$; distribution errors are $\delta=1-\sum_i\min(p_i,q_i)$, where $p_i$ and $q_i$ are discrete probabilities normalized to sum to one. Recoil-path errors are the mean absolute deviation from DFT divided by the DFT energy rise.
RDF and bond-angle comparisons in Figs.~3 and~7 use 1200-point grids over 1.0--4.0~\r A and $50^\circ$--$180^\circ$, respectively, with normalization over these intervals.
Figure~7a,b shows category means $100\langle\delta\rangle$. Table~\ref{tab:benchmark} gives absolute errors alongside relative errors: for example, TGAP's 22~meV/atom deviation from the 2.9~meV/atom DFT 2H--3C difference gives 760\%.}

\begin{table*}[ht!]
\caption{\revision{Accuracy and computational cost of the SiC potentials in the categories of main-text Fig.~7. $N$ is the number of quantities in each category. The first line gives the mean absolute deviation from the ab initio reference in physical units, and the second the mean deviation in \% plotted in Fig.~7a,b. The last two lines give the atom-step cost of Fig.~7c,d. The formation-energy means use all 39 configurations. The smallest value in each line is bold; -- denotes unavailable data (N/A in the figure).}}
\label{tab:benchmark}
\footnotesize
\begin{ruledtabular}
\begin{tabular}{l c l c c c c c c}
Category & $N$ & Unit & tabGAP & TGAP & FLARE & UF$^{3}$ & Tersoff/ZBL & Vashishta \\ \hline
\multicolumn{9}{l}{\textit{Fig.~7a}} \\
\revision{Lattice constants $a_0$, $c_0$} & \revision{7} & \revision{\%} & \revision{\textbf{0.02}} & \revision{0.30} & \revision{0.46} & \revision{0.50} & \revision{2.3} & \revision{0.76} \\
\revision{Energy differences $\Delta E$} & \revision{2} & \revision{meV/atom} & \revision{\textbf{0.74}} & \revision{78.1} & \revision{30.2} & \revision{49.2} & \revision{222} & \revision{97.1} \\
 & & \revision{\%} & \revision{\textbf{10.0}} & \revision{389} & \revision{609} & \revision{1055} & \revision{79.7} & \revision{286} \\
\revision{3C--RS pressure $P_\mathrm{t}$} & \revision{1} & \revision{GPa} & \revision{\textbf{2.2}} & \revision{--} & \revision{3.7} & \revision{--} & \revision{--} & \revision{24.4} \\
 & & \revision{\%} & \revision{\textbf{3.4}} & \revision{--} & \revision{5.8} & \revision{--} & \revision{--} & \revision{38.0} \\
\revision{Elastic constants $C_{ij}$, $B$, $G$} & \revision{26} & \revision{GPa} & \revision{\textbf{12.8}} & \revision{19.0} & \revision{19.0} & \revision{25.9} & \revision{17.7} & \revision{53.9} \\
 & & \revision{\%} & \revision{\textbf{5.8}} & \revision{9.1} & \revision{9.5} & \revision{24.1} & \revision{12.6} & \revision{39.6} \\
\revision{Formation energies $E_\mathrm{f}$} & \revision{39} & \revision{eV} & \revision{0.64} & \revision{\textbf{0.58}} & \revision{2.5} & \revision{3.3} & \revision{3.0} & \revision{31.3} \\
 & & \revision{\%} & \revision{10.6} & \revision{\textbf{9.7}} & \revision{40.2} & \revision{48.7} & \revision{39.4} & \revision{614} \\
\hline
\multicolumn{9}{l}{\textit{Fig.~7b}} \\
\revision{Recoil-path energies $E_\mathrm{QSD}$} & \revision{10} & \revision{eV} & \revision{\textbf{0.53}} & \revision{0.77} & \revision{5.0} & \revision{5.4} & \revision{7.1} & \revision{355} \\
 & & \revision{\%} & \revision{\textbf{0.73}} & \revision{0.97} & \revision{6.7} & \revision{6.5} & \revision{10.7} & \revision{380} \\
\revision{Pair distribution $g(r)$} & \revision{3} & \revision{\%} & \revision{8.1} & \revision{\textbf{7.6}} & \revision{11.7} & \revision{17.0} & \revision{17.5} & \revision{25.3} \\
\revision{Bond angles $P(\theta)$} & \revision{3} & \revision{\%} & \revision{\textbf{4.6}} & \revision{5.8} & \revision{6.4} & \revision{9.3} & \revision{15.7} & \revision{12.9} \\
\revision{Ring sizes $R(n)$} & \revision{3} & \revision{\%} & \revision{\textbf{4.9}} & \revision{5.2} & \revision{5.8} & \revision{7.7} & \revision{13.9} & \revision{31.9} \\
\revision{Like neighbours $P(n_\mathrm{homo})$} & \revision{3} & \revision{\%} & \revision{\textbf{8.6}} & \revision{9.1} & \revision{16.7} & \revision{32.0} & \revision{15.5} & \revision{69.1} \\
\revision{Si coordination $P(z_\mathrm{Si})$} & \revision{3} & \revision{\%} & \revision{\textbf{7.4}} & \revision{11.8} & \revision{25.0} & \revision{24.8} & \revision{18.7} & \revision{30.6} \\
\hline
\multicolumn{9}{l}{\textit{Fig.~7c,d}} \\
\revision{CPU, one core} & \revision{--} & \revision{$\mu$s/atom-step} & \revision{46.6} & \revision{2313} & \revision{134} & \revision{200} & \revision{\textbf{2.1}} & \revision{5.9} \\
\revision{GPU, one A800} & \revision{--} & \revision{ns/atom-step} & \revision{141} & \revision{--} & \revision{320} & \revision{1939} & \revision{\textbf{5.7}} & \revision{16.6} \\
\end{tabular}
\end{ruledtabular}
\end{table*}

\clearpage 
\begin{table}[htbp]
\centering
\caption{Surviving defects in paired cascade simulations with the current and previous tabGAP (360 trajectories each). Each pair has the same recoil energy, primary knock-on atom (PKA), direction and random seed. Counts are evaluated after 20~ps of cascade evolution, 2~ps of cooling and energy minimisation. Values are mean $\pm$ SEM over $n$ directions. $V_\mathrm{C}$, $V_\mathrm{Si}$ and $N_\mathrm{FP}=N(V_\mathrm{C})+N(V_\mathrm{Si})$ refer to the current tabGAP; the final column gives $N_\mathrm{FP}$ for the previous tabGAP. C and Si PKAs are listed separately; the energy-dependent means quoted in the text give the two species equal weight.}
\label{tab:cascade_full}
\small
\setlength{\tabcolsep}{3.5pt}
\begin{ruledtabular}
\begin{tabular}{l c r r r r r}
$E$ (keV) & PKA & $n$ & $N(V_\mathrm{C})$ & $N(V_\mathrm{Si})$ & $N_\mathrm{FP}$ & $N_\mathrm{FP}$ (previous) \\
\hline
0.1 & C & 60 & $1.22\pm0.07$ & $0.23\pm0.06$ & $1.45\pm0.07$ & $1.43\pm0.08$ \\
0.1 & Si & 60 & $0.62\pm0.07$ & $0.65\pm0.06$ & $1.27\pm0.08$ & $1.47\pm0.10$ \\
0.25 & C & 30 & $2.33\pm0.22$ & $0.93\pm0.14$ & $3.27\pm0.21$ & $3.37\pm0.21$ \\
0.25 & Si & 30 & $2.17\pm0.20$ & $1.13\pm0.11$ & $3.30\pm0.20$ & $3.33\pm0.19$ \\
0.5 & C & 30 & $4.87\pm0.27$ & $1.67\pm0.18$ & $6.53\pm0.28$ & $6.93\pm0.30$ \\
0.5 & Si & 30 & $4.93\pm0.25$ & $2.37\pm0.19$ & $7.30\pm0.32$ & $6.23\pm0.32$ \\
1 & C & 30 & $9.90\pm0.47$ & $3.53\pm0.22$ & $13.43\pm0.41$ & $11.93\pm0.42$ \\
1 & Si & 30 & $10.63\pm0.48$ & $3.47\pm0.22$ & $14.10\pm0.49$ & $12.87\pm0.46$ \\
2 & C & 30 & $20.10\pm0.65$ & $7.47\pm0.36$ & $27.57\pm0.72$ & $23.33\pm0.59$ \\
2 & Si & 30 & $20.77\pm0.64$ & $7.20\pm0.49$ & $27.97\pm0.63$ & $26.60\pm0.79$ \\
\end{tabular}
\end{ruledtabular}
\end{table}

 \clearpage

\phantomsection\label{SIReferences}
\begingroup
\makeatletter
\let\SI@original@label\label
\let\SI@original@ref\ref
\renewcommand{\label}[1]{%
  \def\SI@candidate{#1}%
  \def\SI@lastbib{LastBibItem}%
  \ifx\SI@candidate\SI@lastbib
    \SI@original@label{SILastBibItem}%
  \else
    \SI@original@label{#1}%
  \fi
}
\renewcommand{\ref}[1]{%
  \def\SI@candidate{#1}%
  \def\SI@lastbib{LastBibItem}%
  \ifx\SI@candidate\SI@lastbib
    \SI@original@ref{SILastBibItem}%
  \else
    \SI@original@ref{#1}%
  \fi
}
\makeatother

\endgroup

\end{document}